\documentclass[a4paper,12pt]{article}
\usepackage[left=2cm,right=2cm,top=2.5cm,bottom=2.5cm,]{geometry}
\usepackage[T1]{fontenc}

\usepackage{amsmath}
\usepackage{
    accents, adjustbox, afterpage, amssymb, amsfonts, amsthm, array, booktabs, calc, cancel, caption, color, comment, datetime, dcolumn, dsfont, eurosym, float, geometry, graphicx, lipsum, longtable, mathtools, mathbbol, multirow, pdflscape, pdfpages, placeins, ragged2e, rotating, setspace, subcaption, subfiles, textpos, threeparttable, soul, xcolor, zref-savepos
}

\usepackage[hang, flushmargin]{footmisc}

\usepackage{enumitem}
\usepackage{tikz}
\usetikzlibrary{calc,arrows.meta}

\usepackage[authoryear]{natbib}
\setcitestyle{authoryear,aysep={},open={(},close={)}}
\bibpunct{(}{)}{;}{a}{,}{,}

\usepackage[hypertexnames=false,colorlinks]{hyperref}
\hypersetup{
  pdffitwindow=false,
  colorlinks=true,
  linkcolor={red!50!black},
  citecolor={blue!50!black},
  urlcolor={black},
  linkbordercolor={white},
}

\usepackage{etoolbox}
\input{citepos.tex}

\usepackage{xparse}
\NewDocumentCommand{\hyref}{m O{}O{}}{\hyperref[#1]{#2 \ref{#1}#3}}
\newdateformat{DDMonthYYYY}{%
  \THEDAY\, \monthname[\THEMONTH] \THEYEAR
}

\DeclareMathOperator*{\argmax}{arg\,max}

\theoremstyle{plain}

\theoremstyle{definition}

\newtheorem{hypothesis}{Hypothesis}

\usepackage{dsfont}
\usepackage{fouriernc}
\usepackage{charter}
\usepackage{libertine}

\newcommand\blfootnote[1]{
  \begingroup
  \renewcommand\thefootnote{}\footnote{#1}
  \addtocounter{footnote}{-1}
  \endgroup
}

\usepackage[ruled,vlined]{algorithm2e}
\usepackage{bookmark}

\usepackage[sf,sl,outermarks]{titlesec}
\newcommand{\addperiod}[1]{#1.}

\titleformat{\section}[block]
{\normalfont\Large\bfseries}{\thesection.}{.5em}{\Large\bfseries}
\titlespacing*{\section}{0pt}{*1.3}{*0.2}

\titleformat{\subsection}[block]
{\normalfont\large\bfseries}{\thesubsection.}{.5em}{\large\bfseries}
\titlespacing*{\subsection}{0pt}{*1}{*0}

\titleformat{\subsubsection}[runin]
{\normalfont\bfseries}{}{0em}{\normalsize\bfseries\addperiod}
\titlespacing*{\subsubsection}{0pt}{*1}{*1}

\definecolor{mathematica1}{rgb}{0.368417, 0.506779, 0.709798}
\definecolor{mathematica2}{rgb}{0.880722, 0.611041, 0.142051}

\setlist[enumerate]{leftmargin=*,wide=0pt,itemsep=0pt,topsep=2pt}
\AtBeginDocument{
  \setlength\abovedisplayskip{4pt}
  \setlength\belowdisplayskip{4pt}
}
\newcounter{NoTableEntry}
\renewcommand*{\theNoTableEntry}{NTE-\the\value{NoTableEntry}}

\begin{document}
\thispagestyle{empty}
\setcounter{page}{0}

\setcounter{footnote}{0}
\renewcommand{\thefootnote}{\fnsymbol{footnote}}
\begin{center}\Large
    {\noindent\bfseries
    Incentives and Strategic Behaviour: \\
    An Experiment
    }
\end{center}
\vspace*{1em}

\makebox[\textwidth][c]{
    \begin{minipage}{1.2\linewidth}
        \Large\centering
        Teresa Esteban-Casanelles\footnotemark
        \quad \quad
        Duarte Gonçalves\footnotemark
    \end{minipage}
}
\setcounter{footnote}{1}\footnotetext{
    \setstretch{1} Department of Political Economy, King's College London; \hyperlink{mailto:teresa.estebancasanelles@kcl.ac.uk}{\color{black}teresa.estebancasanelles@kcl.ac.uk}.
}
\setcounter{footnote}{2}\footnotetext{
    \setstretch{1} Department of Economics, University College London; \hyperlink{mailto:duarte.goncalves@ucl.ac.uk}{\color{black}duarte.goncalves@ucl.ac.uk}.
}
\blfootnote{
    We thank
    Larbi Alaoui,
    Marina Agranov,
    Guy Aridor,
    Alessandra Casella,
    Kfir Elias,
    Evan Friedman,
    Navin Kartik,
    Rosemarie Nagel,
    Ryan Oprea,
    Antonio Penta,
    Jacopo Perego,
    Silvio Ravaioli,
    Evan Sadler,
    Karl Schlag,
    Leeat Yariv,
    Michael Woodford,
    and the participants at
    Columbia,
    Essex, 
    SWEET,
    the SAE Symposium,
    the EEA-ESEM Congress,
    the Barcelona Summer Forum, 
    the ESA Conference, and 
    the EAYE Annual Meeting for valuable comments.
    We are particularly grateful to
    Yeon-Koo Che,
    Terri Kneeland,
    Mark Dean for their many insightful conversations.
    This material is based upon work supported by the National Science Foundation under Grant Number 1949395.\\
	\emph{First posted draft}: 27 February 2020. \emph{This draft}: \DDMonthYYYY\today.
}
\setcounter{footnote}{0} \renewcommand{\thefootnote}{\arabic{footnote}}

\begin{center} {\bfseries\large Abstract} \vspace*{1em} \end{center}

\noindent\makebox[\textwidth][c]{
    \begin{minipage}{.85\textwidth}
        \noindent
        How do incentive levels affect strategic behaviour? We address this with an experiment that separately identifies own- and opponent-incentive effects in two dominance-solvable games that differ in strategic complexity. Higher own incentives favour more strategically sophisticated actions and increase best responding to stated beliefs. Beliefs shift in a parallel direction and participants expect more sophisticated opponent actions. Furthermore, while higher own incentives increase belief accuracy, opponents with higher incentives are harder to predict. Higher incentives also increase response times, and longer response times are associated with better performance and more sophisticated actions and beliefs. Taken together, the evidence suggests that incentives affect strategic behaviour through two main channels: by reducing payoff-dependent mistakes and by increasing the effort devoted to reasoning, with the returns to that effort shaped by strategic complexity of the environment.

        \vspace*{1em}
    \end{minipage}
}
\newpage
\section{Introduction}
\label{section:introduction}

A large experimental literature documents persistent deviations from Nash equilibrium and even from rationalisable behaviour, and shows that behaviour in strategic settings responds to changes in relative incentives.\footnote{
    This includes, among many others,
    \citet{Nagel1995AER},
    \citet{Ochs1995GEB},
    \citet{CapraGoereeGomezHolt1999AER},
    \citet{McKelveyPalfreyWeber2000JEBO},
    \citet{Costa-GomesCrawford2006AER},
    \citet{Kneeland2015Ecta}, 
    \citet{AlaouiJanezicPenta2020JET}, and
    \citet{FriedmanWard2022WP}.
}
These findings establish that strategic behaviour departs systematically from standard equilibrium benchmarks and that both choices and beliefs react when the payoff attractiveness of actions changes --- oftentimes against Nash equilibrium predictions. 
They have, in turn, brought to the fore a variety of behavioural models of play in strategic settings, featuring limited reasoning ability \citep{Nagel1995AER,StahlWilson1995GEB,CamererHoChong2004QJE}, payoff-dependent mistakes \citep{McKelveyPalfrey1995GEB}, random beliefs \citep{FriedmanMezzetti2005GEB,Friedman2022AEJMicro}, and costly cognition \citep{AlaouiPenta2016REStud,Goncalves2022WP}. 
More broadly, these models have been used to explain otherwise-puzzling patterns in settings as diverse as auctions \citep{BajariHortacsu2005JPE}, financial bubbles \citep{MoinasPouget2013Ecta}, and responses to monetary policy \citep{FarhiWerning2019AER}.

While a large literature shows that behaviour in games responds to changes in relative incentives, much less is known about how the \emph{level} of incentives itself affects strategic behaviour. 
Strategic settings such as auctions, elections, negotiations, and pricing often differ markedly in the stakes the parties face. 
Understanding how incentive levels affect strategic behaviour matters for both theory and policy. 
On the theory side, evidence on how incentive levels affect behaviour can help distinguish whether the key driver is limited reasoning, payoff-dependent mistakes, random beliefs, or costly cognition. 
On the policy and design side, many interventions work either by changing the strength of incentives or by changing the information available to decision-makers, and the relative effectiveness of these approaches depends on which channel is operative. 

Existing evidence on incentive levels in strategic settings is limited, focuses mostly on whether choices change with incentive levels, and is itself hard to interpret.\footnote{
    A significant share of the literature is dedicated to ultimatum bargaining \citep[see, e.g.,][]{StraubMurnighan1995JEBO,HoffmanMcCabeSmith1996IJGT,SlonimRoth1998Ecta,AndersenErtacGneezyHoffmanList2011AER}, which is prone to other-regarding preferences and cultural norms intervening as a confounding factor \citep{Camerer2003Book}.
    Evidence on other settings includes 
    centipede games \citep{McKelveyPalfrey1992Ecta,RapoportSteinParcoNicholas2003GEB}, 
    coordination games \citep{ParravanoPoulsen2015GEB}, 
    and beauty-contest games \citep[p. 211]{Camerer2003Book}.
} 
The evidence is mixed, with choices responding to incentive levels to different extents across strategic environments, which can reflect multiple channels operating at once. 
Higher incentives make decision mistakes more costly, which may or may not reduce mistake-proneness, but they may also induce more reasoning and affect the beliefs players form. 
Moreover, existing designs confound a player's own-incentive effect with indirect effects operating through higher-order beliefs. 
This is because a player's behaviour may react directly to their own incentive level, through reduced payoff-dependent mistakes or greater effort devoted to reasoning, but it may also react indirectly through beliefs about how opponents respond. 

This paper studies experimentally how incentive levels affect strategic behaviour.
Our design relies on recent advances in experimental methodology to causally identify how own and opponent incentive levels affect actions, beliefs, and response times in strategic settings.
Focusing on dominance-solvable games, we show that higher own incentive levels lead to more strategically sophisticated actions (of higher order of rationality), fewer mistakes, and better responses to stated beliefs. 
Moreover, beliefs are also affected: higher incentives lead participants to expect opponents to play more sophisticated actions, but while higher own incentives improve belief accuracy, opponents facing higher incentives become harder to predict precisely. 
Finally, higher incentive levels significantly increase response time, which, in turn, we find to be positively associated with expected payoffs and both action and belief sophistication. 

Three considerations motivate separating own- and opponent-incentive effects. 
First, different behavioural models make different predictions about which incentive components should matter.\footnote{
    In many standard models, including Nash equilibrium, level-$k$, cognitive hierarchy, and random beliefs, scaling payoffs leaves predicted behaviour unchanged. 
    By contrast, in models with payoff-dependent mistakes, choices react to own incentives when holding fixed the opponent choices, but beliefs do not. 
    In models with costly reasoning, own incentives matter for both choices and beliefs, which should also react to opponents' incentives. 
} 
Moreover, if one varies only overall incentives, then changes in beliefs and higher-order reasoning naturally confound the direct effect of own incentives on behaviour. 
Finally, since many experimental and applied environments involve asymmetric stakes, understanding separately how behaviour responds to own and opponents' incentive levels is useful both for interpretation and for design.

To study these issues, we provide an experimental design that separately identifies the effects of own and opponents' incentive levels on actions, beliefs, and response times.
Participants are randomly assigned to treatments that vary own incentive level, opponents' incentive level, and the game played.
Our identification strategy relies on a simple opponent-matching protocol adapted from the replacement method \citep{AlaouiPenta2016REStud,AlaouiJanezicPenta2020JET}.
For any participant, the action they choose does not enter the payoff determination of the opponent whose behaviour is payoff-relevant for them.
This allows us to vary a participant's own incentive level while holding fixed the behaviour relevant for their decision problem, and conversely to vary opponents' incentive levels while holding fixed the participant's own incentives.
The design also fixes higher-order beliefs, in the sense that participants know not only the incentive level faced by their opponent, but also the incentive level faced by that opponent's opponent, and so on.
We vary incentive levels by randomly assigning participants to a high- or low-bonus treatment, with choices affecting the probability of receiving the corresponding payment bonus.
To focus on initial responses and keep per-decision incentives high-powered, participants complete two unincentivised practice rounds without feedback and then make a single incentivised choice.

We implement this design in two symmetric dominance-solvable normal-form games that differ in strategic complexity. 
We focus on dominance-solvable games because they provide a clean, theory-grounded ranking of actions and beliefs by strategic sophistication. 
In the simpler game, the dominance solution is reached after two rounds of maximal iterated elimination of strictly dominated strategies; in the more complex game, it is reached only after three rounds. 
In both games, level-$k$ actions are identified in a way that is robust to risk attitudes. 
The games share a number of features with standard environments used to study strategic sophistication --- including beauty-contest, undercutting, ring, and related dominance-solvable games \citep{Nagel1995AER,Costa-GomesCrawford2006AER,Kneeland2015Ecta} --- which enables us to relate our results to the existing literature. 
This makes them especially useful for studying whether incentives affect mistakes, higher-order reasoning, and belief formation in a tightly controlled strategic setting. 
This results in a 2$\times$2$\times$2 between-participant design with random assignment, in which over 1600 participants recruited online play only one game and only once.

The paper features four main sets of results. 

First, higher own incentives make actions more sophisticated, but the relevant margin depends on game complexity. 
In both games, higher own incentives reduce dominated play and increase play of the level-2 action. 
The level-2 action corresponds to the dominance-solution action in the simpler game.
In the more complex game, the level-3 action is the dominance-solution action, but its frequency does not change with higher incentives.  
Opponents' incentive levels have qualitatively similar but substantially weaker effects.

Second, the data support a payoff-sensitivity channel consistent with a subjective version of quantal response. 
Actions with higher \emph{subjective} expected payoff (as per stated beliefs) are chosen more often, but this payoff-rank monotonicity does not carry over to \emph{objective} expected payoffs (as given by empirical frequency of opponents' actions). 
Higher own incentives not only substantially increase best-response rates to stated beliefs, but also shift choices toward better responses to participants' reported beliefs. 
A subjective logit quantal-response specification fits the data much better than an analogous specification based on empirical opponent frequencies, underscoring the importance of elicited beliefs for understanding observed choices.

Third, incentives affect beliefs in a way that mirrors actions. 
Higher own incentives reduce the probability assigned to dominated play and shift beliefs toward more sophisticated opponent behaviour. 
In the 2-Step game, this shift extends to the dominance-solution action. 
In the 3-Step game, however, beliefs move primarily toward intermediate level-1 and level-2 actions rather than toward the level-3 dominance solution. 
The effects of opponent incentive level are qualitatively similar, but sharply muted. 
Higher own incentives also improve several dimensions of belief accuracy, whereas opponents facing higher incentives become harder to predict precisely.

Fourth, response times provide suggestive evidence on mechanism. 
Higher own incentives increase response times over 40\% in both games. 
Additionally, controlling for incentives and individual characteristics, longer response times are associated with better performance: higher best-response rates, higher expected payoffs, fewer dominated choices, and more sophisticated beliefs. 
Finally, we explore how response times relate to both actions and beliefs and find that deliberating for longer is associated with more sophisticated actions being chosen and with more sophisticated beliefs being reported. 
Specifically, in both games, the frequency of dominated actions monotonically decreases with time, the level-1 action is the modal action at short response times and its frequency decreases with deliberation, while the level-2 action frequency increases.
Beliefs track these patterns consistently: longer response times are also associated with greater belief mass on both the level-1 and level-2 actions.
Interestingly, the frequency of and the belief about the dominance-solution action in the more complex game, the level-3 action, are fairly invariant with respect to response time. 
We interpret these patterns as suggestive evidence that higher incentives affect behaviour through both reduced payoff-dependent mistakes and greater effort devoted to reasoning, while emphasising that response time is only an informative, though imperfect, proxy for cognitive effort \citep{CaplinCsabaLeahyNov2020QJE,GillProwse2022EJ}.

Our robustness analysis shows that the main patterns are stable within our setup.
The results replicate across participant pools, remain qualitatively similar under alternative specifications, and are not driven by participants with weaker attention or poorer comprehension.
We also discuss explicitly the extent to which alternative explanations based on other-regarding preferences may still matter, while showing that the data are difficult to reconcile with such channels being the main driver of the treatment effects.

These findings should nevertheless be interpreted within the scope of a tightly controlled benchmark environment, precisely one in which direct confounds from feedback, learning, and payoff externalities are reduced.
We therefore do not claim to speak to all strategic environments.
More generally, it is well known that very high or unfamiliar incentive levels can sometimes backfire, so our conclusions should not be extrapolated mechanically to settings with extreme stakes or to environments in which additional motivational channels are likely to be important \citep{CamererHogarth1999JRU,ArielyGneezyLoewensteinMazar06RES,EnkeGneezyHallMartinNelidovOffermanvandeVan2023REStat}.

Within those limits, this paper contributes well-identified evidence on how incentive levels affect reasoning, mistakes, beliefs, and response times in a strategic setting.
By separately identifying own- and opponent-incentive effects, the paper clarifies through which channels higher incentives change behaviour and how those effects depend on strategic complexity.
Taken together, our results suggest that, at familiar incentive levels, higher incentives increase action sophistication both by reducing the propensity to make mistakes and by increasing the incentive to reason further and form better beliefs.
At the same time, the fact that action sophistication and belief accuracy increase in the simpler game, but less so or not at all in the more complex game, implies that one should not expect changes in incentives to have uniform effects across strategic environments that differ in complexity.
Our findings therefore caution both against inferring universally low levels of sophistication from evidence obtained under low-powered incentives and against relying on predictions that require high levels of sophistication in complex but low-stakes environments.

The remainder of the paper proceeds as follows. 
\hyref{section:hypotheses-design}[Section] presents the hypotheses, experimental design, and identification strategy. 
\hyref{section:sophistication}[Sections], \ref{section:mistakes}, \ref{section:beliefs}, and \ref{section:response-time} report the results on actions, payoff-dependent mistakes, beliefs, and response times. 
\hyref{section:robustness}[Section] discusses alternative explanations, robustness checks, and limitations. 
\hyref{section:conclusion}[Section] concludes.

\subsection{Related Literature}
\label{section:introduction:literature}

Our paper contributes to four related strands of literature.

\subsubsection{Strategic Sophistication and Bounded Rationality}

A large literature documents persistent deviations from Nash equilibrium and substantial heterogeneity in strategic sophistication.
This is true in dominance-solvable environments such as beauty-contest, undercutting, and ring games \citep{Nagel1995AER,Costa-GomesCrawford2006AER,Kneeland2015Ecta}, as well as in many other normal-form games \citep{AradRubinstein2012AER,FudenbergLiang2019AER}.
Experimental work also shows that sophistication is not stable across games and depends importantly on what players believe about their opponents \citep{AgranovPotamitesSchotterTergiman2012GEB,GeorganasHealyWeber2015JET}.
These findings motivated several influential models of bounded rationality in games, the common theme of which is that observed play reflects a combination of shallow reasoning, payoff-dependent mistakes, or costly cognition.\footnote{
Prominent examples include level-$k$ and cognitive hierarchy \citep{Stahl1993GEB,StahlWilson1994JEBO,StahlWilson1995GEB,Nagel1995AER,CamererHoChong2004QJE}, cursed equilibrium \citep{EysterRabin2005Ecta}, quantal response equilibrium \citep{McKelveyPalfrey1995GEB,GoereeHoltPalfrey2005EE}, random- and noisy-belief equilibrium \citep{FriedmanMezzetti2005GEB,Friedman2022AEJMicro}, endogenous depth of reasoning \citep{AlaouiPenta2016REStud,AlaouiPentaJPE2022}, and sequential models of costly reasoning \citep{Goncalves2022WP}.
}

Our paper contributes to this literature by studying whether and how incentive \emph{levels} affect choices, beliefs, and response times in an environment where actions and beliefs admit a clean ranking by strategic sophistication.
In particular, the design allows us to distinguish shifts away from dominated play from shifts toward intermediate and fully sophisticated actions, and therefore to separate changes in mistake propensity from changes in the depth of reasoning.
This distinction is central to interpreting the results in the 2-Step and 3-Step games.

\subsubsection{Incentives and Mistakes in Strategic Settings}

A substantial literature studies how behaviour reacts to changes in \emph{relative} incentives across actions.
This includes work on matching-pennies games \citep{Ochs1995GEB,McKelveyPalfreyWeber2000JEBO}, traveler's dilemma \citep{CapraGoereeGomezHolt1999AER}, joker games \citep{MeloPogorelskiyShum2019IER}, and the modified 11--20 game \citep{AlaouiPenta2016REStud,AlaouiJanezicPenta2020JET}.
The broad lesson from this literature is that players react strongly when one action becomes more attractive than another, and that beliefs about opponents often play a central role in mediating these responses.

Closest to our paper is \citet{AlaouiJanezicPenta2020JET}.
Their paper shows that changing \emph{relative} incentives in a replacement design shifts behaviour toward lower numbers and highlights the role of beliefs about opponents' sophistication.
Our paper differs in two key respects.
First, we study \emph{absolute} incentive levels while holding relative incentives between actions fixed.
Second, we separately identify own- and opponent-incentive effects on both choices and beliefs.
Our findings are consistent with theirs in highlighting the central role of beliefs, but they also show that, even holding relative incentives fixed, higher own incentives operate through reduced payoff-dependent mistakes and greater effort devoted to reasoning.

The literature on \emph{absolute} incentive levels in strategic settings is much thinner. 
Existing evidence shows that higher stakes sometimes can move behaviour toward equilibrium benchmarks, for instance in centipede games \citep{McKelveyPalfrey1992Ecta,RapoportSteinParcoNicholas2003GEB}. 
The broader methodological literature has long stressed that stake variation is useful both for assessing robustness and for understanding the trade-off between incentives and decision costs \citep{SmithWalker1993EconInq,CamererHogarth1999JRU}. 
Our contribution is to provide clean evidence on absolute incentives in a strategic setting while jointly observing choices, elicited beliefs, and response times.

Our results also speak to the literature on stochastic choice in games.
Beginning with quantal response equilibrium, a large body of work models strategic behaviour as probabilistic choice --- as in 
discrete choice \citep{Luce1959Book,McFadden1974Chapter} and costly control models \citet{FudenbergIijimaStrzalecki2015Ecta} --- in which better actions are chosen more often, but not always \citep{McKelveyPalfrey1995GEB,MattssonWeibull2002GEB,GoereeHoltPalfrey2005EE}. 
In this class of models, higher incentives should, under suitable conditions, sharpen the relation between expected payoffs and choice probabilities.
We provide evidence in line with this logic, but with an important qualification: monotonicity and best-response improvements are much clearer when expected payoffs are computed using participants' \emph{stated beliefs} than when they are computed using empirical opponent frequencies.
This places beliefs at the centre of any explanation based on payoff-dependent mistakes.

\subsubsection{Beliefs in Strategic Settings}

Our paper also contributes to the literature that elicits beliefs in games and studies how they vary systematically with the strategic environment.
A large empirical literature shows that stated beliefs are informative about behaviour, though far from perfectly aligned with observed choices \citep{NyarkoSchotter2002Ecta,Costa-GomesWeizsacker2008REStud,Rey-Biel2009GEB}.
Related work shows that eliciting beliefs can itself affect play, including in dominance-solvable environments \citep{Croson1999OBHDP,Croson2000JEBO,GachterRenner2010EE}, and that beliefs respond systematically to features of the game and to players' environments \citep{AgranovPotamitesSchotterTergiman2012GEB,FriedmanWard2022WP}.

Our paper contributes to this literature in two ways.
First, we examine directly how own and opponents' incentive levels affect the beliefs participants form about others' actions.
Second, we relate elicited beliefs to observed choices and show that subjective expected payoffs explain behaviour much better than expected payoffs computed from empirical opponent frequencies.
In this sense, the paper supports the view that beliefs are informative about strategic behaviour, but also that understanding how incentives affect beliefs is essential for interpreting observed play.

\subsubsection{Response Times in Strategic Settings}

Finally, our paper relates to the literature using response times to study strategic behaviour.
Response times have been used to classify participants, to predict choices and sophistication, and as an informative proxy for cognitive effort \citep{Rubinstein2007EJ,Rubinstein2016QJE,ProtoRustichiniSofianos2019JPE,SchotterTrevino2021EE,GillProwse2022EJ,FrydmanNunnari2023WP}.
Reviews include \citet{SpiliopoulosOrtmann2018EE} and \citet{Clithero2018JEconPsy}.
Particularly relevant is \citet{Alos-FerrerBuckenmaier2021EE}, who study response times in beauty-contest and 11--20-type games. 

We differ by providing a design that separately identifies how own and opponents' incentive levels affect response times, and by relating response times jointly to actions, elicited beliefs, and payoff performance.
More broadly, our paper connects the strategic response-time literature to the growing evidence on incentives and effort in non-strategic tasks, including \citet{ArielyGneezyLoewensteinMazar06RES}, \citet{CaplinCsabaLeahyNov2020QJE}, \citet{DeanNeligh2023JPE}, and \citet{EnkeGneezyHallMartinNelidovOffermanvandeVan2023REStat}.
Our findings highlight that higher incentives increase response times, but the performance gains are limited by the complexity of the strategic environment.

In short, these connections place the paper at the intersection of the literatures on strategic sophistication, incentives in games, belief formation, and costly reasoning.

\section{Hypotheses and Experimental Design}
\label{section:hypotheses-design}

\subsection{Hypotheses}
\label{section:hypotheses-design:hypotheses}

In this section, we present our hypotheses.
Throughout, we consider finite normal-form games, $\Gamma=\langle I, A, u \rangle$, where $I$ denotes the set of players, $A_i$ the action set of player $i$, $A_{-i}$ the action profiles of player $i$'s opponents, $A:=\times_i A_i$ the set of action profiles, and $u_i:A\to \mathbb R$ player $i$'s payoff function, with $u:={u_i}_{i\in I}$.
As is conventional, we extend payoffs to the space of probabilities over actions, and write $-i$ to denote player $i$'s opponents.

Our main interest is in how incentive levels affect strategic behaviour. 
For any two games $\Gamma$ and $\tilde\Gamma$ that are identical in all except that player $i$'s payoffs are scaled up --- so that $\tilde u_i=\lambda u_i$ for some $\lambda>1$ --- we say that player $i$ has a higher incentive level in $\tilde\Gamma$ than in $\Gamma$.

Our first hypothesis relates to the effect of incentives on action sophistication. 
We use rationalisability to formalise this concept.
Recall that an action is {$k$-rationalisable} if it is a best response to some distribution over opponents' $(k-1)$-rationalisable actions, and that an action is rationalisable if it is $k$-rationalisable for all $k$.\footnote{
    As is well-known \citep[Lemma 3]{Pearce1984Ecta}, in two-player games we can equivalently define player $i$'s $k$-rationalisable actions $R_i^k$ as those surviving $k$ rounds of iterated elimination of strictly dominated strategies:
    $R_i^k:=\{a_i \in A_i\mid \nexists \sigma_i \in \Delta(A_i): u_i(\sigma_i,a_{-i})>u_i(a_i,a_{-i}),\forall a_{-i} \in \times_{j\ne i} R_{j}^{k-1} \}$, with $R_i^0:=A_i$.
} 
We say that action $a_i$ is more sophisticated than action $a_i'$ if (i) there exists some $k$ such that $a_i$ is $k$-rationalisable whereas $a_i'$ is not, or (ii) if these are both $k$-rationalisable but $a_i$ iteratively strictly dominates $a_i'$.

Although experiments routinely find positive choice frequencies for non-rationalisable actions \citep[e.g.][]{Costa-GomesWeizsacker2008REStud,Rey-Biel2009GEB}, it remains unclear whether and how incentive levels affect the sophistication of play. 
Differently from payoff distortions, increasing incentive levels leaves relative incentives unchanged and thus the sets of $k$-rationalisable action distributions unaffected. 
Existing models also yield different predictions. 
In many models --- including Nash equilibrium, level-$k$ \citep{StahlWilson1994JEBO,StahlWilson1995GEB,Nagel1995AER}, and cognitive hierarchy \citep{CamererHoChong2004QJE} --- increasing a player's incentive level has no effect on the predicted distribution of actions. 
In other models, as endogenous depth of reasoning in symmetric games \citep{AlaouiPenta2016REStud}, increasing a player's incentive level may affect that player's actions but not other players.\footnote{
    For instance, if $\Gamma$ is a symmetric two-player game and cognitive costs and anchors are symmetric, scaling up a player's incentive level increases would lead the player to take more steps of reasoning but would not affect predictions about their opponent's actions. 
} 
And other models still make no clear prediction on how action sophistication is affected by incentive levels, as is the case of regular quantal response equilibrium \citep{GoereeHoltPalfrey2005EE}.

Intuition would suggest that by increasing a player's incentive level, not only that player's action sophistication increases, but also that opponents can, to some extent, anticipate this and adjust their behaviour accordingly. 
We formulate this as our first hypothesis:

\begin{hypothesis}[Action Sophistication]
    \label{hypothesis:action-sophistication}
    Action sophistication increases in (a) own incentive level and (b) opponents' incentive level.
\end{hypothesis}

While testing \hyref{hypothesis:action-sophistication}[Hypothesis] is in itself a contribution of this paper, we seek to shed light on mechanisms, that is, \emph{why} incentive levels may affect choices.

One possible channel is that higher incentives make mistakes more costly and therefore reduce the frequency of suboptimal choices. 
This is the logic underlying structural quantal response models, in which the probability of choosing an action depends on the vector of expected payoffs. 
Both models of stochastic choice based on random utility with additive i.i.d. payoff disturbances 
\citep{McFadden1974Chapter,McKelveyPalfrey1995GEB} 
and models based on additive perturbed utility or costly precision \citep{MattssonWeibull2002GEB,FudenbergIijimaStrzalecki2015Ecta} 
predict, under mild conditions, that higher incentives lead players to choose better actions more often. 
More generally, regular quantal response equilibrium implies that, within a given game, actions with higher expected payoffs should be chosen more frequently \citep{GoereeHoltPalfrey2005EE}. 
To assess support for this class of models, we test the following hypothesis: 

\begin{hypothesis}[Quantal Responding]
    \label{hypothesis:best-responses}
    (a) Best-response rates increase in own incentive level.
    (b) Actions with higher expected payoff are chosen more often.
\end{hypothesis}

Another possible channel through which incentives may affect choices is by affecting players' beliefs. 
Previous work shows that beliefs about opponents' sophistication can help explain the choice of non-rationalisable actions. 
For example, sophistication of play changes depending on whether players are informed about opponents' characteristics associated with sophistication, such as chess ratings \citep{Palacios-HuertaVolij2009AER} or their educational background \citep{AgranovPotamitesSchotterTergiman2012GEB,AlaouiJanezicPenta2020JET}. 
In our setting, higher incentives may induce greater effort in understanding the game, leading participants to think more carefully about opponents' payoffs and to assign lower probability to opponents choosing non-rationalisable actions.

The effect of incentive level on beliefs about the opponent may depend on both own and opponent incentive levels. 
On the one hand, if higher incentives induce more sophisticated play, then participants should expect more sophisticated behaviour from opponents facing higher incentives.
On the other hand, if reasoning about the game is costly, then higher own incentives raise the value of forming more accurate beliefs. 
This motivates our next hypothesis:

\begin{hypothesis}[Belief Sophistication]
    \label{hypothesis:belief-sophistication}
    Beliefs about opponents' strategic sophistication increase with (a) own incentive level and (b) opponents' incentive level. 
    Furthermore, (c) belief accuracy increases with own incentive level.
\end{hypothesis}

Finally, we consider whether higher incentives induce greater cognitive effort and whether such effort is associated with better decisions. 
In line with recent models of sequential reasoning in games \citep{AlaouiPenta2016REStud,AlaouiPentaJPE2022,Goncalves2022WP}, we conjecture that increasing a player's incentive level induces greater cognitive effort and may, in turn, result in better choices. 
Although greater effort need not always translate into longer deliberation, as it may instead take the form of more focused thinking, we follow existing literature in treating response time as an informative, even if imperfect, proxy for cognitive effort --- see \citet{GillProwse2022EJ} and references therein. 
Our final hypothesis is: 

\begin{hypothesis}[Response Time]
    \label{hypothesis:response-time}
    (a) Response time increases in one's own incentive level and (b) expected payoff increases with response time.
\end{hypothesis}

\subsection{Experimental Design and Identification Strategy}
\label{section:hypotheses-design:design}

\subsubsection{Identification Strategy}
\label{section:hypotheses-design:design:identification}

Identifying the effect of incentives on behaviour in strategic settings poses a fundamental challenge. 
A higher incentive level for player $i$ may have a first-order, direct effect on the player's behaviour by motivating them to choose better actions, make fewer mistakes, or forming better beliefs. 
But it may also generate indirect effects through higher-order reasoning: player $i$ may expect opponents to respond to $i$'s higher incentives, and may adjust behaviour accordingly. 
To study whether and how incentive levels matter for strategic behaviour, it is therefore important to separate first-order, direct effects from higher-order, indirect effects operating through beliefs about others' responses. 

\begin{figure}[t]
    \small\singlespacing
    \begin{center}
        \begin{tikzpicture}[xscale=2,yscale=2]
    \node[rectangle, draw=black, fill=white, inner sep=5pt, minimum size=3pt, align=left, label=below:{(Low, Low)}] (LL) at (0,1) {
        \parbox{4.5cm}{
            \centering
            Low Own Incentives\\[5pt] 
            \centering
            Low Opponent Incentives\\[3pt]
        }
    };
    \node[rectangle, draw=black, fill=white, inner sep=5pt, minimum size=3pt, align=left, label=above:{(High, Low)}] (HL) at (0,2) {
        \parbox{4.5cm}{
            \centering
            High Own Incentives\\[5pt] 
            \centering
            Low Opponent Incentives\\[3pt]
        }
    };
    \node[rectangle, draw=black, fill=white, inner sep=5pt, minimum size=3pt, align=left, label=below:{(Low, High)}] (LH) at (3,1) {
        \parbox{4.5cm}{
            \centering
            Low Own Incentives\\[5pt] 
            \centering
            High Opponent Incentives\\[3pt]
        }
    };
    \node[rectangle, draw=black, fill=white, inner sep=5pt, minimum size=3pt, align=left, label=above:{(High, High)}] (HH) at (3,2) {
        \parbox{4.5cm}{
            \centering
            High Own Incentives\\[5pt] 
            \centering
            High Opponent Incentives\\[3pt]
        }
    };            
    \draw [-{Stealth[width=15pt, length=15pt,]}, line width=.75pt] (4.25,2) arc (-90:180:4.25mm);
    \draw [-{Stealth[width=15pt, length=15pt,]}, line width=.75pt] (4.25,1) arc (-90:90:4.25mm);
    \draw [-{Stealth[width=15pt, length=15pt,]}, line width=.75pt] (-1.25,1) arc (90:360:4.25mm);
    \draw [-{Stealth[width=15pt, length=15pt,]}, line width=.75pt] (-1.25,2) arc (90:270:4.25mm);
\end{tikzpicture}
    \end{center}
    \begin{minipage}{1\linewidth}
        \caption{Opponent Matching Procedure}
        \label{figure:matching-procedure}
        \emph{Notes}:
        This diagram describes the matching procedure implemented.
        Participants in incentive level groups (High, High) and (Low, Low) have their payoffs depend on an action taken by another random participant in the same group.
        Participants in incentive level groups (High, Low) and (Low, High) have their payoffs depend on an action taken by a random participant in the groups (Low, Low) and (High, High), respectively.
    \end{minipage}
\end{figure}

Our identification strategy relies on a simple opponent-matching protocol, summarised in \hyref{figure:matching-procedure}[Figure]. 
Participants are assigned both a role in the game and an incentive group, where the group determines their own incentive level and the incentive level of the opponent whose behaviour is relevant for their decision problem. 
First, we have two incentive groups --- (High, High) and (Low, Low)  in \hyref{figure:matching-procedure}[Figure] --- in which all players face the same incentive level and know their opponents do as well. 
Second, we have another two incentive groups --- (High, Low) and (Low, High) in \hyref{figure:matching-procedure}[Figure] --- in which participants are matched with opponents with a different incentive level. 
The opponents themselves are randomly drawn from the other two groups, so that participants in (High, Low) are matched with opponents from (Low, Low), and participants in (Low, High) are matched with opponents from (High, High). 
This design allows us (i) to hold fixed the opponent's behaviour while varying own incentives, thereby identifying the effect of own incentives on choices and beliefs; and (ii) to hold fixed own incentives while varying opponents' incentives, thereby identifying how beliefs and behaviour respond to the incentive level faced by opponents. 

Importantly, this procedure also fixes higher-order incentives: it determines not only the incentive level faced by a participant's opponent, but also the incentive level faced by that opponent's opponent, and so on.\footnote{
    Note that simply informing player $i$ about the opponent's incentive level would not suffice, because player $i$ may also care about the incentive level faced by that opponent's opponent when forming beliefs about behaviour. 
    By contrast, our matching procedure fixes the entire higher-order incentive structure relevant for strategic reasoning.
}

This approach corresponds to a version of the replacement method featured in \citet{AlaouiPenta2016REStud} and \citet{AlaouiJanezicPenta2020JET}, and is also related to observer-based designs such as \citet{HuckWeizsaecker2002JEBO}. 
A key feature of the design is that, for participants in the mixed-incentive groups (High, Low) and (Low, High), the action they choose does not enter the payoff determination of the opponent whose behaviour is payoff-relevant for them. 
To avoid making this feature unique to those groups, we implemented the matching protocol so that the same property holds throughout the design: the participant whose action is payoff-relevant for player $i$ is never someone whose own payoff is determined by $i$'s action.\footnote{
    Related designs with this property include ring games \citep{Kneeland2015Ecta}.
}

Given this matching procedure and random assignment across incentive groups, identification of causal effects of incentive levels is obtained by comparing outcomes --- choices, beliefs, and response times --- across participants who differ in own incentives or in opponents' incentives.

\subsubsection{Games}
\label{section:hypotheses-design:design:games}

Participants faced one of the two normal-form games shown in \hyref{figure:games}[Figure].
Both are dominance-solvable and have similar payoff structures. 
The game in panel (a) requires two rounds of (maximal and simultaneous) iterated elimination of strictly dominated strategies to reach the dominance-solution action, whereas the game in panel (b) requires three rounds. 

\begin{figure}[t]
    \centering\small\singlespacing
    \begin{subfigure}{.49\linewidth}
        \begin{tabular}{lc|cccc} 
    & &\multicolumn{4}{c}{Player 2}\\
    \multicolumn{2}{r|}{Actions} & $a_1$ & $a_2$ & $a_3$ & $a_4$ \\
    \hline
    \multirow{4}{*}{Player 1}
    & $a_1$ & 40, 40 & 70, 30 & 80, 20 & 10, 10\\
    & $a_2$ & 30, 70 & 40, 40 & 70, 30 & 80, 20 \\
    & $a_3$ & 20, 80 & 30, 70 & 40, 40 & 70, 30\\
    & $a_4$ & 10, 10 & 20, 80 & 30, 70 & 40, 40\\
\end{tabular}
        \caption{2-Step Game}
        \label{figure:game-iesds2}
    \end{subfigure}
    \begin{subfigure}{.05\linewidth}
    \end{subfigure}
    \begin{subfigure}{.49\linewidth}
        \begin{tabular}{lc|cccc} 
    & &\multicolumn{4}{c}{Player 2}\\
    \multicolumn{2}{r|}{Actions} & $a_1$ & $a_2$ & $a_3$ & $a_4$ \\
    \hline
    \multirow{4}{*}{Player 1}
    & $a_1$ & 40, 40 & 70, 30 & 10, 20 & 10, 10\\
    & $a_2$ & 30, 70 & 40, 40 & 70, 30 & 10, 20 \\
    & $a_3$ & 20, 10 & 30, 70 & 40, 40 & 70, 30\\
    & $a_4$ & 10, 10 & 20, 10 & 30, 70 & 40, 40\\
\end{tabular}
        \caption{3-Step Game}
        \label{figure:game-iesds3}
    \end{subfigure}
    \begin{minipage}{1\linewidth}
        \caption{Games}
        \label{figure:games}
        \emph{Notes}:
        This figure exhibits the games used in the experiment.
        Both are symmetric, two-player dominance-solvable games, with the game in panel (a) taking 2 steps of iterated (maximal and simultaneous) elimination of strictly dominated strategies to obtain the strategy prescribed by the dominance solution, whereas the game in panel (b) takes 3 steps.
    \end{minipage}
\end{figure}

The games have the useful property that actions can be ranked naturally by strategic sophistication. 
First, if $a_{n+1}$ is $k$-rationalisable, then so is $a_n$.
Second, if both $a_{n}$ and $a_{n+1}$ are $k$- but not $(k+1)$-rationalisable, then $a_n$ (iteratedly) strictly dominates $a_{n+1}$.\footnote{
    For instance, in the 2-Step game (\hyref{figure:games}[Figure][(a)]), (i) $a_4$ is strictly dominated by $a_3$, which, in turn, is strictly dominated by $a_2$, and (ii) after eliminating $a_3$ and $a_4$, $a_2$ is iteratedly strictly dominated by $a_1$.
    Similarly, in 3-Step game (\hyref{figure:games}[Figure][(b)]), (i) $a_4$ is strictly dominated by $a_3$; (ii) once $a_4$ is deleted, $a_3$ becomes iteratedly strictly dominated by $a_2$; and (iii) after deleting $a_3$, again $a_2$ is iteratedly strictly dominated by $a_1$.
} 
Accordingly, we will throughout treat $a_n$ as more strategically sophisticated than $a_{n+1}$. 

We note that, assuming a level-0 player uniformly randomises, level-$k$ actions are also uniquely pinned down in a way that does not depend on participants' risk attitudes. 
Specifically, in the 2-Step game, if level 0 uniformly randomises, $a_2$ corresponds to the level-1 action regardless of risk preferences, and $a_1$ to level-2 action; both $a_3$ and $a_4$ are strictly dominated.
In the 3-Step game, the level-1 action is now $a_3$, $a_2$ is level 2, and $a_1$ the level-3 action.

The preceding discussion shows why these games are useful for our purposes. 
Because they admit a clean, theory-grounded ranking of actions and beliefs by strategic sophistication, they allow us to study whether incentives affect mistakes, higher-order reasoning, and belief formation in a tightly controlled environment. 

We specifically chose four-action games because this is the smallest action space that allows us to vary the number of elimination rounds with only minor payoff changes, and we imposed symmetry to improve statistical power and reduce data requirements. 
The games are related to 11--20 \citep{AradRubinstein2012AER} and other undercutting games \citep{Nagel1995AER,Costa-GomesCrawford2006AER,GeorganasHealyWeber2015JET} in that they induce ordered steps of strategic reasoning, but differ in that our games are dominance-solvable and admit a clean ranking of actions by strategic sophistication.

\subsubsection{Other Design Details and Logistics}
\label{section:hypotheses-design:design:logistics}

The experiment implemented a $2\times2\times2$ design, corresponding to own incentive level, opponents' incentive level, and the game played (2-Step or 3-Step game). 
Participants were assigned uniformly at random to one of the eight treatments and played only once. 

The experiment used binary-lottery incentives, so that game payoffs corresponded to the probability of receiving a prize of \$$x$ rather than \$2.00.
We presented exactly the same game representation across incentive conditions.
This allows us to isolate the effect of incentives while holding the strategic structure fixed and avoids potential confounds arising from presenting different payoff numbers, such as perceptual or salience effects. 
In the high-incentive treatment, \$$x$=\$22.00; in the low-incentive treatment, \$$x$=\$2.50. 
We chose the high incentive level to maximise treatment variation, while the low incentive level still exceeds the expected payment per game in several related experiments.\footnote{
    For instance, the average payments per game in \citet{AlaouiJanezicPenta2020JET} and \citet{FudenbergLiang2019AER} were
    \texteuro 0.88 and \$0.93, respectively.
} 

The experimental design induces incentive-compatible elicitation of both actions and beliefs under weak assumptions. 
To this end, we (i) restricted the experiment to a single paid round, thereby also ruling out non-feedback learning \citep{Weber2003GEB}, and (ii) paid either the action task or the belief task, but not both, to avoid hedging concerns \citep{BlancoEngelmannKochNormann2010EE}. 
Actions selected determine game payoffs, which correspond to a probability of getting the prize. 
The order of rows and columns was randomised and participants were informed of this. 
Beliefs $b_i=(b_{i,1},b_{i,2},b_{i,3},b_{i,4}) \in \Delta(A_i)$, interpreted as reported probabilities that the opponent chooses each action, were incentivised using a binarised scoring rule \citep{HossainOkui2013REStud}. 
Choices and beliefs were elicited simultaneously to reduce the mismatch between choices and beliefs as suggested by existing evidence \citep[see][Figure 4]{Costa-GomesWeizsacker2008REStud}, and one of them was randomly selected for payment. 
In the instructions, we also explicitly explained that the payment rule was designed to reward choosing the action that maximises expected points given one's beliefs, as well as to report beliefs truthfully. 

We took several steps to reduce the scope for focal-point effects and other-regarding motives to mechanically account for the observed treatment differences. 
First, we chose games in which the dominance-solution outcome is neither Pareto dominant nor Pareto dominated, as is true for most outcomes in the games. 
Second, for most outcomes, the same payoff pair is associated with multiple action profiles. 
Third, because higher incentives can increase coordination on focal points \citep{ParravanoPoulsen2015GEB}, we randomly shuffled the order of rows and columns, reducing the salience of visually symmetric payoff configurations and action labels. 
And fourth, the matching protocol was random and asymmetric, and this was explicitly explained to participants. 
We return to these concerns in \hyref{section:robustness:alternative-explanations}[Section], where we discuss the extent to which the treatment patterns are consistent with alternative explanations based on other-regarding preferences.

The experiment proceeded as follows:
(i) participants received instructions and completed comprehension and attention checks;
(ii) participants played two unincentivised practice rounds without feedback;\footnote{
    Both practice rounds were four-action two-player games, one with a strictly dominant action and one with no pure-strategy Nash equilibrium. 
}
(iii) own and opponent incentive levels were revealed;\footnote{
    Only 36 participants --- fewer than 1.9\% of the sample --- dropped out after learning their incentive condition, and attrition was nearly balanced across own-incentive treatments: 20 in low incentives and 16 in high incentives.
}
(iv) actions and beliefs were elicited simultaneously; and
(v) participants completed a short socio-demographic questionnaire and received payment information.
Screenshots of the interface and instructions are provided in
\hyref{section:o-appendix:screenshots}.

We recruited 834 participants on Amazon Mechanical Turk (MTurk) in 8-10 and 15 January 2020 and, for replication and robustness purposes, 880 participants on Prolific in 22-23 January 2026, ensuring 100 participants in each treatment. 
This additional sample also allows us to assess whether the main treatment patterns are robust to a less salient emphasis on incentives.\footnote{
    Following the editor's suggestion, we revised the instructions and interface to reduce the salience of incentive levels and simplified the explanation of the matching procedure.
} 
In both samples, participants were adults residing in the United States with approval rates of at least 95\%. 
We find only minor differences across platforms and therefore report results for the pooled sample; we expand on this in \hyref{section:robustness:robustness}[Section]. 
Participants faced no time limit. 
Average duration and average hourly earnings were approximately 22 minutes and \$24.97 on MTurk, and 17 minutes and \$23.43 on Prolific. 
We also collected socio-demographic information, including age, education, and field of study; see \hyref{section:appendix:additional-figures-tables:sample-demographics} for details.

\section{Incentives and Action Sophistication}
\label{section:sophistication}

We first examine whether and how incentive levels affect observed action sophistication (\hyref{hypothesis:action-sophistication}[Hypothesis]).

Our first observation is that participants do not always play the dominance-solution action, $a_1$.
\hyref{table:sample-avg}[Table] shows substantial heterogeneity in action frequencies across the eight treatments. 
Dominance play ranges from 13.7\% to 56.6\%, while dominated play --- corresponding to actions $a_3$ and $a_4$ in the 2-Step game and action $a_4$ in the 3-Step game --- ranges from 5.9\% to 28.4\%. 
The raw frequencies already suggest that own incentives matter most strongly in the 2-Step game, whereas in the 3-Step game dominance play remains low across all treatments.

\begin{table}[t]\setstretch{1.1}
	\centering\small\singlespacing
	\begin{tabular}{@{\extracolsep{4pt}}cccccccccccc@{}}
\hline\hline
\multicolumn{1}{c}{Game}  &  \multicolumn{2}{c}{Incentives}  &  \multicolumn{4}{c}{Action Frequency}  &  \multicolumn{4}{c}{Average Belief Distribution}  &  \multicolumn{1}{c}{Time} \\
\cline{1-1} \cline{2-3} \cline{4-7} \cline{8-11}  \cline{12-12}
Steps & Own & Opponent & 1 & 2 & 3 & 4 & 1 & 2 & 3 & 4 & (secs) \\
 (1)& (2)& (3)& (4)& (5)& (6)& (7)& (8)& (9)& (10)& (11)& (12)\\
\hline
2 & High & High & 56.59 & 28.29 & 8.78 & 6.34 & 34.02 & 36.11 & 17.51 & 12.35 & 97.1 \\
2 & High & Low & 49.76 & 35.07 & 9.48 & 5.69 & 30.46 & 34.74 & 19.91 & 14.90 & 82.1 \\
2 & Low & High & 37.33 & 36.41 & 14.75 & 11.52 & 28.43 & 32.89 & 21.63 & 17.06 & 62.2 \\
2 & Low & Low & 31.86 & 39.71 & 17.16 & 11.27 & 27.12 & 31.11 & 23.97 & 17.80 & 65.5 \\
3 & High & High & 15.32 & 48.20 & 30.18 & 6.31 & 20.36 & 29.78 & 35.17 & 14.70 & 113.5 \\
3 & High & Low & 13.70 & 42.92 & 37.44 & 5.94 & 21.45 & 26.80 & 33.37 & 18.37 & 93.8 \\
3 & Low & High & 16.38 & 33.62 & 37.93 & 12.07 & 22.70 & 27.99 & 30.55 & 18.77 & 69.9 \\
3 & Low & Low & 14.22 & 32.35 & 42.65 & 10.78 & 22.79 & 26.77 & 31.39 & 19.05 & 54.6 \\
\hline\hline
\end{tabular}
    \begin{minipage}{1\linewidth}
        \caption{Action Frequency, Average Beliefs Distributions, and Response Time}
        \label{table:sample-avg}
        \emph{Notes}:
        This table reports treatment-level summary statistics. 
        Each participant is assigned to one treatment, defined by the game (2-Step or 3-Step game), own incentive level (high or low), and opponents' incentive level (high or low).
        Columns (4)--(7) report action frequencies, columns (8)--(11) report mean belief distributions, and column (12) reports mean response time in seconds.
        Participants play one game and make one incentivised decision.
    \end{minipage}
\end{table}

\begin{table}[!ht]\setstretch{1.1}
	\centering\small\singlespacing
	\begin{tabular}{l@{\extracolsep{4pt}}cccc@{}}
\hline\hline
& \multicolumn{2}{c}{Dominance Play} & \multicolumn{2}{c}{Dominated Play}  \\
\cline{2-3} \cline{4-5} 
& 2 Steps & 3 Steps & 2 Steps & 3 Steps \\
& (1) & (2) & (3) & (4) \\
\hline
Own Incent. & 19.67$^{***}$ & -1.10 & -12.98$^{***}$ & -5.37$^{***}$ \\
 & (3.44) & (2.46) & (2.84) & (2.02) \\ [.25em]
Opp. Incent. & 5.78$^{*}$ & 1.74 & -0.58 & 0.56 \\
 & (3.43) & (2.43) & (2.81) & (1.93) \\ [.25em]

Controls    & Yes & Yes & Yes & Yes \\
\hline
R$^2$ & 0.06 & 0.02 & 0.05 & 0.03 \\
N & 837 & 877 & 837 & 877 \\
\hline\hline
\end{tabular}
    \begin{minipage}{1\linewidth}
        \caption{Incentive Level, Dominance, and Dominated Play (\hyref{hypothesis:action-sophistication}[Hypothesis])}
        \label{table:action-dominance}
        \emph{Notes}:
        This table reports estimates from the regression specified in equation (\ref{equation:dominance}), where the dependent variable is an indicator for whether the participant chose the dominance-solution action (columns (1) and (2)) or a strictly dominated action (columns (3) and (4)). 
        High Own/Opponent Incentives are indicators for whether the participant and their opponent face a high incentive level. 
        2 Steps and 3 Steps denote the two games in the experiment; see \hyref{figure:games}[Figure]. 
        Controls refer to participants' socio-demographic characteristics.
        Heterocedasticity-robust standard errors are reported in parentheses; $^{*}$, $^{**}$, and $^{***}$ denote $p<0.10$, $p<0.05$, and $p<0.01$, respectively.
    \end{minipage}
\end{table}

\begin{figure}[!t]
    \centering\small\singlespacing
    \begin{tabular}{@{\extracolsep{0pt}}cc@{}}
        \begin{subfigure}{.45\linewidth}
            \includegraphics[width=\linewidth]{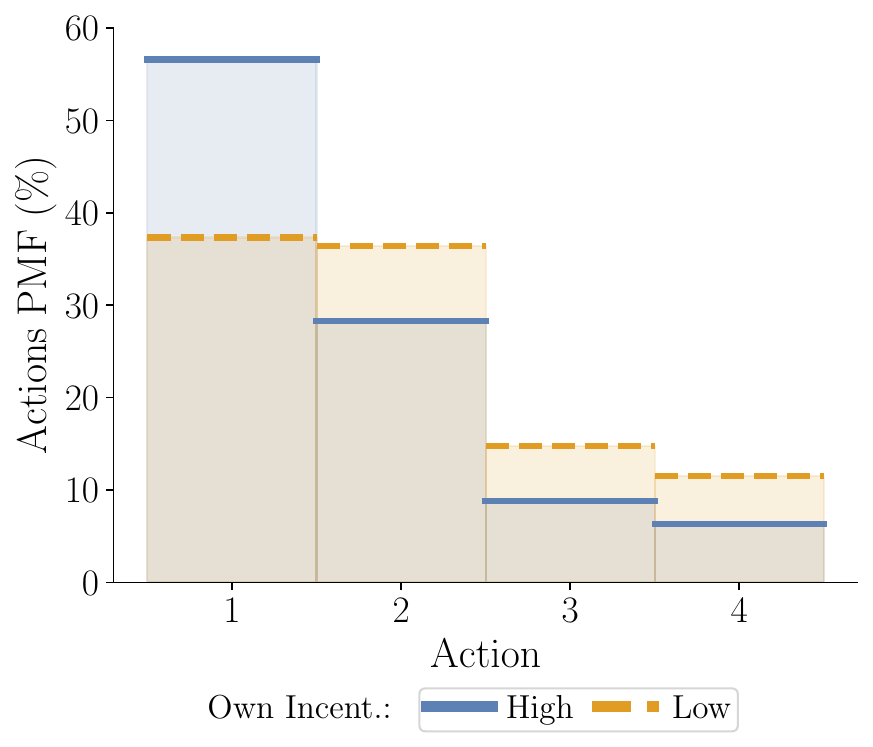}
            \caption{2-Step Game; High Opp. Incentive Level}
            \label{figure:action-sophistication-pmf-iesds2-oppi1}
        \end{subfigure}
        &
        \begin{subfigure}{.45\linewidth}
            \includegraphics[width=\linewidth]{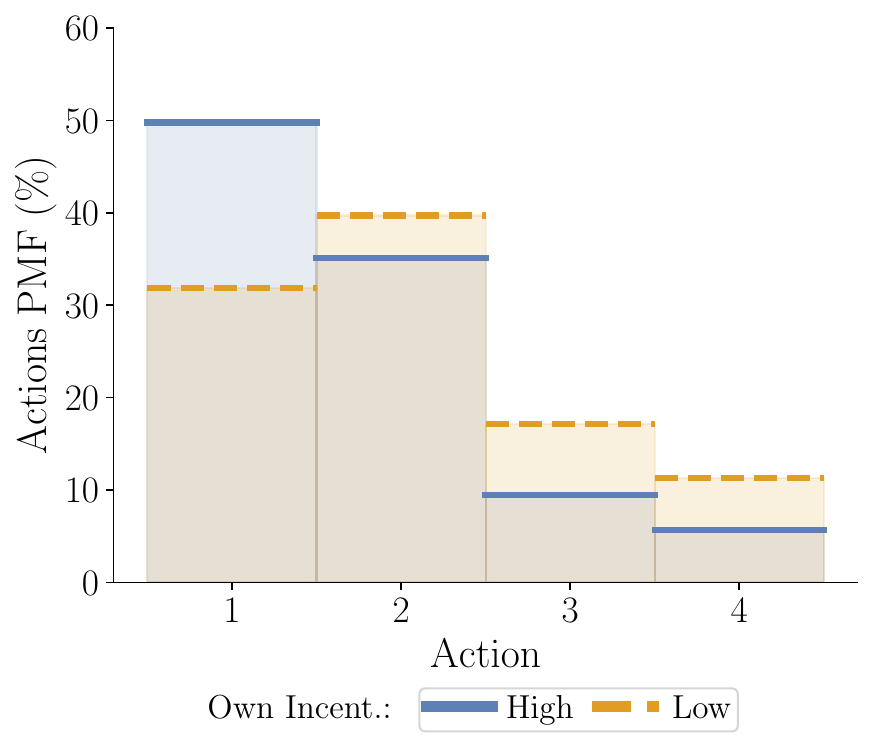}
            \caption{2-Step Game; Low Opp. Incentive Level}
            \label{figure:action-sophistication-pmf-iesds2-oppi0}
        \end{subfigure}
        \\
        \begin{subfigure}{.45\linewidth}
            \includegraphics[width=\linewidth]{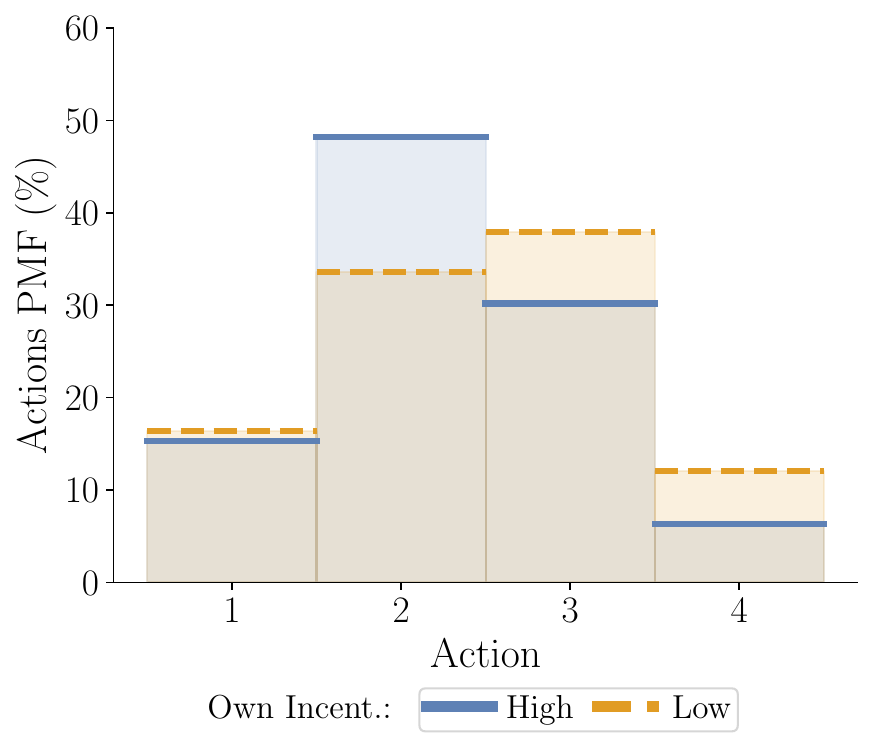}
            \caption{3-Step Game; High Opp. Incentive Level}
            \label{figure:action-sophistication-pmf-iesds3-oppi1}
        \end{subfigure}
        &
        \begin{subfigure}{.45\linewidth}
            \includegraphics[width=\linewidth]{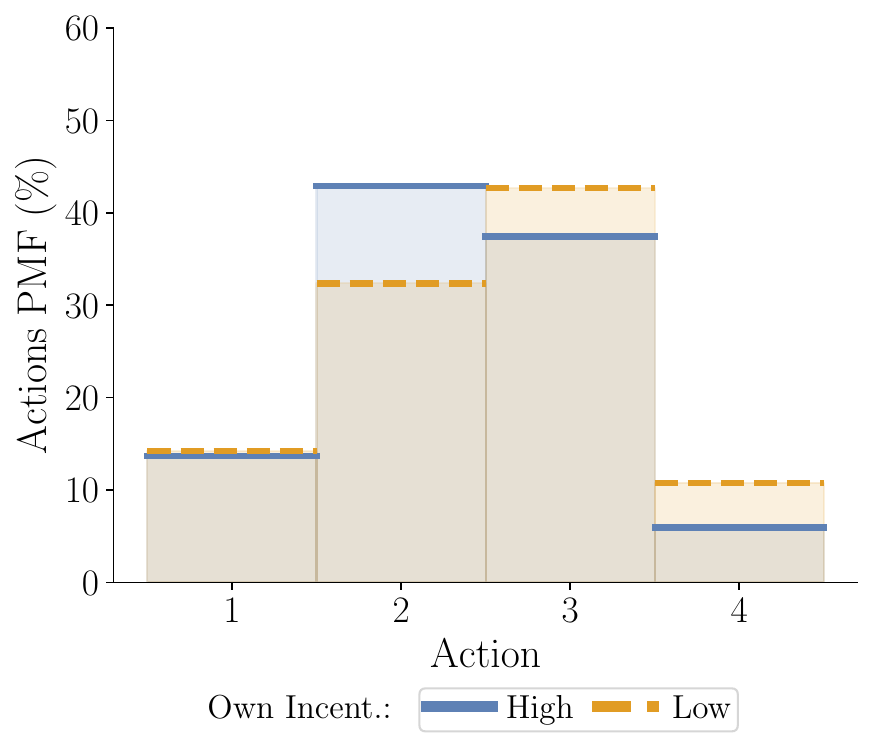}
            \caption{3-Step Game; Low Opp. Incentive Level}
            \label{figure:action-sophistication-pmf-iesds3-oppi0}
        \end{subfigure}
    \end{tabular}
    \begin{minipage}{1\linewidth}
        \caption{Incentive Level and Action Frequency: Own Incentives}
        \label{figure:action-sophistication-pmf-own}
        \emph{Notes}:
        The panels exhibit the action frequency for high and low own incentive levels, for different games (2-Step game, (a) and (b), and 3-Step game, (c) and (d)) and holding fixed opponent incentive level (High, (a) and (c), or Low, (b) and (d)).
        2-Step and 3-Step games denote the different games in the experiment (see \hyref{figure:games}[Figure]).
    \end{minipage}
\end{figure}

We next test whether incentive levels affect the frequency of dominance and dominated play. 
We consider the following specification:
\begin{equation}
    \label{equation:dominance}
    y_i=\beta_0+\beta_1 H_i +\beta_2 H_{-i}+\text{Controls}_i+\epsilon_i
\end{equation}
where $H_i$ is an indicator variable equal to 1 if participant $i$ faces a high own incentive level, and $H_{-i}$ is an analogous indicator equal to 1 if participant $i$'s opponent faces a high incentive level.  
Controls refer to participants' socio-demographic characteristics.\footnote{
    Throughout, we use the same set of controls. 
    Results are robust to omitting controls and to including the interaction between own and opponents' incentive levels. 
    We report the full regression tables with the interaction term in \hyref{section:appendix:additional-figures-tables:interaction}[Appendix]. 
    Where relevant, we also discuss the full treatment-level conditioning shown in the figures.
} 
When testing for the effect of incentive levels on dominance play, the dependent variable $y_i$ is an indicator equal to 1 if participant $i$ chose the dominance-solution action and 0 otherwise. 
When testing the effect of incentives on dominated play, we use the analogous indicator for whether the participant chose a strictly dominated action. 
\hyref{table:action-dominance}[Table] summarises the results.

\begin{figure}[!t]
    \centering\small\singlespacing
    \begin{tabular}{@{\extracolsep{0pt}}cc@{}}
        \begin{subfigure}{.45\linewidth}
            \includegraphics[width=\linewidth]{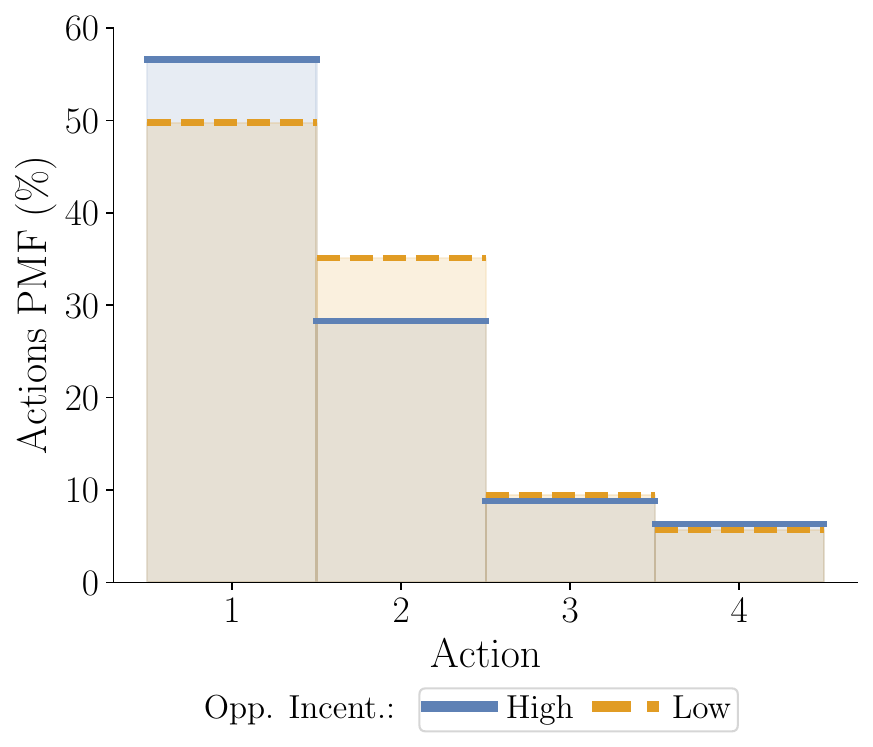}
            \caption{2-Step Game; High Own Incentive Level}
            \label{figure:action-sophistication-pmf-iesds2-owni1}
        \end{subfigure}
        &
        \begin{subfigure}{.45\linewidth}
            \includegraphics[width=\linewidth]{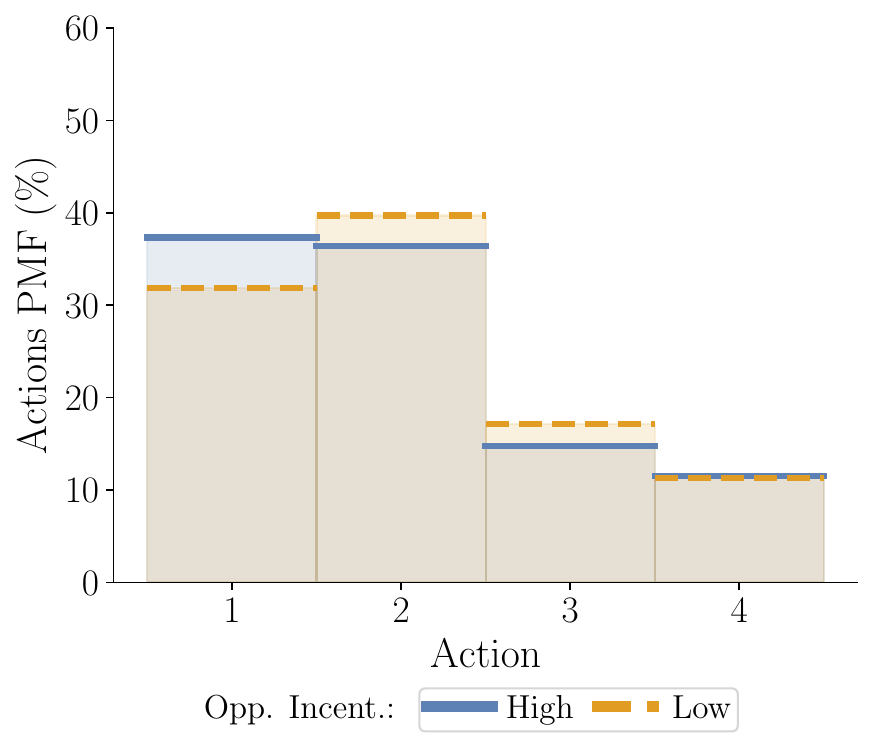}
            \caption{2-Step Game; Low Own Incentive Level}
            \label{figure:action-sophistication-pmf-iesds2-owni0}
        \end{subfigure}
        \\
        \begin{subfigure}{.45\linewidth}
            \includegraphics[width=\linewidth]{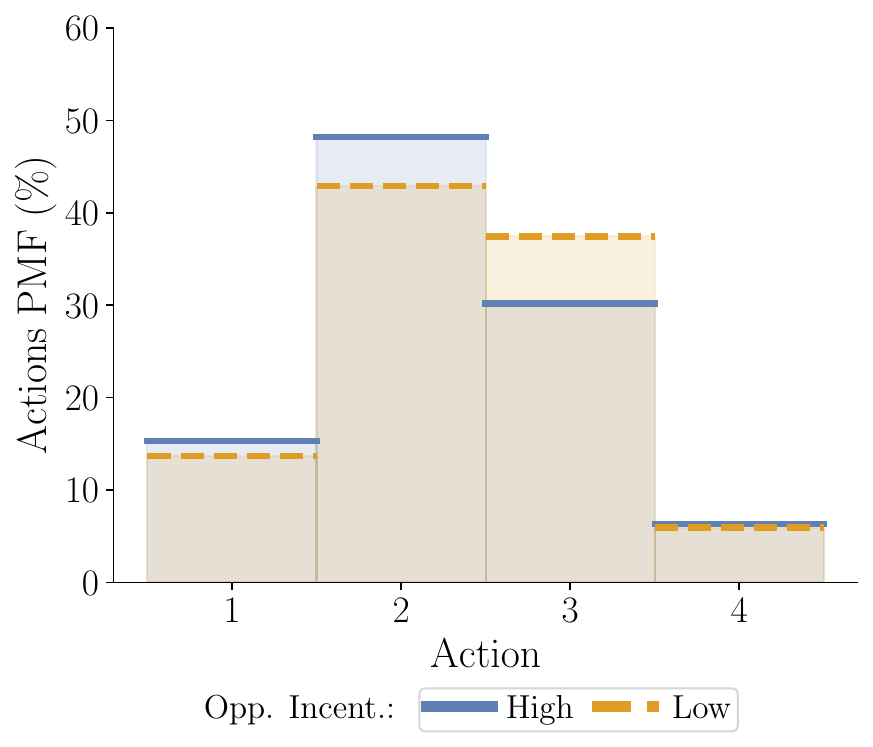}
            \caption{3-Step Game; High Own Incentive Level}
            \label{figure:action-sophistication-pmf-iesds3-owni1}
        \end{subfigure}
        &
        \begin{subfigure}{.45\linewidth}
            \includegraphics[width=\linewidth]{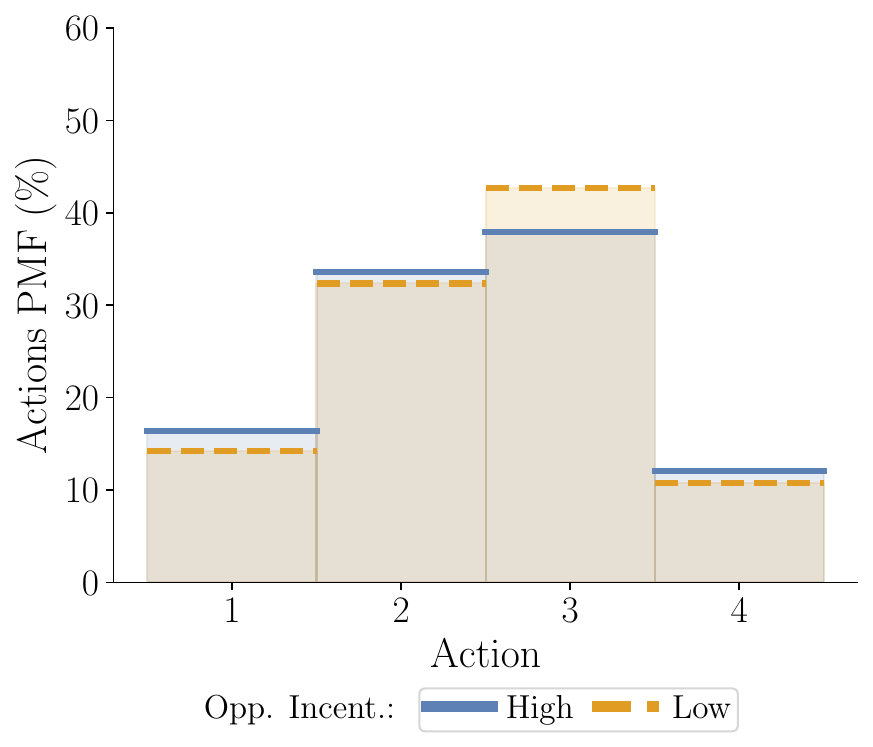}
            \caption{3-Step Game; Low Own Incentive Level}
            \label{figure:action-sophistication-pmf-iesds3-owni0}
        \end{subfigure}
    \end{tabular}
    \begin{minipage}{1\linewidth}
        \caption{Incentive Level and Action Frequency: Opponent Incentives}
        \label{figure:action-sophistication-pmf-opp}
        \emph{Notes}:
        The panels exhibit the action frequency for high and low opponent incentive levels, for different games (2-Step game, (a) and (b), and 3-Step game, (c) and (d)) and holding fixed own incentive level (High, (a) and (c), or Low, (b) and (d)).
        2-Step and 3-Step games denote the different games in the experiment (see \hyref{figure:games}[Figure]).
    \end{minipage}
\end{figure}

Higher own incentives increase action sophistication, but the margin along which this occurs depends on the complexity of the game. 
In the 2-Step game, higher own incentives increase dominance play by 19.67 percentage points (pp) and reduce dominated play by 12.98 pp. 
In the 3-Step game, higher own incentives do not increase dominance play, but still reduce dominated play by 5.37 pp. 
Opponents' incentive levels have much weaker effects: they are associated with a small increase in dominance play in the 2-Step game, but otherwise do not have statistically significant effects. 
One interpretation is that higher incentives help participants avoid clearly inferior actions even in the more complex game, but are not sufficient to increase dominance play in the 3-Step game.

We next consider the full distribution of actions, shown in \hyref{figure:action-sophistication-pmf-own}[Figures] and \hyref{figure:action-sophistication-pmf-opp}.\footnote{
    We report the empirical CDF in \hyref{section:appendix:additional-figures-tables:cdf}[Appendix] --- see \hyref{figure:action-sophistication-cdf-own}[Figures] and \hyref{figure:action-sophistication-cdf-opp}.
} 
For each game and opponent-incentive condition, higher own incentives shift the distribution of actions toward more strategically sophisticated actions (Mann-Whitney U rank tests, $p<.05$). 
In the 2-Step game, this shift also satisfies first-order stochastic dominance, whereas in the 3-Step game the shift is still toward greater sophistication but not in an FOSD sense. 
When considering opponents' incentive levels, the figures suggest a shift away from the level-1 action ($a_2$ in the 2-Step game and $a_3$ in the 3-Step game) toward more sophisticated actions, but these distributional differences are not statistically significant overall.

Overall, the evidence provides clear support for \hyref{hypothesis:action-sophistication}[Hypothesis][a]. 
Higher own incentives make actions more sophisticated: they substantially increase dominance play in the simpler 2-Step game and reduce dominated play in both games. 
This suggests that observed strategic sophistication --- as captured, for example, by types in level-$k$ and cognitive hierarchy models --- depends on the incentive \emph{level} players face. 
By contrast, the evidence for \hyref{hypothesis:action-sophistication}[Hypothesis][b] is much weaker. 
While the treatment distributions are suggestive of greater sophistication when opponents face higher incentives, the corresponding regression estimates are generally small and statistically insignificant.

\section{Incentives and Quantal Responding}
\label{section:mistakes}

One possible reason for the behavioural changes documented above is that, when own incentives are higher, mistakes become more costly and participants therefore choose better responses more often. 
This is the logic underlying \hyref{hypothesis:best-responses}[Hypothesis].

\begin{table}[!ht]\setstretch{1.1}
    \centering\small\singlespacing
	\begin{tabular}{l@{\extracolsep{4pt}}cccc@{}}
\hline\hline
& \multicolumn{2}{c}{Subjective BR} & \multicolumn{2}{c}{Objective BR}  \\
\cline{2-3} \cline{4-5} 
& 2 Steps & 3 Steps & 2 Steps & 3 Steps \\
& (1) & (2) & (3) & (4) \\
\hline
Own Incent. & 20.83$^{***}$ & 15.77$^{***}$ & 19.67$^{***}$ & 12.81$^{***}$ \\
 & (3.44) & (3.43) & (3.44) & (3.33) \\ [.25em]
Opp. Incent. & 5.58 & 3.51 & 5.78$^{*}$ & 2.70 \\
 & (3.44) & (3.39) & (3.43) & (3.33) \\ [.25em]

Controls    & Yes & Yes & Yes & Yes \\
\hline
R$^2$ & 0.07 & 0.05 & 0.06 & 0.04 \\
N & 837 & 877 & 837 & 877 \\
\hline\hline
\end{tabular}
    \begin{minipage}{1\linewidth}
        \caption{Incentive Level and Best-Response Rate (\hyref{hypothesis:best-responses}[Hypothesis][a])}
        \label{table:action-br}
        \emph{Notes}:
        This table reports the results of the regression specified in equation (\ref{equation:dominance}), where the dependent variable is an indicator for whether the participant best responds to their reported beliefs --- that is, chooses the action that maximises subjective expected payoff according to those beliefs (columns (1) and (2)) --- or best responds to the empirical frequency of opponents' actions --- that is, an objective best response (columns (3) and (4)). 
        High Own/Opponent Incentives correspond to indicators for whether the participant and their opponent face a high incentive level.
        2 Steps and 3 Steps denote the different games in the experiment (see \hyref{figure:games}[Figure]).
        Controls refer to the participants' socio-demographic characteristics.
        Heterocedasticity-robust standard errors are given in parentheses; $^*$, $^{**}$, and $^{***}$ denote p-values $< 0.1$, $<0.05$, and $<0.01$, respectively.
    \end{minipage}
\end{table}

\begin{figure}[!ht]
    \centering\small\singlespacing
    \begin{tabular}{@{\extracolsep{0pt}}cc@{}}
        \begin{subfigure}{.45\linewidth}
            \includegraphics[width=\linewidth]{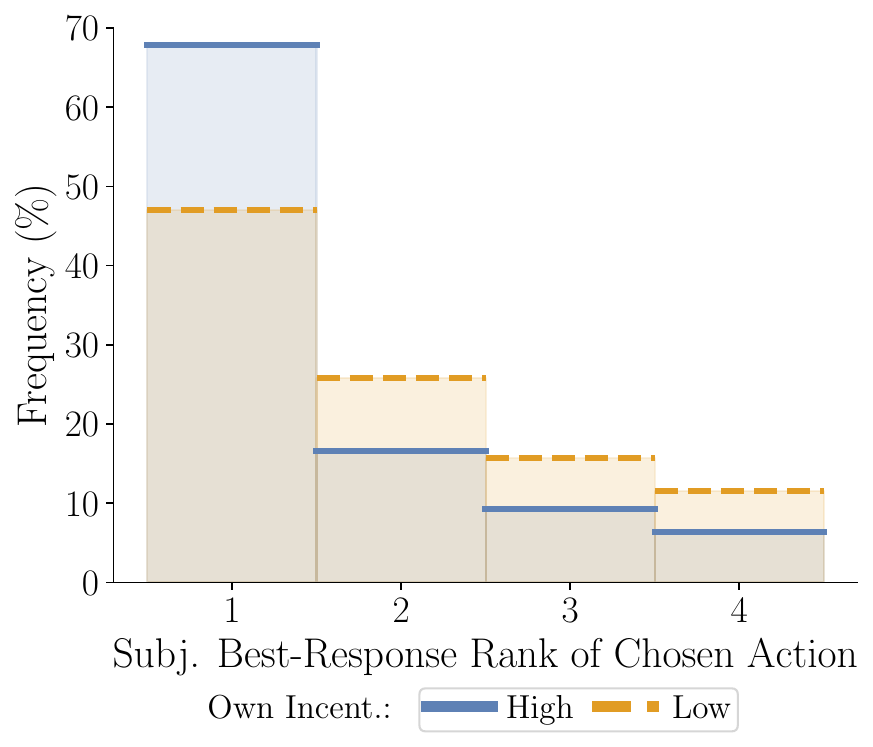}
            \caption{2-Step Game; High Opp. Incentive Level}
            \label{figure:sbr-pmf-iesds2-oppi1}
        \end{subfigure}
        &
        \begin{subfigure}{.45\linewidth}
            \includegraphics[width=\linewidth]{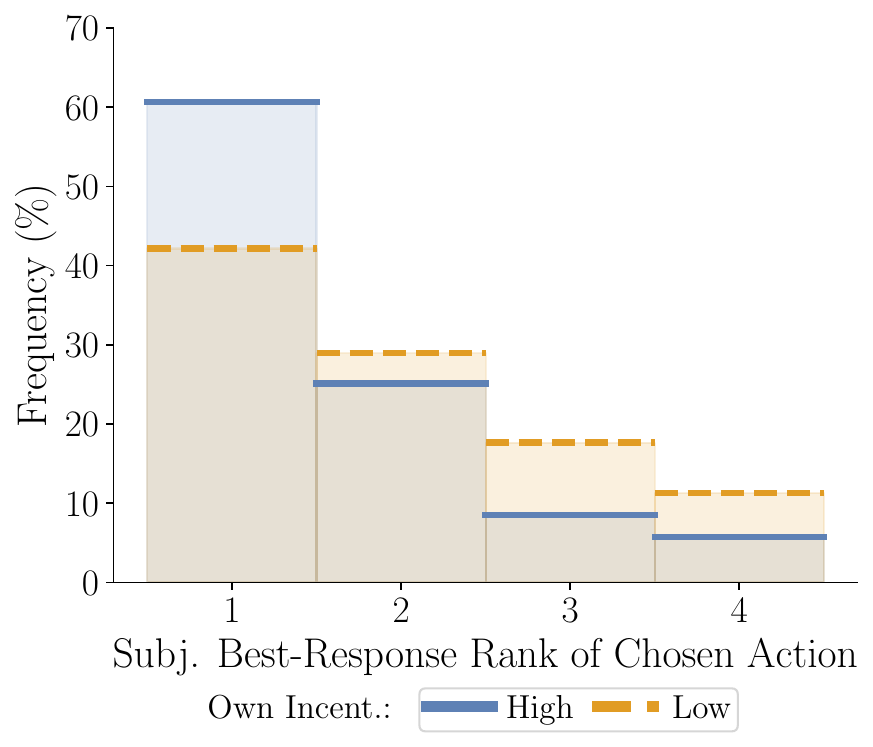}
            \caption{2-Step Game; Low Opp. Incentive Level}
            \label{figure:sbr-pmf-iesds2-oppi0}
        \end{subfigure}
        \\
        \begin{subfigure}{.45\linewidth}
            \includegraphics[width=\linewidth]{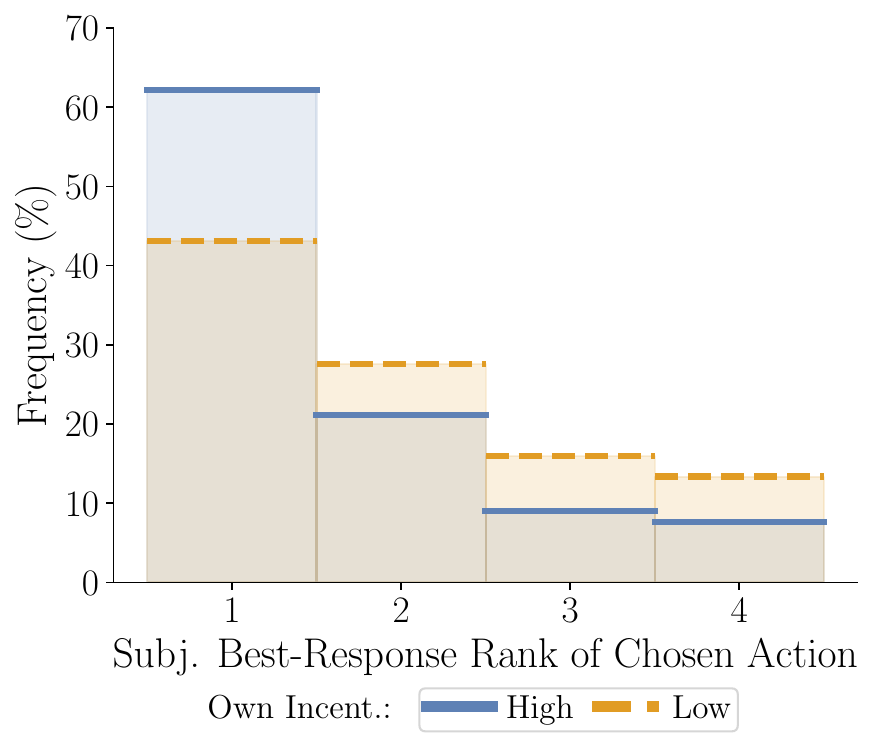}
            \caption{3-Step Game; High Opp. Incentive Level}
            \label{figure:sbr-pmf-iesds3-oppi1}
        \end{subfigure}
        &
        \begin{subfigure}{.45\linewidth}
            \includegraphics[width=\linewidth]{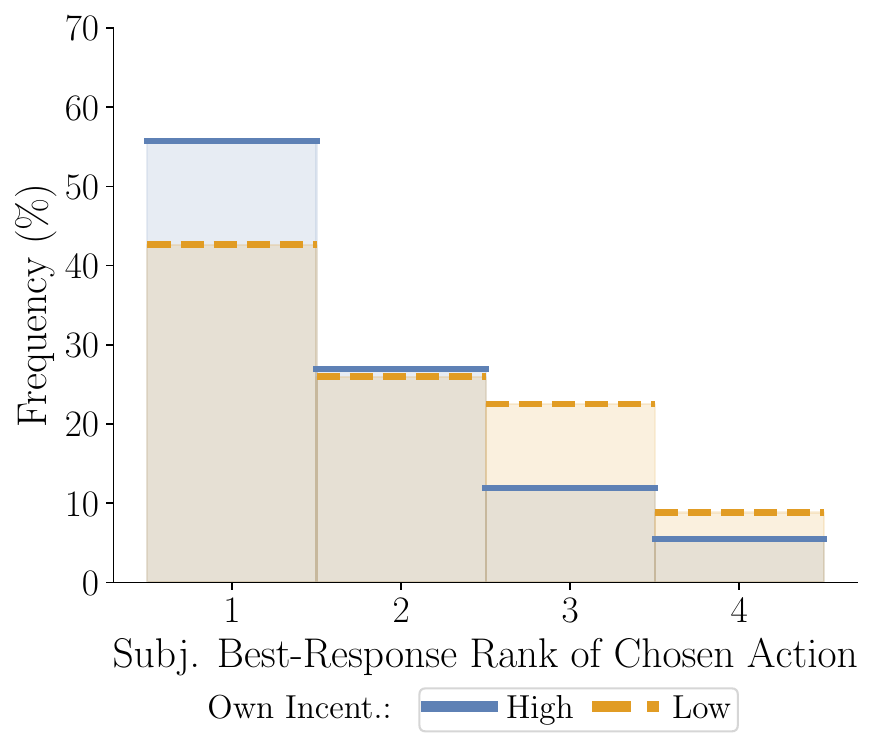}
            \caption{3-Step Game; Low Opp. Incentive Level}
            \label{figure:sbr-pmf-iesds3-oppi0}
        \end{subfigure}
    \end{tabular}
    \begin{minipage}{1\linewidth}
        \caption{Incentive Level and Subjective Best-Response Rank (\hyref{hypothesis:best-responses}[Hypothesis][b])}
        \label{figure:sbr-pmf}
        \emph{Notes}:
        The panels show the distribution of the subjective rank of the chosen action for high and low own incentive levels, separately by game (2-Step game, (a) and (b), and 3-Step game, (c) and (d)) and holding fixed opponent incentive level (High, (a) and (c), or Low, (b) and (d)).
        The subjective rank of an action is $n$ if the action entails the $n$-th highest subjective expected payoffs according to the reported beliefs; e.g. actions with subjective rank 1 are those that maximise subjective expected payoffs.
        2-Step and 3-Step games denote the different games in the experiment (see \hyref{figure:games}[Figure]).
    \end{minipage}
\end{figure}

We first test whether incentive levels affect the likelihood that participants (i) best respond to their stated beliefs (subjective best responses), and (ii) best respond to the observed frequency of opponents' play (objective best responses). 
In \hyref{table:action-br}[Table], we report estimates from a linear probability model using a specification analogous to equation (\ref{equation:dominance}).
The results show that higher own incentives substantially increase subjective best-response rates in both games: by 20.83 pp in the 2-Step game and by 15.77 pp in the 3-Step game.\footnote{
    Best-response rates to stated beliefs are around 40-50\% under low own incentives and 60-70\% under high own incentives; this corresponds to the frequency of choosing the action with subjective rank 1 in \hyref{figure:sbr-pmf}[Figure]. 
    These figures are comparable to those observed in earlier experiments with two-player three-action games, such as \citet{Costa-GomesWeizsacker2008REStud} and \citet{Rey-Biel2009GEB}.
} 
The effect on objective best-response rates is similar, though somewhat smaller, at 19.67 and 12.81 pp, respectively. 
Opponents' incentive levels do not significantly affect subjective best-response rates; they have only a modest positive effect on objective best-response rates in the simpler 2-Step game.

A central property of quantal response is monotonicity: actions with higher expected payoffs should be chosen more often. 
We therefore next examine how incentives affect the distribution of choices by subjective expected-payoff rank. 
Specifically, we define the subjective best-response rank of the chosen action to be $n$ if the chosen action has the $n$-th highest subjective expected payoff, according to the participant's reported beliefs, among the four available actions. 
\hyref{figure:sbr-pmf}[Figure] displays the resulting choice frequencies by subjective rank.

Participants do choose actions with higher \emph{subjective} expected payoff more often, but the same is not true when choices are ranked instead by \emph{objective} expected payoff. 
In every treatment, actions with higher subjective expected payoff are chosen more frequently than actions with lower subjective expected payoff. 
Moreover, higher own incentives shift choices toward better responses to stated beliefs (Mann-Whitney U rank tests, $p<.01$); in addition, the data do not reject a monotone likelihood-ratio shift, consistent with broad classes of quantal-response models (see \hyref{section:appendix:theory-predictions:quantal-response}[Appendix]).
By contrast, when one defines an analogous rank using empirical opponent frequencies and objective expected payoffs, monotonicity is visibly violated; see \hyref{figure:obr-pmf}[Figure] in \hyref{section:appendix:additional-figures-tables:obr}[Appendix]. 
Overall, the evidence provides strong support for \hyref{hypothesis:best-responses}[Hypothesis], especially when best responding is evaluated relative to participants' stated beliefs rather than to empirical opponent frequencies.

To assess how far quantal responding can account for the data, we fit logit quantal-response models and compare their fitted choice frequencies with the observed data. 
In \hyref{section:appendix:additional-figures-tables:qre}[Table], we report the predicted action frequencies obtained by maximum likelihood estimation.\footnote{
    Since treatments in which participants and their opponents have different incentive levels correspond to off-equilibrium play, we restrict attention to treatments in which own and opponent incentive levels are the same.
}
We compare the data with two fitted specifications: an objective logit quantal-response model, in which expected payoffs are computed using empirical opponent choice frequencies, and a subjective logit model, in which expected payoffs are computed using stated beliefs. 
In both specifications, the estimated payoff-sensitivity parameter is higher when own incentives are higher, as one would expect. 
However, the subjective model fits the data much better: its mean squared error is more than 40\% lower than that of the objective model, underscoring the importance of beliefs in accounting for observed choices. 
In this sense, our findings are consistent with \citet{AlaouiJanezicPenta2020JET} in pointing to a central role for beliefs --- which we also directly elicit --- but differ in showing that, even holding relative incentives fixed, higher own incentives also operate through a reduction in payoff-dependent mistakes. 
These findings also point to beliefs as a key part of the mechanism, which motivates the next section's direct analysis of how incentives affect the beliefs participants form.

\section{Incentives and Belief Sophistication}
\label{section:beliefs}

The previous section showed that, while payoff-dependent mistakes are an important channel through which incentives affect choices, beliefs are also central to understanding observed behaviour. 
In this section, we examine whether incentive levels affect the beliefs participants form (\hyref{hypothesis:belief-sophistication}[Hypothesis]).

\begin{figure}[!ht]
    \centering\small\singlespacing
    \begin{tabular}{@{\extracolsep{0pt}}cc@{}}
        \begin{subfigure}{.45\linewidth}
            \includegraphics[width=\linewidth]{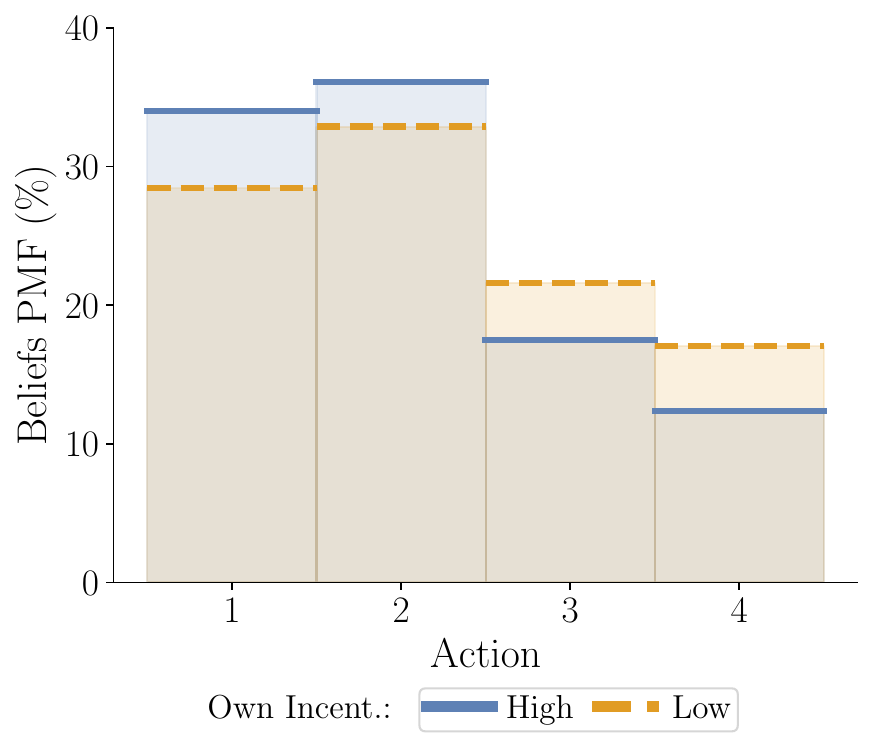}
            \caption{2-Step Game; High Opp. Incentive Level}
            \label{figure:belief-sophistication-pmf-iesds2-oppi1}
        \end{subfigure}
        &
        \begin{subfigure}{.45\linewidth}
            \includegraphics[width=\linewidth]{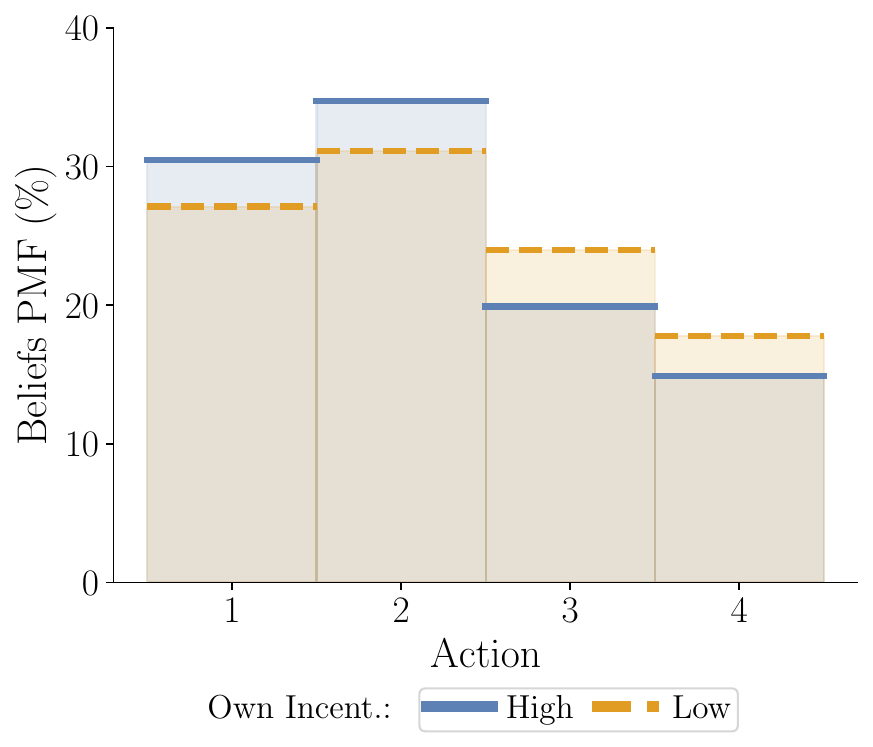}
            \caption{2-Step Game; Low Opp. Incentive Level}
            \label{figure:belief-sophistication-pmf-iesds2-oppi0}
        \end{subfigure}
        \\
        \begin{subfigure}{.45\linewidth}
            \includegraphics[width=\linewidth]{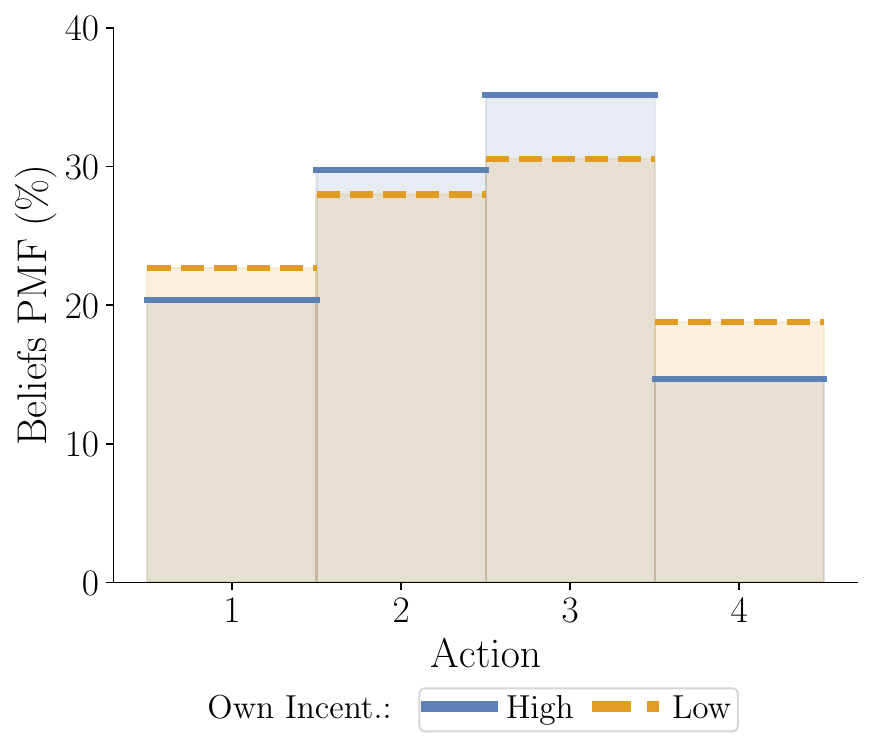}
            \caption{3-Step Game; High Opp. Incentive Level}
            \label{figure:belief-sophistication-pmf-iesds3-oppi1}
        \end{subfigure}
        &
        \begin{subfigure}{.45\linewidth}
            \includegraphics[width=\linewidth]{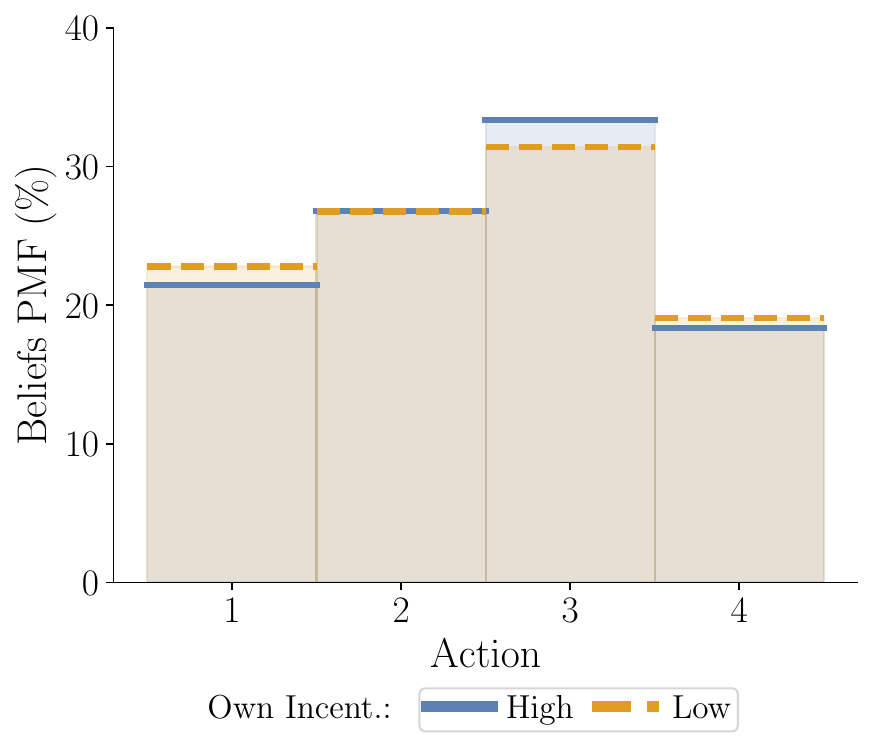}
            \caption{3-Step Game; Low Opp. Incentive Level}
            \label{figure:belief-sophistication-pmf-iesds3-oppi0}
        \end{subfigure}
    \end{tabular}
    \begin{minipage}{1\linewidth}
        \caption{Incentive Level and Beliefs: Own Incentives}
        \label{figure:belief-sophistication-pmf-own}
        \emph{Notes}:
        The different panels exhibit the mean of the reported belief distributions for high and low own incentive levels, for different games (2-Step game, (a) and (b), and 3-Step game, (c) and (d)) and holding fixed opponent incentive level (High, (a) and (c), or Low, (b) and (d)).
        2-Step and 3-Step games denote the different games in the experiment (see \hyref{figure:games}[Figure]).
    \end{minipage}
\end{figure}

\begin{figure}[!ht]
    \centering\small\singlespacing
    \begin{tabular}{@{\extracolsep{0pt}}cc@{}}
        \begin{subfigure}{.45\linewidth}
            \includegraphics[width=\linewidth]{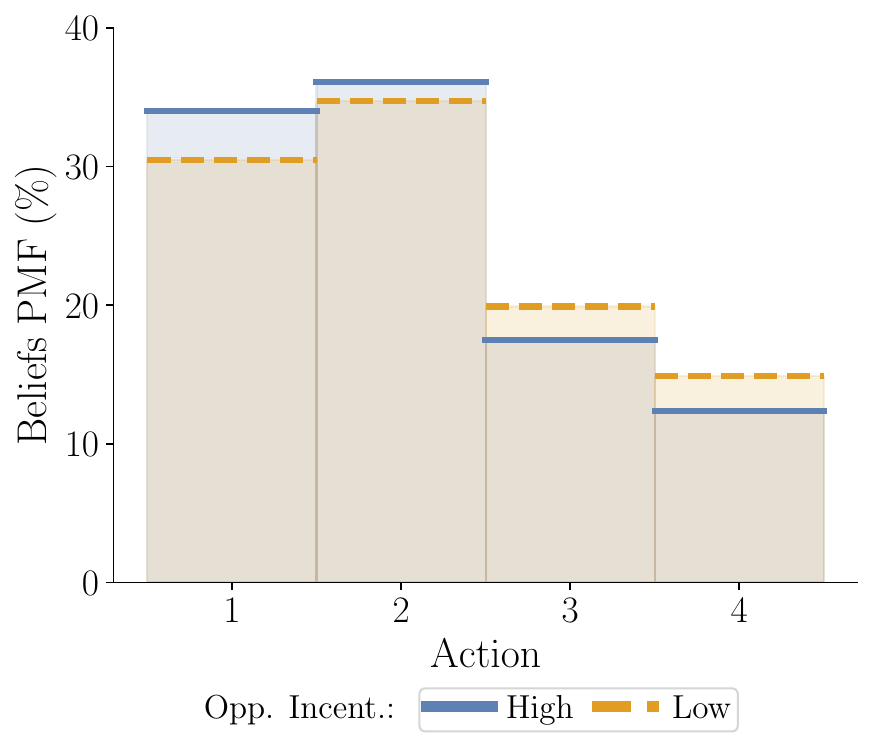}
            \caption{2-Step Game; High Own Incentive Level}
            \label{figure:belief-sophistication-pmf-iesds2-owni1}
        \end{subfigure}
        &
        \begin{subfigure}{.45\linewidth}
            \includegraphics[width=\linewidth]{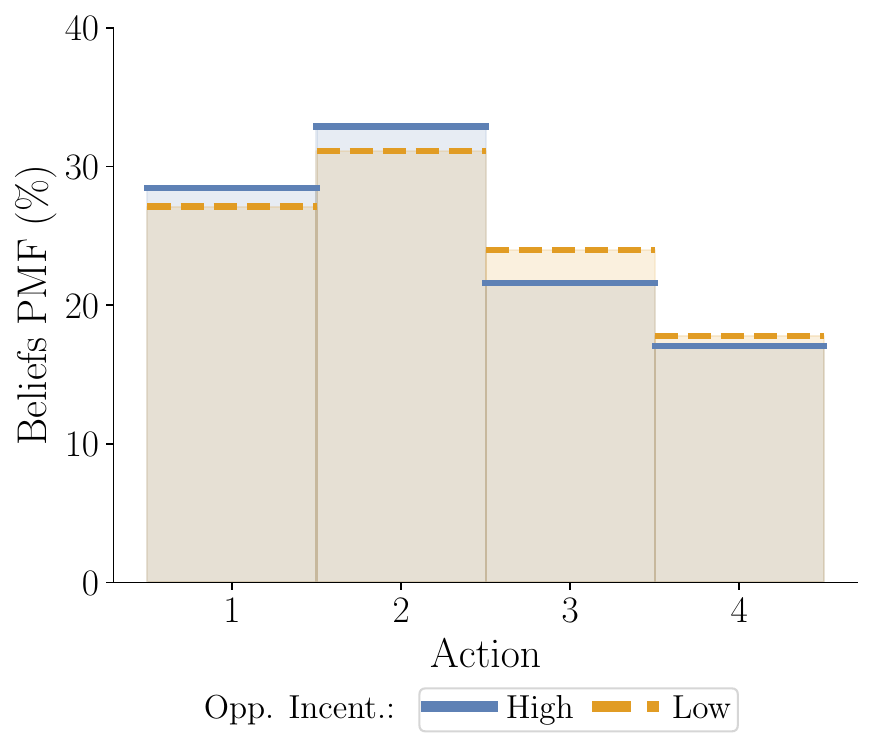}
            \caption{2-Step Game; Low Own Incentive Level}
            \label{figure:belief-sophistication-pmf-iesds2-owni0}
        \end{subfigure}
        \\
        \begin{subfigure}{.45\linewidth}
            \includegraphics[width=\linewidth]{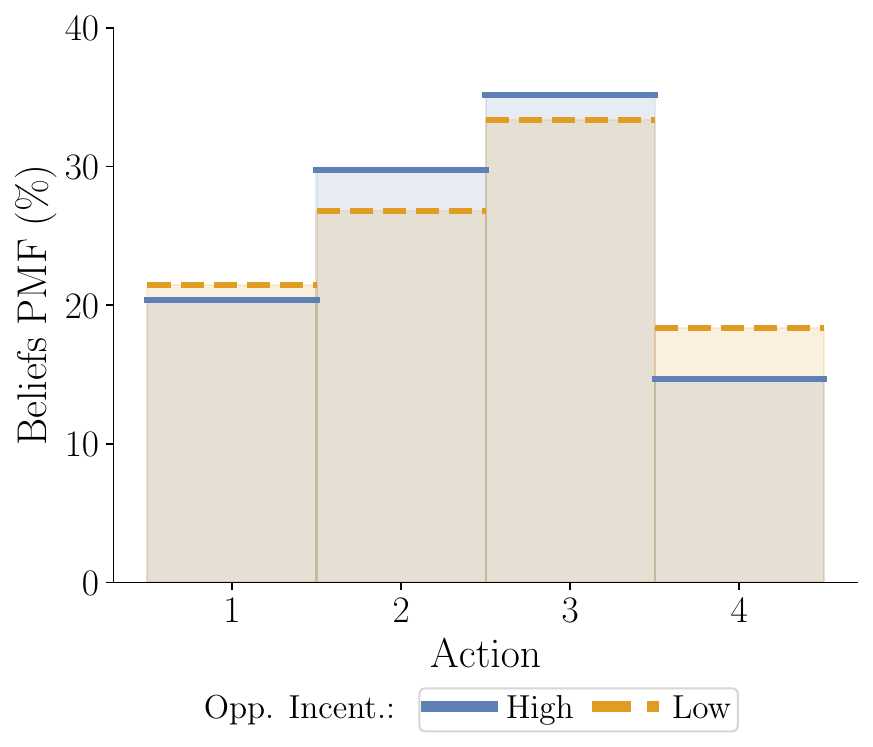}
            \caption{3-Step Game; High Own Incentive Level}
            \label{figure:belief-sophistication-pmf-iesds3-owni1}
        \end{subfigure}
        &
        \begin{subfigure}{.45\linewidth}
            \includegraphics[width=\linewidth]{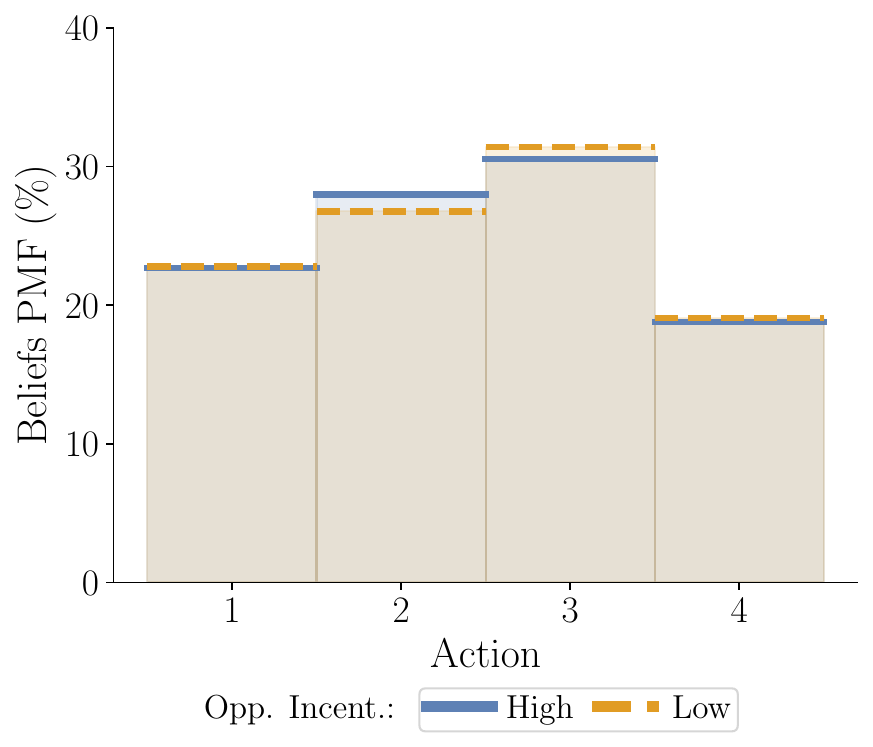}
            \caption{3-Step Game; Low Own Incentive Level}
            \label{figure:belief-sophistication-pmf-iesds3-owni0}
        \end{subfigure}
    \end{tabular}
    \begin{minipage}{1\linewidth}
        \caption{Incentive Level and Beliefs: Opponent Incentives}
        \label{figure:belief-sophistication-pmf-opp}
        \emph{Notes}:
        The different panels exhibit the mean of the reported belief distributions for high and low opponent incentive levels, for different games (2-Step game, (a) and (b), and 3-Step game, (c) and (d)) and holding fixed own incentive level (High, (a) and (c), or Low, (b) and (d)).
        2-Step and 3-Step games denote the different games in the experiment (see \hyref{figure:games}[Figure]).
    \end{minipage}
\end{figure}

Mean reported belief distributions by treatment are shown in \hyref{figure:belief-sophistication-pmf-own}[Figures] and \hyref{figure:belief-sophistication-pmf-opp}.\footnote{
    We report the corresponding mean cumulative belief distributions in \hyref{section:appendix:additional-figures-tables} --- see \hyref{figure:belief-sophistication-cdf-own}[Figures] and \hyref{figure:belief-sophistication-cdf-opp}.
}
In both games, similarly to what occurred with action frequencies, higher own incentives shift mean beliefs toward level-1 and level-2, and reduce the perceived likelihood of dominated play. 
In the simpler 2-Step game, where the dominance solution is the level-2 action, higher own incentives shift mean beliefs toward greater opponent sophistication in a first-order stochastic dominance sense. 
In the more complicated 3-Step game, however, higher own incentives reduce the perceived likelihood of dominated play without increasing the perceived likelihood of the dominance-solution action, which is there the level-3 action. 
Instead, beliefs shift toward assigning greater probability to the intermediate level-1 and -2 actions. 
More generally, higher incentives also make beliefs less uniform.\footnote{
    This is confirmed by regressing the L1 distance of individual belief reports from the uniform distribution on incentive levels; see \hyref{table:belief-uniform}[Table] in \hyref{section:appendix:additional-figures-tables:beliefs-less-uniform}[Appendix].
}
The effects of opponents' incentive levels on reported beliefs are qualitatively similar, but less pronounced. 
Importantly, beliefs seem less responsive to opponents' incentive levels when own incentives are low.

\begin{table}[!t]\setstretch{1.1}
    \centering\small\singlespacing
	\begin{tabular}{l@{\extracolsep{4pt}}cccc@{}}
\hline\hline
& \multicolumn{2}{c}{Belief in}  &  \multicolumn{2}{c}{Belief in} \\
& \multicolumn{2}{c}{Dominance Play} & \multicolumn{2}{c}{Dominated Play}  \\
\cline{2-3} \cline{4-5} 
& 2 Steps & 3 Steps & 2 Steps & 3 Steps \\
& (1) & (2) & (3) & (4) \\
\hline
Own Incent. & 4.68$^{***}$ & -1.82$^{*}$ & -7.67$^{***}$ & -2.46$^{***}$ \\
 & (1.16) & (0.96) & (1.27) & (0.82) \\ [.25em]
Opp. Incent. & 2.04$^{*}$ & -0.71 & -3.76$^{***}$ & -1.82$^{**}$ \\
 & (1.15) & (0.97) & (1.24) & (0.82) \\ [.25em]

Controls    & Yes & Yes & Yes & Yes \\
\hline
R$^2$ & 0.06 & 0.02 & 0.09 & 0.05 \\
N & 837 & 877 & 837 & 877 \\
\hline\hline
\end{tabular}
    \begin{minipage}{1\linewidth}
        \caption{Incentive Level and Belief in Opponent Sophistication (\hyref{hypothesis:belief-sophistication}[Hypotheses][a-b])}
        \label{table:belief-dominance}
        \emph{Notes}:
        This table reports the results of the regression specified in equation (\ref{equation:dominance}), where the dependent variable is the reported belief that the opponent chooses the dominance-solution action, $b_{i,1}$, in columns (1) and (2), or the reported belief that the opponent chooses a strictly dominated action, $b_{i,3}+b_{i,4}$ in column (3) and $b_{i,4}$ in column (4). 
        High Own/Opponent Incentives correspond to indicators for whether the participant and their opponent face a high incentive level.
        2 Steps and 3 Steps denote the different games in the experiment (see \hyref{figure:games}[Figure]).
        Controls refer to the participants' socio-demographic characteristics.
        Heterocedasticity-robust standard errors are given in parentheses; $^*$, $^{**}$, and $^{***}$ denote p-values $< 0.1$, $<0.05$, and $<0.01$, respectively.
    \end{minipage}
\end{table}

To assess the effect of incentives on belief sophistication more formally, we estimate the effect of own and opponents' incentive levels on the reported probability that the opponent chooses the dominance-solution action and on the reported probability that the opponent chooses a strictly dominated action. 
\hyref{table:belief-dominance}[Table] shows that higher own incentives lead participants to attribute greater sophistication to their opponents, but in a nuanced way. 
In the 2-Step game, higher own incentives increase the reported probability that the opponent chooses the dominance-solution action and decrease the reported probability of dominated play. 
In the 3-Step game, by contrast, higher own incentives reduce the perceived probability of dominated play, but do not increase the perceived probability of dominance play. 
Opponents' incentive levels have weaker effects. 
Participants tend to believe that opponents facing higher incentives are less likely to choose dominated actions, but the effects on beliefs about dominance play are small and statistically insignificant. 
Overall, the evidence provides partial support for \hyref{hypothesis:belief-sophistication}[Hypotheses][a-b].

\begin{table}[!ht]\setstretch{1.1}
    \centering\small\singlespacing
	\begin{tabular}{l@{\extracolsep{4pt}}cccccc@{}}
\hline\hline
& \multicolumn{4}{c}{|Belief - Opponent Action Frequency|} & \multicolumn{2}{c}{|Subj. - Obj. Payoff|} \\
& \multicolumn{2}{c}{Dominance Play}  &  \multicolumn{2}{c}{Dominated Play} & & \\
\cline{2-3} \cline{4-5} \cline{6-7} 
& 2 Steps & 3 Steps & 2 Steps & 3 Steps & 2 Steps & 3 Steps \\
& (1) & (2) & (3) & (4) & (5) & (6) \\
\hline
Own Incent. & -1.82$^{**}$ & -0.66 & -5.35$^{***}$ & -1.75$^{***}$ & -4.89$^{***}$ & -1.81$^{**}$ \\
 & (0.85) & (0.73) & (0.89) & (0.64) & (0.91) & (0.82) \\ [.25em]
Opp. Incent. & 16.37$^{***}$ & -1.07 & 5.74$^{***}$ & 1.49$^{**}$ & 8.75$^{***}$ & 6.09$^{***}$ \\
 & (0.85) & (0.74) & (0.87) & (0.64) & (0.90) & (0.82) \\ [.25em]

Controls    & Yes & Yes & Yes & Yes & Yes & Yes \\
\hline
R$^2$ & 0.33 & 0.03 & 0.13 & 0.04 & 0.16 & 0.09 \\
N & 837 & 877 & 837 & 877 & 837 & 877 \\
\hline\hline
\end{tabular}
    \begin{minipage}{1\linewidth}
        \caption{Incentive Level and Belief Accuracy (\hyref{hypothesis:belief-sophistication}[Hypothesis][c])}
        \label{table:belief-accuracy}
        \emph{Notes}:
        This table shows the results for the regression specified in equation (\ref{equation:dominance}), considering as dependent variable the absolute difference between reported belief in dominance play by opponents and its realised frequency, $|b_{i,1}-\overline \sigma_{-i,1}|$, (columns (1) and (2)), or the analogous absolute difference for strictly dominated play, $|b_{i,3}+b_{i,4}-\overline \sigma_{-i,3}-\overline \sigma_{-i,4}|$ in column (3) and $|b_{i,4}-\overline \sigma_{-i,4}|$ in column (4). 
        In columns (5) and (6), the dependent variable is the L1 distance between subjective expected payoffs (computed using reported beliefs) and objective expected payoffs (computed using observed action frequencies). 
        High Own/Opponent Incentives correspond to indicators for whether the participant and their opponent face a high incentive level.
        2 Steps and 3 Steps denote the different games in the experiment (see \hyref{figure:games}[Figure]).
        Controls refer to the participants' socio-demographic characteristics.
        Heterocedasticity-robust standard errors are given in parentheses; $^*$, $^{**}$, and $^{***}$ denote p-values $< 0.1$, $<0.05$, and $<0.01$, respectively.
    \end{minipage}
\end{table}

We next test \hyref{hypothesis:belief-sophistication}[Hypothesis][c], namely whether higher incentives lead participants to form more accurate beliefs. 
\hyref{table:belief-accuracy}[Table] reports regressions for three measures of belief bias. 
Columns (1) and (2) consider the absolute difference between the participant's reported belief that the opponent chooses the dominance-solution action and the corresponding empirical frequency. 
Columns (3) and (4) use the analogous measure for strictly dominated play. 
Columns (5) and (6) consider the absolute difference between subjective expected payoffs, computed using reported beliefs, and objective expected payoffs, computed using observed opponent frequencies: $\sum_{n=1}^4|u_i(a_n,b_i)-u_i(a_n,\overline \sigma_{-i})|$, where $u_i(a_n,b_i)$ denotes participant $i$'s expected payoff from action $a_n$ under their reported beliefs $b_i$ and $u_i(a_n,\overline \sigma_{-i})$ denotes the corresponding expected payoff under the empirical frequency of opponents' actions $\bar\sigma_{-i}$. 
Higher own incentives generally improve belief accuracy. 
They significantly reduce errors in predicting dominated play in both games and also reduce the overall discrepancy between subjective and objective expected payoffs in both games. 
For beliefs about dominance play, the improvement is significant only in the simpler 2-Step game. 
Interestingly, opponents facing higher incentives seem, however, harder to predict: bias in beliefs tends to increase. 
Taken together with \hyref{table:belief-dominance}[Table], this suggests that participants broadly understand the direction in which opponents' behaviour changes with incentives, but misjudge the magnitude of those changes.

Overall, higher own incentives tend to make beliefs both more sophisticated and more accurate, but these effects differ across games and across belief measures. 
Higher own incentives tend to shift beliefs toward attributing greater sophistication to opponents, especially in the simpler 2-Step game, and they improve several dimensions of belief accuracy. 
At the same time, beliefs about opponents facing higher incentives become harder to predict precisely. 
Thus, beliefs are an important part of the mechanism through which incentives affect choices.
This motivates our next analysis of whether higher incentives also affect cognitive effort, as proxied by response times.

\section{Incentives and Response Time}
\label{section:response-time}

Finally, we examine whether incentive levels affect response time. 
We view response time as an informative, though imperfect, proxy for cognitive effort. 
Higher own incentives raise the marginal value of reducing mistakes and refining beliefs, and may therefore induce participants to devote more effort to the decision problem. 
If that additional effort takes the form of longer deliberation, one would expect response times to increase, as conjectured in \hyref{hypothesis:response-time}[Hypothesis][a]. 
We examine the mechanism underlying this hypothesis by regressing log response time on the incentive treatments, using the same baseline specification as before. 
Since greater effort need not always manifest itself in longer deliberation, we therefore interpret the analysis in this section as providing suggestive evidence on mechanism rather than a direct test of cognitive effort.

As \hyref{table:rt}[Table] shows, higher own incentives increase response times substantially in both games: the log-point estimates are 0.38 in the 2-Step game and 0.37 in the 3-Step game, corresponding to increases of roughly 40-45\%.
Opponents' incentives have a mixed effect. 
In the 2-Step game they have essentially no effect on response time, whereas in the 3-Step game they significantly increase response time by almost 20\%. 
These patterns are consistent with the view that higher incentives increase the value of deliberation, although they should not be interpreted as showing that response time is itself a direct measure of cognitive effort. 
Relative to the existing literature, our findings are in line with the data in \citet{Alos-FerrerBuckenmaier2021EE}, where scaling up payoffs tends to increase response time in most variants of the 11-20 games discussed.\footnote{
    \citet{Alos-FerrerBuckenmaier2021EE} use several different versions of the 11-20 game and, for each of these, they scale up payoffs and subtract a constant. 
    We used the published dataset to explore the issue of the effect of incentive levels on response time via linear regression of the latter on the former. 
    The overall average effect is also positive and significant. 
    Separating by type variant, we find that a higher bonus (1) decreases response time only in the low-cost version of their ``baseline'' game and the high-cost version of the ``extreme'' variant, (2) has no effect in the low-cost version of the ``flat'' variant there is no effect, and (3) has a positive and significant effect on the other five versions in their data --- high-cost ``baseline'', low-cost ``extreme'', high-cost ``flat'', and low- and high-cost ``SOCP''. 
    These conclusions are unaffected by controlling or not for participant socio-demographic characteristics and using response time in levels or logs.
} as well as with non-strategic evidence in \citet{CaplinCsabaLeahyNov2020QJE,EnkeGneezyHallMartinNelidovOffermanvandeVan2023REStat}, where higher incentives increase response times, even though performance gains depend on task complexity.

\begin{table}[!ht]\setstretch{1.1}
    \centering\small\singlespacing
	\begin{tabular}{l@{\extracolsep{4pt}}cc@{}}
\hline\hline
& \multicolumn{2}{c}{Log(Time)}  \\
\cline{2-3} 
& 2 Steps & 3 Steps \\
& (1) & (2) \\
\hline
Own Incent. & 0.38$^{***}$ & 0.37$^{***}$ \\
 & (0.05) & (0.05) \\ [.25em]
Opp. Incent. & 0.02 & 0.18$^{***}$ \\
 & (0.05) & (0.05) \\ [.25em]

Controls    & Yes & Yes \\
\hline
R$^2$ & 0.10 & 0.13 \\
N & 837 & 877 \\
\hline\hline
\end{tabular}
    \begin{minipage}{1\linewidth}
        \caption{Incentive Level and Response Time (\hyref{hypothesis:response-time}[Hypothesis][a])}
        \label{table:rt}
        \emph{Notes}:
        This table reports the results of the regression specified in equation (\ref{equation:dominance}), where the dependent variable is the participant's log response time (in seconds).
        High Own/Opponent Incentives correspond to indicators for whether the participant and their opponent face a high incentive level.
        2 Steps and 3 Steps denote the different games in the experiment (see \hyref{figure:games}[Figure]).
        Controls refer to the participants' socio-demographic characteristics.
        Heterocedasticity-robust standard errors are given in parentheses; $^*$, $^{**}$, and $^{***}$ denote p-values $< 0.1$, $<0.05$, and $<0.01$, respectively.
    \end{minipage}
\end{table}

\begin{table}[!ht]\setstretch{1.1}
    \centering\small\singlespacing
	\begin{tabular}{l@{\extracolsep{4pt}}cccc@{}}
\hline\hline
& \multicolumn{2}{c}{Objective BR} & \multicolumn{2}{c}{Log(Obj. Exp. Payoff}  \\
\cline{2-3} \cline{4-5} 
& 2 Steps & 3 Steps & 2 Steps & 3 Steps \\
& (1) & (2) & (3) & (4) \\
\hline
Log(Time) & 15.76$^{***}$ & 11.30$^{***}$ & 0.10$^{***}$ & 0.06$^{***}$ \\
 & (2.18) & (2.05) & (0.01) & (0.01) \\ [.25em]
Own Incent. & 13.76$^{***}$ & 8.66$^{**}$ & 0.08$^{***}$ & 0.03$^{**}$ \\
 & (3.51) & (3.39) & (0.02) & (0.01) \\ [.25em]
Opp. Incent. & 5.41 & 0.65 & -0.14$^{***}$ & -0.03$^{**}$ \\
 & (3.32) & (3.30) & (0.02) & (0.01) \\ [.25em]

Controls    & Yes & Yes & Yes & Yes \\
\hline
R$^2$ & 0.12 & 0.08 & 0.16 & 0.10 \\
N & 837 & 877 & 837 & 877 \\
\hline\hline
\end{tabular}
    \begin{minipage}{1\linewidth}
        \caption{Response Time, Best Responses, and Payoffs  (\hyref{hypothesis:response-time}[Hypothesis][b])}
        \label{table:obr-oeu}
        \emph{Notes}:
        This table examines the relation between, on the one hand, log response times (in seconds), and on the other objective best responses to the empirical frequency of opponents' actions (columns (1) and (2)) and log expected payoffs, where expectations are taken also with respect to the empirical frequency of opponents' actions (columns (3) and (4)).
        High Own/Opponent Incentives correspond to indicators for whether the participant and their opponent face a high incentive level.
        2 Steps and 3 Steps denote the different games in the experiment (see \hyref{figure:games}[Figure]).
        Controls refer to the participants' socio-demographic characteristics.
        Heterocedasticity-robust standard errors are given in parentheses; $^*$, $^{**}$, and $^{***}$ denote p-values $< 0.1$, $<0.05$, and $<0.01$, respectively.
    \end{minipage}
\end{table}

Whether longer response times are associated with better performance is, however, an empirical question. 
We therefore examine whether longer response times are associated with (i) higher objective best-response rates and (ii) higher expected payoffs. 
As reported in \hyref{table:obr-oeu}[Table], longer response times are indeed associated with a higher rate of best responses to opponent action frequencies (columns (1) and (2)) as well as an increase in expected payoffs (columns (3) and (4)). 
The association with expected payoffs is statistically strong but economically modest. 
These regressions control for incentive treatments and the same socio-demographic characteristics used throughout. 
The findings support \hyref{hypothesis:response-time}[Hypothesis][b], although it is important to emphasise that the estimates here are associative rather than causal. 
Moreover, as in the existing literature, response time is an informative but imperfect proxy for cognitive effort, so the evidence in this section should be read as suggestive of mechanism rather than as a direct causal test of cognitive effort.

\begin{figure}[!t]
    \centering\small\singlespacing
    \begin{subfigure}{.45\linewidth}
        \includegraphics[width=\linewidth]{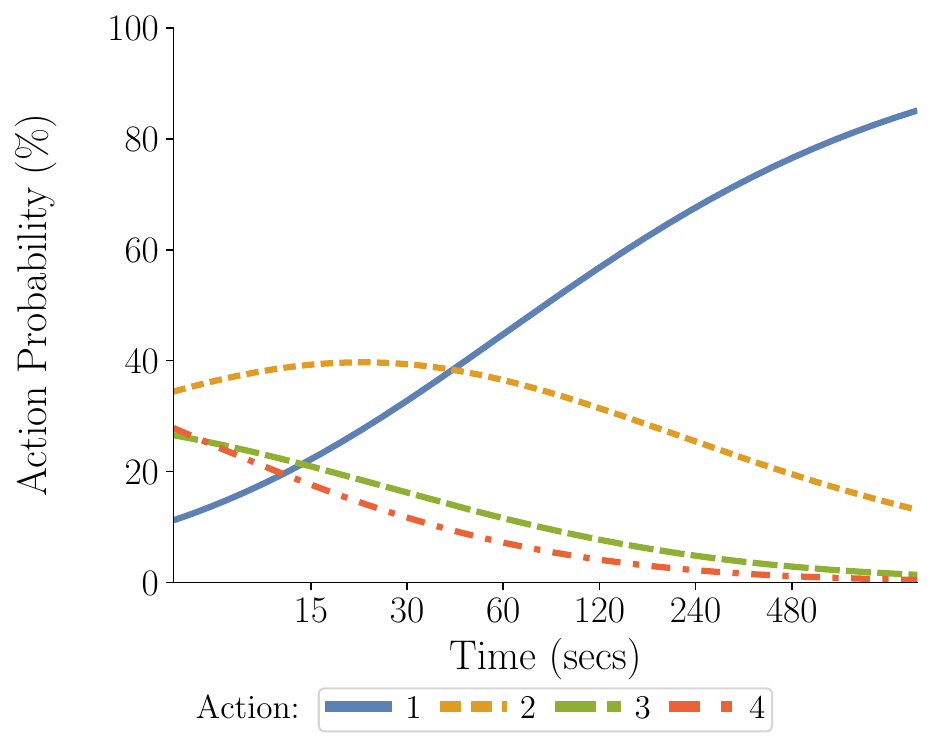}
        \caption{2-Step Game}
        \label{figure:action-rt-mnl-iesds2-own01-opp01}
    \end{subfigure}
    \begin{subfigure}{.45\linewidth}
        \includegraphics[width=\linewidth]{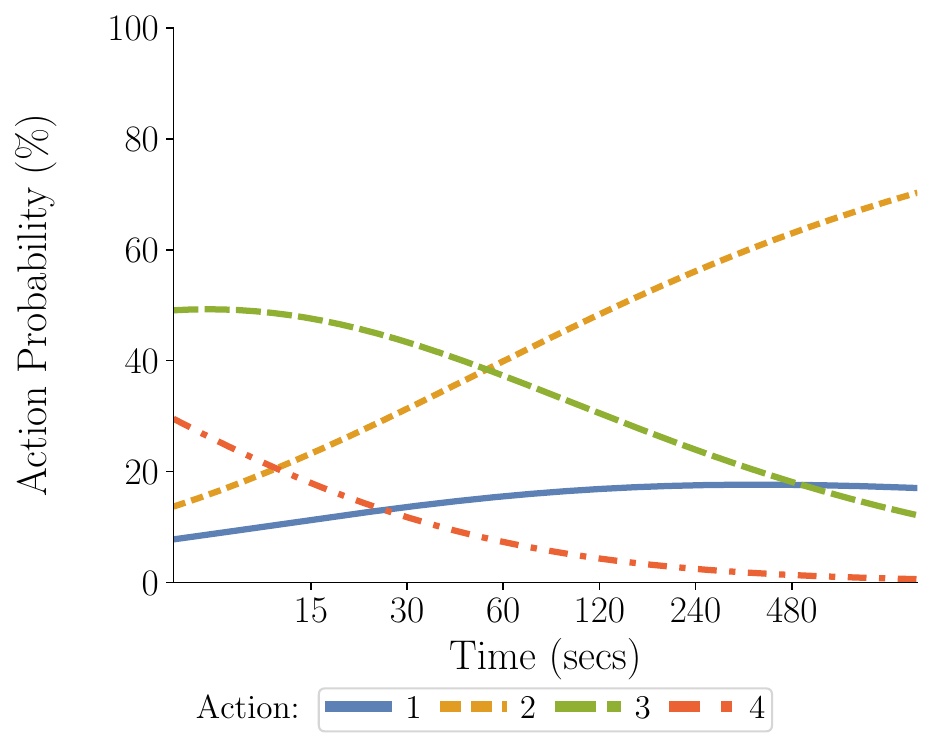}
        \caption{3-Step Game}
        \label{figure:action-rt-mnl-iesds3-own01-opp01}
    \end{subfigure}
    \begin{minipage}{1\linewidth}
        \caption{Action Frequency and Response Time}
        \label{figure:action-rt-mnl}
        \emph{Notes}:
        This figure shows the predicted relation between action frequency and response time (in seconds).
        The predictions are given by multinomial logit estimation of the relation between participants' choices, on the one hand, and response time, incentive treatments, and individual characteristics, estimated separately for each game.
        The action frequency conditional on response times is given by marginalising over the other regressors.
        2-Step and 3-Step games denote the different games in the experiment (see \hyref{figure:games}[Figure]).
        Individual characteristics refer to the participants' age, sex, education, and prior exposure to game theory.
    \end{minipage}
\end{figure}

\begin{figure}[!h]
    \centering\small\singlespacing
    \begin{subfigure}{.45\linewidth}
        \includegraphics[width=\linewidth]{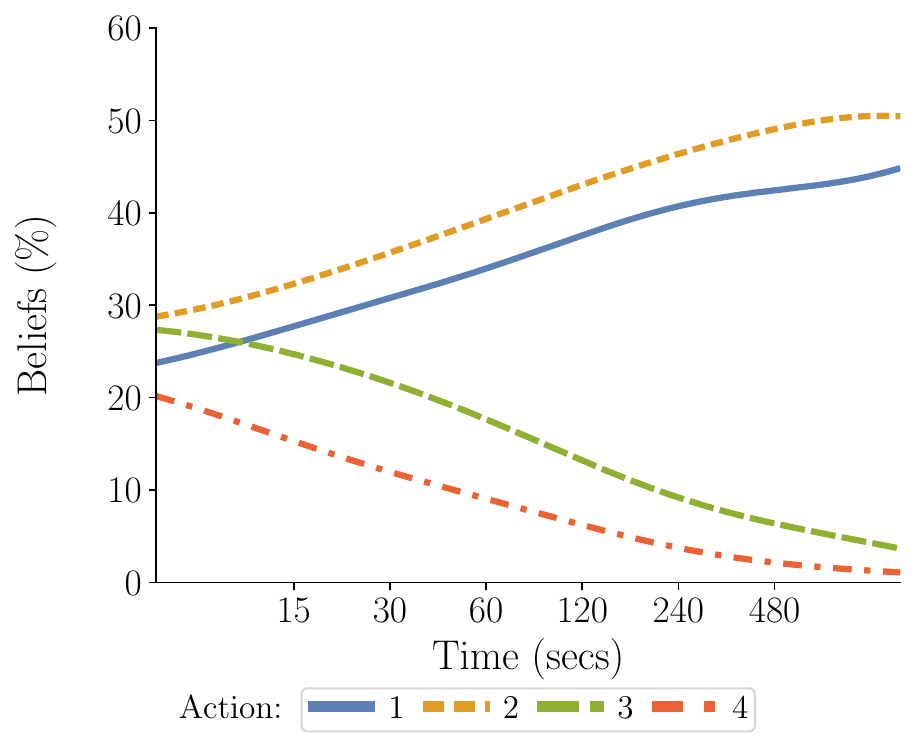}
        \caption{2-Step Game}
        \label{figure:belief-rt-kernel-iesds2-own01-opp01}
    \end{subfigure}
    \begin{subfigure}{.45\linewidth}
        \includegraphics[width=\linewidth]{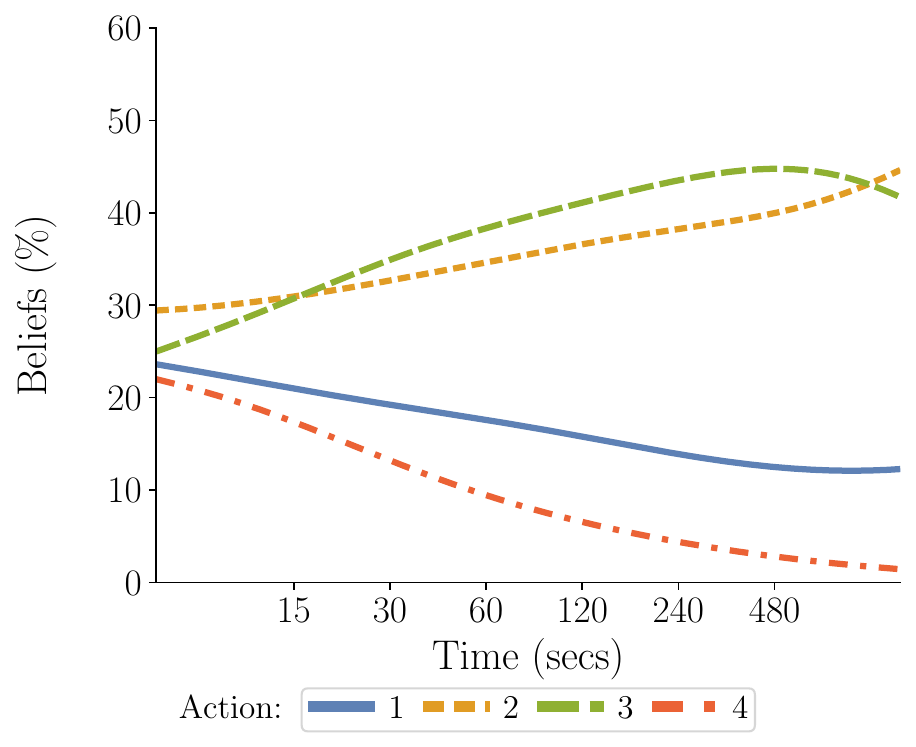}
        \caption{3-Step Game}
        \label{figure:belief-rt-kernel-iesds3-own01-opp01}
    \end{subfigure}
    \begin{minipage}{1\linewidth}
        \caption{Beliefs and Response Time}
        \label{figure:belief-rt-kernel}
        \emph{Notes}:
        This figure shows the predicted relation between reported belief distributions and response time (in seconds).
        The predictions are given by kernel estimation of the relation between participants' reported beliefs, on the one hand, and response time, incentive treatments, and individual characteristics, estimated separately for each game.
        The expected beliefs conditional on response times are obtained by using the estimated joint conditional probability distribution over a grid on $\Sigma_{-i}$ and marginalising over covariates other than response times.
        2-Step and 3-Step games denote the different games in the experiment (see \hyref{figure:games}[Figure]).
        Individual characteristics refer to the participants' age, sex, education, and prior exposure to game theory.
    \end{minipage}
\end{figure}

We conclude with a descriptive analysis of how actions and beliefs vary with response time. 
For actions, we estimate a multinomial logit model relating choices to response time, incentive treatments, and individual characteristics, and then recover fitted choice frequencies as a function of response time by marginalising over the remaining covariates. 
For beliefs, we use nonparametric kernel regression to estimate the mean belief vector, $b_i=(b_{i,1},b_{i,2},b_{i,3},b_{i,4})$, conditional on response time, incentive treatments, and individual characteristics. 
We then use the estimated conditional distribution to recover predicted mean belief distribution as a function of response time, holding the remaining covariates fixed at their sample means. 

\hyref{figure:action-rt-mnl}[Figures] and \ref{figure:belief-rt-kernel} show how action probabilities and belief distributions change with response time, revealing two main patterns.\footnote{
    In \hyref{section:appendix:additional-figures-tables:rt}[Appendix], we also provide linear regressions for each action $a_{i,n}$ and belief about a given action $b_{i,n}$ individually on the same set of covariates confirming the trends depicted --- see \hyref{table:action-rt}[Tables] and \ref{table:belief-rt}.
} 
First, in the simpler 2-Step game, longer response times are associated with greater sophistication in both actions and beliefs. 
Panel (a) of \hyref{figure:action-rt-mnl}[Figure] shows that the predicted frequency of dominated actions, $a_3$ and $a_4$, declines quickly with response time, while $a_2$ --- the level-1 action --- first mildly increases and then declines. 
Throughout, the frequency of choice of the dominance solution action $a_1$ increases monotonically in response time. 
Panel (a) of \hyref{figure:belief-rt-kernel}[Figure] similarly shows that longer decisions are associated with lower beliefs in opponents choosing dominated actions. 
At the same time, longer response times are associated with higher beliefs in both the level-1 action ($a_2$) and the dominance solution, which here is the level-2 action ($a_1$). 

Second, in the 3-Step game, we observe analogous patterns between actions and beliefs that correspond to the same level of sophistication.
Panel (b) of \hyref{figure:action-rt-mnl}[Figure] shows that dominated actions again become less frequent as response time increases. 
Moreover, the relation between response time and the frequencies of $a_2$ and $a_3$ in the 3-Step game closely mirrors the relation between response time and the frequencies of $a_1$ and $a_2$ in the 2-Step game. 
In each case, these are the level-2 and level-1 actions, respectively. 
As in the 2-Step game, the frequency of the level-1 action ($a_3$ in 3-Step game, $a_2$ in the 2-Step game) declines slowly, while the level-2 action ($a_2$ in 3-Step game, $a_1$ in the 2-Step game) becomes the most likely at longer response times. 
The choice frequency of the dominance solution is, however, rather invariant with respect to response time, remaining low.
Beliefs also follow similar patterns (\hyref{figure:belief-rt-kernel}[Figure][b]): belief in opponent dominated play is lower and belief in both level-1 and level-2 play is higher at later response times.
Again a notable difference is that in the 3-Step game belief in dominance play --- which corresponds to the level-3 action $a_1$ --- is lower for longer response times.

In short, the descriptive evidence suggests that faster decisions are concentrated on the intuitive level-1 action and on beliefs closer to uniform, whereas longer decisions are associated with fewer dominated choices, more weight on level-1 and level-2 opponent play, and a shift toward level-2 actions.\footnote{ 
    This is consistent with the data in \citet{Alos-FerrerBuckenmaier2021EE}, where longer response times are associated with more sophisticated choices. 
    This is clear for the beauty contest game from their Figure 2. 
    For the 11-20 games, the paper does not report the effect of response time on choices; using the published data, we regress choices on response time and find that longer response times are associated with more strategically sophisticated choices, in the level-$k$ sense. 
} 
What does not increase with response time in the 3-Step game is the frequency of the level-3 dominance solution.

The estimation exercise and the positive relation between own incentives and response time point to a coherent mechanism. 
When own incentives are higher, the expected gain from reducing mistakes and refining beliefs is larger, so participants have a stronger incentive to invest in further reasoning. 
Consistent with this interpretation, higher own incentives are associated with longer response times, and longer response times are associated with more strategically sophisticated beliefs and actions. 
This is in line with models of sequential reasoning \citep[e.g.][]{AlaouiPenta2016REStud,Goncalves2022WP}, in which players decide whether to exert additional cognitive effort by weighing its expected benefits against its cognitive costs.

\section{Alternative Explanations, Robustness, and Limitations of the Findings}
\label{section:robustness}
In this section, we discuss alternative explanations, robustness checks, and the natural limitations of our findings.

\subsection{Alternative Explanations}
\label{section:robustness:alternative-explanations}

Many prominent models do not predict changes in behaviour in response to incentive \emph{levels} alone. 
This is true of Nash equilibrium \citep{Nash1951AnnMath}, level-$k$ \citep{Nagel1995AER,StahlWilson1994JEBO,StahlWilson1995GEB}, cognitive hierarchy \citep{CamererHoChong2004QJE}, regret minimisation \citep{LoomesSugden1982EJ,Bell1982OR}, uncertainty-aversion equilibrium \citep{Klibanoff1996WP,Bade2011GEB}, random and noisy-belief equilibria \citep{FriedmanMezzetti2005GEB,Friedman2022AEJMicro}, and sampling equilibrium \citep{OsborneRubinstein1998AER,OsborneRubinstein2003GEB}, among others. 
In these models, behaviour is invariant to scaling of a player's payoffs.

One possible confounding factor is other-regarding preferences. 
As discussed in \hyref{section:hypotheses-design:design}[Section], we designed the experiment to reduce the scope for such motives to mechanically account for the treatment effects, through the choice of games, the way they are displayed, and the matching protocol. 
Moreover, many standard models of other-regarding preferences cannot generate behavioural changes from symmetrically scaling \emph{everyone's} payoffs \citep{Levine1998RED,FehrSchmidt1999QJE,CharnessRabin2002QJE},\footnote{
    Exceptions do exist, for example \citet{BoltonOckenfels2000AER}.
} though asymmetric treatments could still matter if participants react to relative payoff differences.

A natural concern is therefore that participants assigned to low own incentives may become disappointed when their opponent faces high incentives, and that this disappointment may lead to more random or less strategic behaviour. 
The data do not support this interpretation. 
Such a disappointment effect should be absent when both participant and opponent face low incentives, yet if anything choices are weakly \emph{more}, not less, sophisticated when the opponent has higher incentives than when both face low incentives; see \hyref{figure:action-sophistication-cdf-iesds2-owni0}[Figures] and \hyref{figure:action-sophistication-cdf-iesds3-owni0}. 
In addition, even under low own incentives, participants best respond to their stated beliefs more than 40\% of the time (cf. \hyref{figure:sbr-pmf}[Figure]), well above the 25\% rate implied by random choice, and this rate is not significantly affected by opponents' incentives (\hyref{table:action-br}[Table]).

More generally, the contrast between the 2-Step and 3-Step games is difficult to reconcile with a simple disappointment-based account. 
If the main effect of asymmetric incentives were merely to induce frustration or randomisation, one would not expect the systematic shift away from dominated actions and toward intermediate non-dominated actions that we observe, especially in the more complex game. 
While we cannot rule out all forms of other-regarding motives, the evidence suggests that they are unlikely to be the main driver of our findings.

\subsection{Robustness Checks}
\label{section:robustness:robustness}

\subsubsection{Participant Pools}
\label{section:robustness:robustness:pools}

MTurk was the benchmark platform for online economics experiments in 2020, when our first experiment was run, whereas Prolific had become the most common platform by the time of the replication. 
Online participant pools are particularly suitable for our purposes for two reasons. 
First, participants face a salient opportunity cost of time, since alternative paid tasks are readily available, which makes variation in incentives and response times especially meaningful. 
Second, online implementation avoids the frictions that a laboratory setting would create for response-time analysis, since participants can complete or leave the task individually without imposing waiting costs or disruption on others. 
Instead, lab participants would (i) need to stay until the end of the experiment, significantly affecting their opportunity cost of time, or (ii) would be able to leave early, disrupting others, or (iii) would need to come one by one, implying very significant time demands to recruit hundreds of participants. 

Following the editor's suggestion, we conducted a robust replication in 2026 on Prolific. 
The design was kept the same except that the instructions were simplified and incentive levels were de-emphasised both in the instructions and in the interface, in order to reduce the scope for treatment-demand effects.

While there are some level differences across the two participant pools,\footnote{
    The most salient difference is that Prolific participants tend to play dominated actions less often and the dominance solution slightly more often, but otherwise react in the same way to incentive levels.
} the treatment effects are broadly consistent across MTurk and Prolific, with mostly statistically insignificant differences; see \hyref{section:appendix:robustness:pool}[Appendix].\footnote{
    Another recent paper, \citet{GoncalvesLibgoberWillis2025REStud}, reports an experiment run on MTurk and a replication on Prolific in similar time periods, 2020 and 2025 respectively, and likewise finds consistent treatment effects across platforms.
} 
We therefore view the main findings as robust both across participant pools and to a less directive presentation of the incentive manipulation.

\subsubsection{Attention and Understanding}
\label{section:robustness:robustness:comp}

One possible concern is that our findings are driven by inattention and misunderstanding of the tasks. 

We took several steps to promote understanding: instructions were designed to be as clear and transparent as possible, participants could familiarise themselves with the interface during the instructions, and incentive levels were revealed only after the instruction stage, so attentiveness to the instructions should not differ systematically across treatments. 
The data also suggest that participants understood the task. 
Best-response rates to stated beliefs range from roughly 40-50\% under low own incentives and 60-70\% under high own incentives. 
These rates are comparable to those observed in laboratory studies with two- and three-action games, such as \citet{NyarkoSchotter2002Ecta}, \citet{Costa-GomesWeizsacker2008REStud}, \citet{Rey-Biel2009GEB}, and \citet{FriedmanWard2022WP}.
Moreover, the median time spent on the paid round was about one minute, while the median time spent on the instruction page was 9 minutes. 
These durations are consistent with participants engaging seriously with both the instructions and the paid task. 
For comparison, participants in \citet{FriedmanWard2022WP} spend less than 20 seconds on average, including a forced minimum of 10 seconds.

At the end of the instructions, participants completed a comprehension quiz designed both to consolidate understanding and to emphasise the key aspects of the experiment; the median number of mistakes was two. 
Because all participants had to complete the quiz correctly before proceeding, the quiz ensured a minimum level of understanding. 
We nevertheless use initial quiz performance to assess whether the main results are driven by participants who at first misunderstood the task. 
In \hyref{section:appendix:robustness:comp}[Appendix], we reproduce the main regressions interacting the treatment indicators with the number of mistakes in the quiz.  
Initial quiz performance makes little difference to the qualitative conclusions and, if anything, the treatment effects are somewhat stronger among participants who made fewer mistakes.

\subsection{Extreme Incentives}
\label{section:robustness:extreme-incentives}

Although higher incentives in our design tend to increase strategic sophistication, caution is warranted when extrapolating beyond the range of incentives we study. 
There may be an optimal level of incentives, deviations from which are detrimental to performance, as posited by \citet{YerkesDodson1908JCNP} in what came to be known as the Yerkes-Dodson `law'. 
As discussed by \citet{CamererHogarth1999JRU}, while for simple or well-learned tasks stronger incentives lead to better performance, extreme incentive levels can generate stress and impair cognition, limiting or even worsening performance. 
Indeed, \citet{ArielyGneezyLoewensteinMazar06RES} and \citet{EnkeGneezyHallMartinNelidovOffermanvandeVan2023REStat} provide evidence that performance in individual decision-making tasks involving memory and reasoning can deteriorate under extreme incentives (monthly salary or more). 
In strategic settings, evidence on this phenomenon is more limited; if anything, extreme incentives have been found to shift behaviour toward benchmark equilibrium predictions in ultimatum bargaining \citep{AndersenErtacGneezyHoffmanList2011AER}, a setting arguably strongly affected by cultural norms and other-regarding preferences.

In our design, the low incentive treatment was broadly in line with payment levels commonly used in related experiments, while the high incentive treatment scaled those payments up substantially but still remained within a realistic range. 
Our findings should therefore be interpreted as evidence about moderate incentive variation rather than about arbitrarily small or arbitrarily large stakes. 
The effect of scaling incentives is likely to depend not only on the strategic environment, as our data suggest, but also on the absolute level of incentives. 
At extremely low stakes, even large proportional changes may remain behaviourally negligible, while at very high stakes further increases may have little additional effect or could even become counterproductive.
That said, it is plausible that the main qualitative patterns we document --- greater action and belief sophistication, better responses to stated beliefs, and longer response times --- would continue to hold for somewhat wider incentive variation within a reasonable range. 

Within these limits, however, the main findings are robust across participant pools, instruction framings, and measures of attention and understanding.

\section{Conclusion}
\label{section:conclusion}
This paper provides well-identified evidence that incentive levels affect behaviour in strategic settings. 
Higher own incentives reduce dominated play in both games. 
In the simpler 2-Step game, they also increase play of the dominance-solution action, corresponding to the level-2 action. 
In the more complex 3-Step game, however, they mainly shift behaviour away from dominated and level-1 actions toward the level-2 action, without increasing play of the dominance-solution action, which there corresponds to the level-3 action. 
Reported beliefs shift in a similar direction and also become more accurate along several dimensions. 
Choices and beliefs also react to opponent incentive levels in a consistent manner, although more weakly. 
At the same time, beliefs about opponents facing higher incentives become less accurate, indicating that although higher opponent incentives shift behaviour in predictable directions, the magnitude of the effect is harder to predict than noisier, more uniform play.

Response times provide suggestive evidence on mechanism. 
Higher own incentives are associated with substantially longer response times in both games, and longer response times are in turn associated with more sophisticated choices and beliefs. 
Faster decisions are concentrated on the intuitive level-1 action and on beliefs closer to uniform, whereas longer decisions are associated with fewer dominated choices, greater weight on level-1 and level-2 opponent play, and a shift toward level-2 actions. 
What longer response times do not deliver in the 3-Step game is an increase in play of the level-3 dominance solution.

In summary, the evidence points to two main channels through which own incentives affect strategic behaviour: a reduction in payoff-dependent mistakes and an increase in effort devoted to reasoning, for which response time provides an informative, though imperfect, proxy.

Two directions for future research seem particularly important. 
First, it would be useful to better understand the limits of incentives in fostering strategic sophistication, and how those limits depend on the complexity of the strategic environment. 
If, on the one hand, there is naturally a limit to what one can achieve in thinking about others' behaviour through introspection alone, on the other hand such a limit surely depends on (or is informative about) how complex the environment is.  
This can shed light on what constitutes a simple strategic environment, which is a crucial consideration in designing successful mechanisms. 

Second, response times deserve further attention in the study of strategic behaviour. 
They are often straightforward to collect and, as our results suggest, can be informative about both beliefs and choices. 
Further work developing game-theoretic models that incorporate response times, and establishing more systematically how response times relate to behaviour across environments, could significantly deepen our understanding of observed strategic behaviour.

\FloatBarrier

\section{References}
{   
    \setstretch{1}
    \bibliographystyle{econ-aea}
    \bibliography{isb.bib}

@article{SmithWalker1993EconInq,
  author  = {V. L. Smith and J. M. Walker},
  title   = {Monetary Rewards and Decision Cost in Experimental Economics},
  journal = {Economic Inquiry},
  year    = {1993},
  volume  = {31},
  pages   = {245--261},
  doi     = {10.1111/j.1465-7295.1993.tb00881.x}
}

@article{Pearce1984Ecta,
  author  = {D. Pearce},
  title   = {Rationalizable Strategic Behavior and the Problem of Perfection},
  journal = {Econometrica},
  year    = {1984},
  volume  = {52},
  number  = {4},
  pages   = {1029--1050},
  doi     = {10.2307/1911197}
}

@article{CharnessRabin2002QJE,
  author  = {G. Charness and M. Rabin},
  title   = {Understanding Social Preferences with Simple Tests},
  journal = {The Quarterly Journal of Economics},
  year    = {2002},
  volume  = {117},
  number  = {3},
  pages   = {817--869},
  doi     = {10.1162/003355302760193904}
}

@article{ParravanoPoulsen2015GEB,
  author  = {M. Parravano and O. Poulsen},
  title   = {Stake size and the power of focal points in coordination games: Experimental evidence},
  journal = {Games and Economic Behavior},
  year    = {2015},
  volume  = {94},
  pages   = {191--199},
  doi     = {10.1016/j.geb.2015.05.001}
}

@article{BoltonOckenfels2000AER,
  author  = {G. E. Bolton and A. Ockenfels},
  title   = {ERC: A Theory of Equity, Reciprocity, and Competition},
  journal = {American Economic Review},
  year    = {2000},
  volume  = {90},
  number  = {1},
  pages   = {166--193},
  doi     = {10.1257/aer.90.1.166}
}

@article{Levine1998RED,
  author  = {D. K. Levine},
  title   = {Modeling Altruism and Spitefulness in Experiments},
  journal = {Review of Economic Dynamics},
  year    = {1998},
  volume  = {1},
  number  = {3},
  pages   = {593--622},
  doi     = {10.1006/redy.1998.0023}
}

@article{LoomesSugden1982EJ,
  author  = {G. Loomes and R. Sugden},
  title   = {Regret Theory: An Alternative Theory of Rational Choice Under Uncertainty},
  journal = {Economic Journal},
  year    = {1982},
  volume  = {92},
  number  = {368},
  pages   = {805--824},
  doi     = {10.2307/2232669}
}

@article{Bell1982OR,
  author  = {D. E. Bell},
  title   = {Regret in Decision Making under Uncertainty},
  journal = {Operations Research},
  year    = {1982},
  volume  = {30},
  number  = {5},
  pages   = {803--1022},
  doi     = {10.1287/opre.30.5.961}
}

@article{Klibanoff1996WP,
  author  = {P. Klibanoff},
  title   = {Uncertainty, Decision and Normal form Games},
  journal = {Working Paper},
  year    = {1996},
  pages   = {1--39}
}

@article{Bade2011GEB,
  author  = {S. Bade},
  title   = {Ambiguous act equilibria},
  journal = {Games and Economic Behavior},
  year    = {2011},
  volume  = {71},
  number  = {2},
  pages   = {246--260},
  doi     = {10.1016/j.geb.2010.04.006}
}

@article{FriedmanMezzetti2005GEB,
  author  = {J. Friedman and C. Mezzetti},
  title   = {“Random belief equilibrium in normal form games},
  journal = {Games and Economic Behavior},
  year    = {2005},
  volume  = {51},
  number  = {2},
  pages   = {296--323},
  doi     = {10.1016/j.geb.2003.08.004}
}

@article{BlancoEngelmannKochNormann2010EE,
  author  = {M. Blanco and D. Engelmann and A. K. Koch and H.-T. Normann},
  title   = {Belief elicitation in experiments: is there a hedging problem?},
  journal = {Experimental Economics},
  year    = {2010},
  volume  = {13},
  pages   = {412--438},
  doi     = {10.1007/s10683-010-9249-1}
}

@article{GoereeHoltPalfrey2005EE,
  author  = {J. Goeree and C. Holt and T. Palfrey},
  title   = {Regular Quantal Response Equilibrium},
  journal = {Experimental Economics},
  year    = {2005},
  volume  = {8},
  number  = {4},
  pages   = {347--367},
  doi     = {10.1007/s10683-005-5374-7}
}

@article{AlaouiJanezicPenta2020JET,
  author  = {L. Alaoui and K. A. Janezic and A. Penta},
  title   = {Reasoning about Others' Reasoning},
  journal = {Journal of Economic Theory},
  year    = {2020},
  volume  = {189},
  pages   = {1--51},
  doi     = {10.1016/j.jet.2020.105091}
}

@article{GeorganasHealyWeber2015JET,
  author  = {S. Georganas and P. J. Healy and R. A. Weber},
  title   = {On the Persistence of Strategic Sophistication},
  journal = {Journal of Economic Theory},
  year    = {2015},
  volume  = {159},
  doi     = {10.1016/j.jet.2015.07.012},
  pages   = {369--400}
}

@article{OsborneRubinstein1998AER,
  author  = {M. Osborne and A. Rubinstein},
  title   = {Games with Procedurally Rational Players},
  journal = {American Economic Review},
  year    = {1998},
  volume  = {88},
  number  = {4},
  pages   = {834--847}
}

@article{Friedman2022AEJMicro,
  author  = {E. Friedman},
  title   = {Stochastic Equilibria: Noise in Actions or Beliefs?},
  journal = {American Economic Journal: Microeconomics},
  year    = {2022},
  volume  = {14},
  number  = {1},
  pages   = {94--142},
  doi     = {10.1257/mic.20190013}
}

@article{OsborneRubinstein2003GEB,
  author  = {M. Osborne and A. Rubinstein},
  title   = {Sampling equilibrium, with an application to strategic voting},
  journal = {Games and Economic Behavior},
  year    = {2003},
  volume  = {45},
  number  = {2},
  pages   = {434--444},
  doi     = {10.1016/S0899-8256(03)00147-7}
}

@article{Kneeland2015Ecta,
  author  = {T. Kneeland},
  title   = {Identifying Higher-Order Rationality},
  journal = {Econometrica},
  year    = {2015},
  volume  = {83},
  number  = {5},
  pages   = {2065--2079},
  doi     = {10.3982/ECTA11983}
}

@article{AradRubinstein2012AER,
  author  = {A. Arad and A. Rubinstein},
  title   = {The 11-20 money request game: A level-k reasoning study},
  journal = {American Economic Review},
  year    = {2012},
  volume  = {102},
  number  = {7},
  pages   = {3561--3573},
  doi     = {10.1257/aer.102.7.3561}
}

@article{Rubinstein2007EJ,
  author  = {A. Rubinstein},
  title   = {Instinctive and Cognitive Reasoning: A Study of Response Times},
  journal = {The Economic Journal},
  year    = {2007},
  volume  = {117},
  number  = {523},
  pages   = {1243--1259},
  doi     = {10.1111/j.1468-0297.2007.02081.x}
}

@article{Rubinstein2016QJE,
  author  = {A. Rubinstein},
  title   = {A Typology of Players: Between Instinctive and Contemplative},
  journal = {Quarterly Journal of Economics},
  year    = {2016},
  volume  = {131},
  number  = {2},
  pages   = {859--90},
  doi     = {10.1093/qje/qjw008}
}

@article{McKelveyPalfrey1995GEB,
  author  = {R. D. McKelvey and T. R. Palfrey},
  title   = {Quantal Response Equilibria for Normal Form Games},
  journal = {Games and Economic Behavior},
  year    = {1995},
  volume  = {10},
  number  = {1},
  pages   = {6--38},
  doi     = {10.1006/game.1995.1023}
}

@incollection{McFadden1974Chapter,
  author    = {D. McFadden},
  title     = {Conditional logit analysis of qualitative choice behavior},
  booktitle = {Frontiers in Econometrics},
  publisher = {Academic Press},
  year      = {1974},
  editor    = {P. Karembka},
  pages     = {105--142}
}

@article{MattssonWeibull2002GEB,
  author  = {L.-G. Mattsson and J. Weibull},
  title   = {Probabilistic choice and procedurally bounded rationality},
  journal = {Games and Economic Behavior},
  year    = {2002},
  volume  = {41},
  number  = {1},
  pages   = {61--78},
  doi     = {10.1016/S0899-8256(02)00014-3}
}

@article{FudenbergIijimaStrzalecki2015Ecta,
  author  = {D. Fudenberg and R. Iijima and T. Strzalecki},
  title   = {Stochastic Choice and Revealed Perturbed Utility},
  journal = {Econometrica},
  year    = {2015},
  volume  = {83},
  number  = {6},
  pages   = {2371--2409},
  doi     = {10.3982/ECTA12660}
}

@article{Nash1951AnnMath,
  author  = {J. F. Nash},
  title   = {Non-Cooperative Games},
  journal = {Annals of Mathematics},
  year    = {1951},
  volume  = {54},
  number  = {2},
  pages   = {286--295},
  doi     = {10.2307/1969529}
}

@article{AlaouiPenta2016REStud,
  author  = {L. Alaoui and A. Penta},
  title   = {Endogenous Depth of Reasoning},
  journal = {Review of Economic Studies},
  year    = {2016},
  volume  = {83},
  number  = {4},
  pages   = {1297--1333},
  doi     = {10.1093/restud/rdv052}
}

@article{Nagel1995AER,
  author  = {R. Nagel},
  title   = {Unraveling in Guessing Games: An Experimental Study},
  journal = {American Economic Review},
  year    = {1995},
  volume  = {85},
  number  = {5},
  pages   = {1313--1326}
}

@article{StahlWilson1994JEBO,
  author  = {D. O. Stahl and P. W. Wilson},
  title   = {Experimental evidence on players' models of other players},
  journal = {Journal of Economic Behavior and Organization},
  year    = {1994},
  volume  = {25},
  number  = {3},
  pages   = {309--327},
  doi     = {10.1016/0167-2681(94)90103-1}
}

@article{StahlWilson1995GEB,
  author  = {D. O. Stahl and P. W. Wilson},
  title   = {On Players' Models of Other Players: Theory and Experimental Evidence},
  journal = {Games and Economic Behavior},
  year    = {1995},
  volume  = {10},
  number  = {1},
  pages   = {218--254},
  doi     = {10.1006/game.1995.1031}
}

@article{CamererHoChong2004QJE,
  author  = {C. Camerer and T.-H. Ho and J.-K. Chong},
  title   = {A Cognitive Hierarchy Model of Games},
  journal = {Quarterly Journal of Economics},
  year    = {2004},
  volume  = {119},
  number  = {3},
  pages   = {861--898},
  issn    = {0033-5533},
  doi     = {10.1162/0033553041502225}
}

@article{Costa-GomesWeizsacker2008REStud,
  author  = {M. A. Costa-Gomes and G. Weizs{\"{a}}cker},
  title   = {Stated Beliefs and Play in Normal Form Games},
  journal = {Review of Economic Studies},
  year    = {2008},
  volume  = {75},
  number  = {3},
  pages   = {729--762},
  doi     = {10.1111/j.1467-937X.2008.00498.x}
}

@article{Rey-Biel2009GEB,
  author  = {P. Rey-Biel},
  title   = {Equilibrium play and best response to (stated) beliefs in normal form games},
  journal = {Games and Economic Behavior},
  year    = {2009},
  volume  = {65},
  number  = {2},
  pages   = {572--585},
  doi     = {10.1016/j.geb.2008.03.003}
}

@article{HuckWeizsaecker2002JEBO,
  author  = {S. Huck and G. Weizs{\"{a}}cker},
  title   = {Do players correctly estimate what others do? Evidence of conservatism in beliefs},
  journal = {Journal of Economic Behavior and Organization},
  year    = {2002},
  volume  = {47},
  number  = {1},
  pages   = {71--85},
  doi     = {10.1016/S0167-2681(01)00170-6}
}

@article{Goncalves2022WP,
  author  = {D. Gon\c{c}alves},
  title   = {Sequential Sampling Equilibrium},
  journal = {Working Paper},
  year    = {2024},
  pages   = {1--53},
  doi     = {10.48550/arXiv.2212.07725}
}

@article{Alos-FerrerBuckenmaier2021EE,
  author  = {C. Al\'{o}s-Ferrer and J. Buckenmaier},
  title   = {Cognitive Sophistication and Deliberation Times},
  journal = {Experimental Economics},
  year    = {2021},
  volume  = {24},
  pages   = {558--592},
  doi     = {10.1007/s10683-020-09672-w}
}

@article{FrydmanNunnari2023WP,
  author  = {C. Frydman and S. Nunnari},
  title   = {Coordination with Cognitive Noise},
  journal = {Working Paper},
  year    = {2023},
  doi     = {10.2139/ssrn.3939522}
}

@article{Clithero2018JEconPsy,
  author  = {J. A. Clithero},
  title   = {Response times in economics: Looking through the lens of sequential sampling models},
  journal = {Journal of Economic Psychology},
  year    = {2018},
  volume  = {69},
  pages   = {61--86},
  doi     = {10.1016/j.joep.2018.09.008}
}

@article{Costa-GomesCrawford2006AER,
  author  = {M. A> Costa-Gomes and V. P. Crawford},
  title   = {Cognition and Behavior in Two-Person Guessing Games: An Experimental Study},
  journal = {American Economic Review},
  year    = {2006},
  volume  = {96},
  number  = {5},
  pages   = {1737--1768},
  doi     = {10.1257/aer.96.5.1737}
}

@article{DeanNeligh2023JPE,
  author  = {M. Dean and N. Neligh},
  title   = {Experimental Tests of Rational Inattention},
  journal = {Journal of Political Economy},
  year    = {2023},
  volume  = {131},
  number  = {12},
  pages   = {3415--3461},
  doi     = {10.1086/725174}
}

@article{AndersenErtacGneezyHoffmanList2011AER,
  author  = {S. Andersen and S. Erta{\c{c}} and U. Gneezy and M. Hoffman and J. A. List},
  title   = {Stakes Matter in Ultimatum Games},
  journal = {American Economic Review},
  year    = {2011},
  volume  = {101},
  number  = {7},
  pages   = {3427--3439},
  doi     = {10.1257/aer.101.7.3427}
}

@article{McKelveyPalfreyWeber2000JEBO,
  author  = {R. D. McKelvey and T. R. Palfrey and R. Weber},
  title   = {The effects of payoff magnitude and heterogeneity on behavior in 2{\texttimes}2 games with unique mixed strategy equilibria},
  journal = {Journal of Economic Behavior {\&} Organization},
  year    = {2000},
  volume  = {42},
  number  = {4},
  pages   = {523--548},
  doi     = {10.1016/s0167-2681(00)00102-5}
}

@article{FudenbergLiang2019AER,
  author    = {D. Fudenberg and A. Liang},
  journal   = {American Economic Review},
  title     = {Predicting and Understanding Initial Play},
  year      = {2019},
  number    = {12},
  pages     = {4112--4141},
  volume    = {109},
  doi       = {10.1257/aer.20180654},
  publisher = {American Economic Association}
}

@article{HossainOkui2013REStud,
  author  = {T. Hossain and R. Okui},
  title   = {The Binarized Scoring Rule},
  journal = {Review of Economic Studies},
  year    = {2013},
  volume  = {80},
  number  = {3},
  pages   = {984--1001},
  doi     = {10.1093/restud/rdt006}
}

@article{Palacios-HuertaVolij2009AER,
  author  = {I. Palacios-Huerta and O. Volij},
  title   = {Field Centipedes},
  journal = {American Economic Review},
  year    = {2009},
  volume  = {99},
  number  = {4},
  pages   = {1619--35},
  doi     = {10.1257/aer.99.4.1619}
}

@article{AgranovPotamitesSchotterTergiman2012GEB,
  author  = {M. Agranov and E. Potamites and A. Schotter and C. Tergiman},
  title   = {Beliefs and endogenous cognitive levels: An experimental study},
  journal = {Games and Economic Behavior},
  year    = {2012},
  volume  = {75},
  number  = {2},
  pages   = {449--63},
  doi     = {10.1016/j.geb.2012.02.002}
}

@article{McKelveyPalfrey1992Ecta,
  author  = {R. D. McKelvey and T. R. Palfrey},
  title   = {An Experimental Study of the Centipede Game},
  journal = {Econometrica},
  year    = {1992},
  volume  = {60},
  number  = {4},
  pages   = {803--36},
  doi     = {10.2307/2951567}
}

@book{Camerer2003Book,
  author    = {C. Camerer},
  title     = {Behavioral Game Theory: Experiments in Strategic Interaction},
  year      = {2003},
  publisher = {Princeton University Press}
}

@article{FriedmanWard2022WP,
  author  = {E. Friedman and J. Ward},
  title   = {Stochastic Choice and Noisy Beliefs in Games: an Experiment},
  journal = {Working Paper},
  year    = {2022},
  pages   = {1-77},
  doi     = {10.2139/ssrn.4190338}
}

@article{Ochs1995GEB,
  author  = {J. Ochs},
  title   = {Games with Unique, Mixed Strategy Equilibria: An Experimental Study},
  journal = {Games and Economic Behavior},
  year    = {1995},
  volume  = {10},
  number  = {1},
  pages   = {202--17},
  doi     = {10.1006/game.1995.1030}
}

@article{RapoportSteinParcoNicholas2003GEB,
  author  = {A. Rapoport and W. E. Stein and J. E. Parco and T. E. Nicholas},
  title   = {Equilibrium play and adaptive learning in a three-person centipede game},
  journal = {Games and Economic Behavior},
  year    = {2003},
  volume  = {42},
  number  = {3},
  pages   = {239--65},
  doi     = {10.1016/S0899-8256(03)00009-5}
}

@article{CaplinCsabaLeahyNov2020QJE,
  author  = {A. Caplin and D. Csaba and J. Leahy and O. Nov},
  title   = {Rational Inattention, Competitive Supply, and Psychometrics},
  journal = {Quarterly Journal of Economics},
  year    = {2020},
  volume  = {135},
  number  = {3},
  pages   = {1681--724},
  doi     = {10.1093/qje/qjaa011}
}

@article{EnkeGneezyHallMartinNelidovOffermanvandeVan2023REStat,
  author  = {B. Enke and U. Gneezy and B. Hall and D. C. Martin and V. Nelidov and T. Offerman and J. van de Ven},
  title   = {Cognitive Biases: Mistakes or Missing Stakes?},
  journal = {The Review of Economics and Statistics},
  year    = {2023},
  volume  = {105},
  number  = {4},
  pages   = {818--832},
  doi     = {10.1162/rest_a_01093}
}

@book{Luce1959Book,
  author    = {D. Luce},
  title     = {Individual choice behavior},
  year      = {1959},
  publisher = {John Wiley}
}

@article{SchotterTrevino2021EE,
  author  = {A. Schotter and I. Trevino},
  title   = {Is Response Time Predictive of Choice? An Experimental Study of Threshold Strategies},
  journal = {Experimental Economics},
  year    = {2021},
  volume  = {24},
  number  = {1},
  pages   = {87--117},
  doi     = {10.1007/s10683-020-09651-1}
}

@article{SpiliopoulosOrtmann2018EE,
  author  = {L. Spiliopoulos and A. Ortmann},
  title   = {The BCD of response time analysis in experimental economics},
  journal = {Experimental
             Economics},
  year    = {2018},
  volume  = {21},
  number  = {2},
  pages   = {383--433},
  doi     = {10.1007/s10683-017-9528-1}
}

@article{ProtoRustichiniSofianos2019JPE,
  author  = {E. Proto and A. Rustichini and A. Sofianos},
  title   = {Intelligence, Personality, and Gains from Cooperation in Repeated Interactions},
  journal = {Journal of Political
             Economy},
  year    = {2019},
  volume  = {127},
  number  = {3},
  pages   = {1351--90},
  doi     = {10.1086/701355}
}

@article{GillProwse2022EJ,
  author  = {Gill, D. and Prowse, V.},
  title   = {Strategic Complexity and the Value of Thinking},
  journal = {Economic Journal},
  year    = {2022},
  volume  = {133},
  pages   = {761--86},
  doi     = {10.1093/ej/ueac070}
}

@article{HoffmanMcCabeSmith1996IJGT,
  author    = {Elizabeth Hoffman and Kevin A. McCabe and Vernon L. Smith},
  title     = {On expectations and the monetary stakes in ultimatum games},
  journal   = {International Journal of Game Theory},
  year      = {1996},
  doi       = {10.1007/BF02425259},
  volume    = {25},
  pages     = {289--301},
}

@article{SlonimRoth1998Ecta,
  author    = {Robert Slonim and Alvin E. Roth},
  title     = {Learning in High Stakes Ultimatum Games: An Experiment in the Slovak Republic},
  journal   = {Econometrica},
  year      = {1998},
  doi       = {10.2307/2998575},
  volume    = {66},
  number    = {3},
  pages     = {569--596},
}

@article{CamererHogarth1999JRU,
  author    = {Colin F. Camerer and Robin M. Hogarth},
  title     = {The Effects of Financial Incentives in Experiments: A Review and Capital-Labor-Production Framework},
  journal   = {Journal of Risk and Uncertainty},
  year      = {1999},
  doi       = {10.1023/A:1007850605129},
  volume    = {19},
  pages     = {7--42},
}

@article{GachterRenner2010EE,
  author  = {Simon Gächter and Elke Renner},
  journal = {Experimental Economics},
  title   = {The effects of (incentivized) belief elicitation in public goods experiments},
  year    = {2010},
  number  = {3},
  pages   = {364--377},
  volume  = {13},
  doi     = {10.1007/s10683-010-9246-4}
}

@article{FehrSchmidt1999QJE,
  author  = {Ernst Fehr and Klaus M. Schmidt},
  title   = {A Theory of Fairness, Competition, and Cooperation},
  journal = {Quarterly Journal of Economics},
  year    = {1999},
  volume  = {114},
  number  = {3},
  pages   = {817--868},
  doi     = {10.1162/003355399556151}
}

@article{Weber2003GEB,
  author  = {Roberto A. Weber},
  journal = {Games and Economic Behavior},
  title   = {'Learning' with no feedback in a competitive guessing game},
  year    = {2003},
  number  = {1},
  pages   = {134-144},
  volume  = {44},
  doi     = {10.1016/S0899-8256(03)00002-2}
}

@article{NyarkoSchotter2002Ecta,
  author  = {Yaw Nyarko and Andrew Schotter},
  journal = {Econometrica},
  title   = {An Experimental Study of Belief Learning Using Elicited Beliefs},
  year    = {2002},
  pages   = {971-1005},
  volume  = {70},
  doi     = {10.1111/1468-0262.00316}
}

@article{AlaouiPentaJPE2022,
  author  = {Larbi Alaoui and Antonio Penta},
  title   = {Cost-Benefit Analysis in Reasoning},
  journal = {Journal of Political Economy},
  year    = {2022},
  volume  = {130},
  number  = {4},
  pages   = {881--925},
  doi     = {10.1086/718378},
  issn    = {0022-3808},
  month   = {4}
}

@article{YerkesDodson1908JCNP,
  author  = {R. M. Yerkes and J. D. Dodson},
  title   = {The relation of strength of stimulus to rapidity of habit-formation},
  journal = {Journal of Comparative Neurology and Psychology},
  year    = {1908},
  volume  = {18},
  number  = {5},
  pages   = {459--482},
  doi     = {10.1002/cne.920180503}
}

@article{ArielyGneezyLoewensteinMazar06RES,
  author  = {D. Ariely and U. Gneezy and G. Loewenstein and N. Mazar},
  title   = {Large Stakes and Big Mistakes},
  journal = {Review of Economic Studies},
  year    = {2009},
  volume  = {76},
  number  = {2},
  pages   = {451--469},
  doi     = {10.1111/j.1467-937X.2009.00534.x}
}

@article{GoncalvesLibgoberWillis2025REStud,
  author  = {D. Gon\c{c}alves and J. Libgober and J. Willis},
  title   = {Retractions: Updating from Complex Information},
  journal = {Review of Economic Studies},
  year    = {2025},
  volume  = {93},
  number  = {1},
  pages   = {476--516},
  doi     = {10.1093/restud/rdaf032}
}

@Article{CapraGoereeGomezHolt1999AER,
  author  = {C. Capra and J. K. Goeree and R. Gomez and C. A. Holt},
  title   = {Anomalous Behavior in a Traveler's Dilemma?},
  journal = {American Economic Review},
  year    = {1999},
  volume  = {89},
  number  = {3},
  pages   = {678--690},
  doi     = {10.1257/aer.89.3.678},
}

@Article{MeloPogorelskiyShum2019IER,
  author  = {E. Melo and K. Pogorelskiy and M. Shum},
  title   = {Testing the Quantal Response Hypothesis},
  journal = {International Economic Review},
  year    = {2019},
  volume  = {60},
  pages   = {53--74},
  doi     = {10.1111/iere.12344},
}

@Article{Stahl1993GEB,
  author  = {D. O. Stahl},
  title   = {Evolution of Smartn Players},
  journal = {Games and Economic Behavior},
  year    = {1993},
  volume  = {5},
  number  = {4},
  pages   = {604--617},
  doi     = {10.1006/game.1993.1033},
}

@Article{EysterRabin2005Ecta,
  author  = {E. Eyster and M. Rabin},
  title   = {Cursed Equilibrium},
  journal = {Econometrica},
  year    = {2005},
  volume  = {73},
  number  = {5},
  pages   = {1623--1672},
  doi     = {10.1111/j.1468-0262.2005.00631.x},
}

@Article{StraubMurnighan1995JEBO,
  author  = {P. G. Straub and J. K. Murnighan},
  title   = {An experimental investigation of ultimatum games: information, fairness, expectations, and lowest acceptable offers},
  journal = {Journal of Economic Behavior \& Organization},
  year    = {1995},
  volume  = {27},
  number  = {3},
  pages   = {345--364},
  doi     = {10.1016/0167-2681(94)00072-M},
}

@article{Croson1999OBHDP,
  author  = {Rachel T. A. Croson},
  title   = {The Disjunction Effect and Reason-Based Choice in Games},
  journal = {Organizational Behavior and Human Decision Processes},
  year    = {1999},
  volume  = {80},
  number  = {2},
  pages   = {118--133},
  doi     = {10.1006/obhd.1999.2846},
}

@article{Croson2000JEBO,
  author  = {Rachel T. A. Croson},
  title   = {Thinking like a game theorist: factors affecting the frequency of equilibrium play},
  journal = {Journal of Economic Behavior \& Organization},
  year    = {2000},
  volume  = {41},
  number  = {3},
  pages   = {299--314},
  doi     = {10.1016/S0167-2681(99)00078-5},
}

@article{BajariHortacsu2005JPE,
  author  = {Patrick Bajari and Ali Horta\c{c}su},
  title   = {Are Structural Estimates of Auction Models Reasonable? Evidence from Experimental Data},
  journal = {Journal of Political Economy},
  year    = {2005},
  volume  = {113},
  number  = {4},
  pages   = {675--741},
  month   = {August},
  doi     = {10.1086/432138}
}

@article{MoinasPouget2013Ecta,
  author  = {Sophie Moinas and S{\'e}bastien Pouget},
  title   = {The Bubble Game: An Experimental Study of Speculation},
  journal = {Econometrica},
  year    = {2013},
  volume  = {81},
  number  = {4},
  pages   = {1507--1539},
  month   = {July},
  doi     = {10.3982/ECTA9433}
}

@article{FarhiWerning2019AER,
  author  = {Emmanuel Farhi and Iv{\'a}n Werning},
  title   = {Monetary Policy, Bounded Rationality, and Incomplete Markets},
  journal = {American Economic Review},
  year    = {2019},
  volume  = {109},
  number  = {11},
  pages   = {3887--3928},
  month   = {November},
  doi     = {10.1257/aer.20171400}
}
}

\FloatBarrier \newpage

\phantomsection
\addcontentsline{toc}{section}{Appendices}

\setcounter{section}{0}
\renewcommand{\thesection}{Appendix \Alph{section}}
\renewcommand{\thesubsection}{\Alph{section}.\arabic{subsection}}

\section{Supporting Tables and Figures}
\label{section:appendix:additional-figures-tables}

\subsection{Objective Best-Response Rank}
\label{section:appendix:additional-figures-tables:obr}

\begin{figure}[!ht]
    \centering\small\singlespacing
    \begin{tabular}{@{\extracolsep{0pt}}cc@{}}
        \begin{subfigure}{.45\linewidth}
            \includegraphics[width=\linewidth]{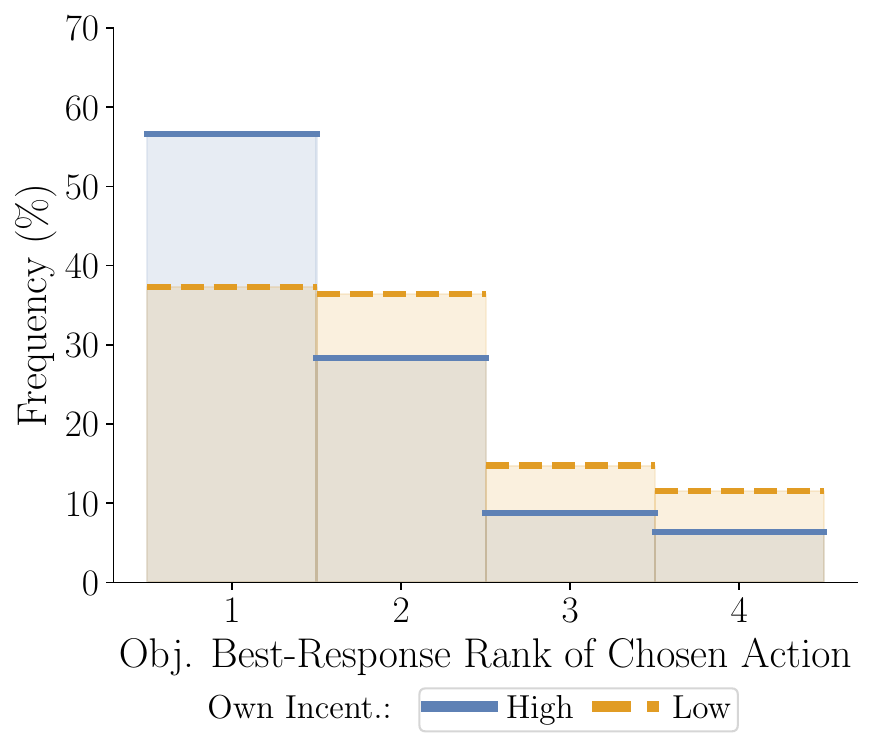}
            \caption{2-Step Game; High Opp. Incentive Level}
            \label{figure:obr-pmf-iesds2-oppi1}
        \end{subfigure}
        &
        \begin{subfigure}{.45\linewidth}
            \includegraphics[width=\linewidth]{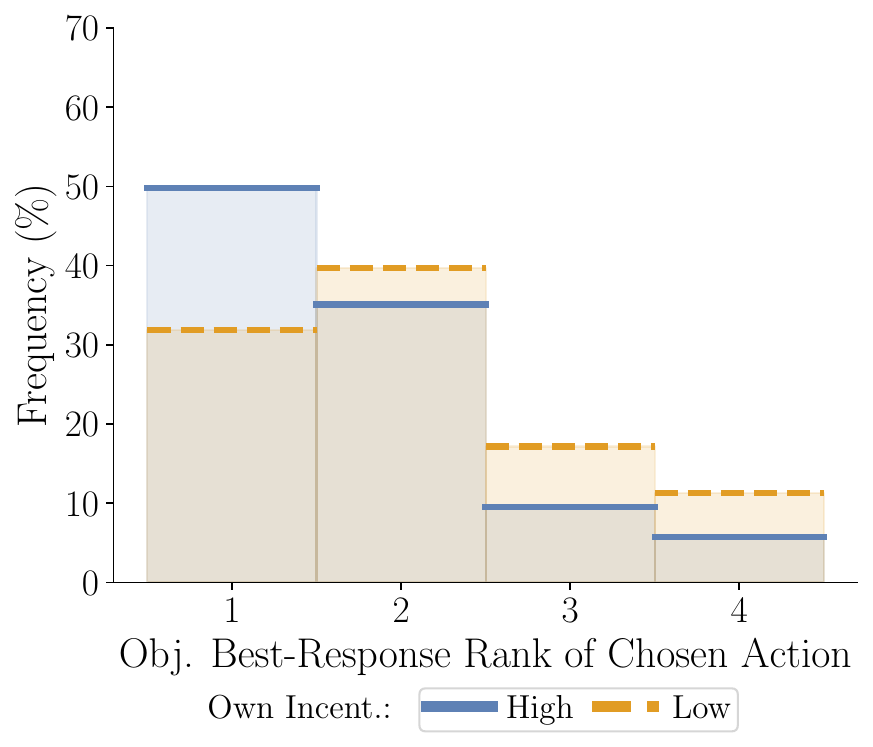}
            \caption{2-Step Game; Low Opp. Incentive Level}
            \label{figure:obr-pmf-iesds2-oppi0}
        \end{subfigure}
        \\
        \begin{subfigure}{.45\linewidth}
            \includegraphics[width=\linewidth]{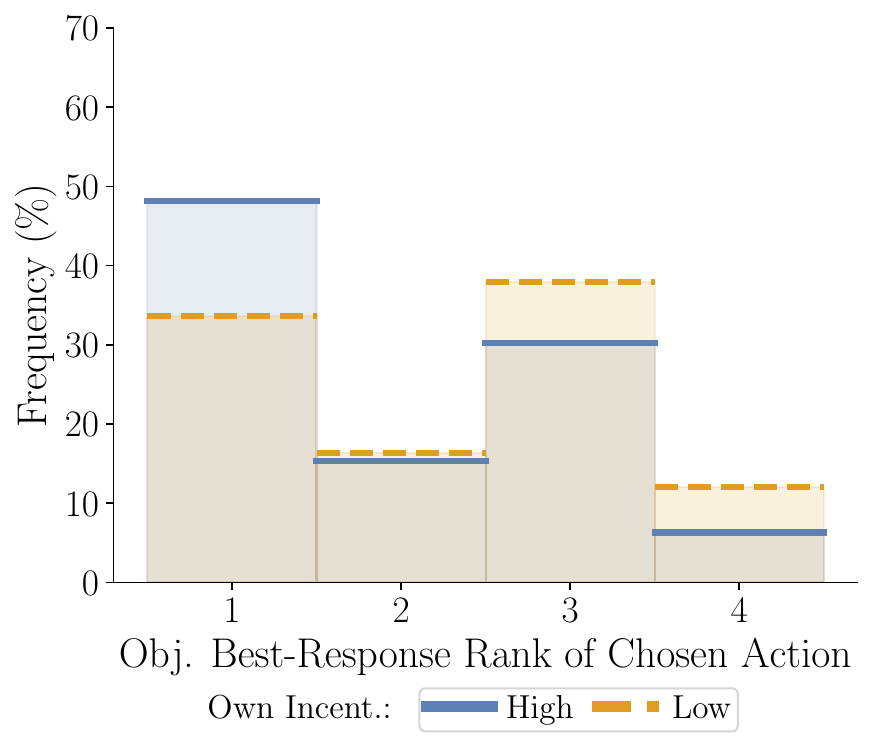}
            \caption{3-Step Game; High Opp. Incentive Level}
            \label{figure:obr-pmf-iesds3-oppi1}
        \end{subfigure}
        &
        \begin{subfigure}{.45\linewidth}
            \includegraphics[width=\linewidth]{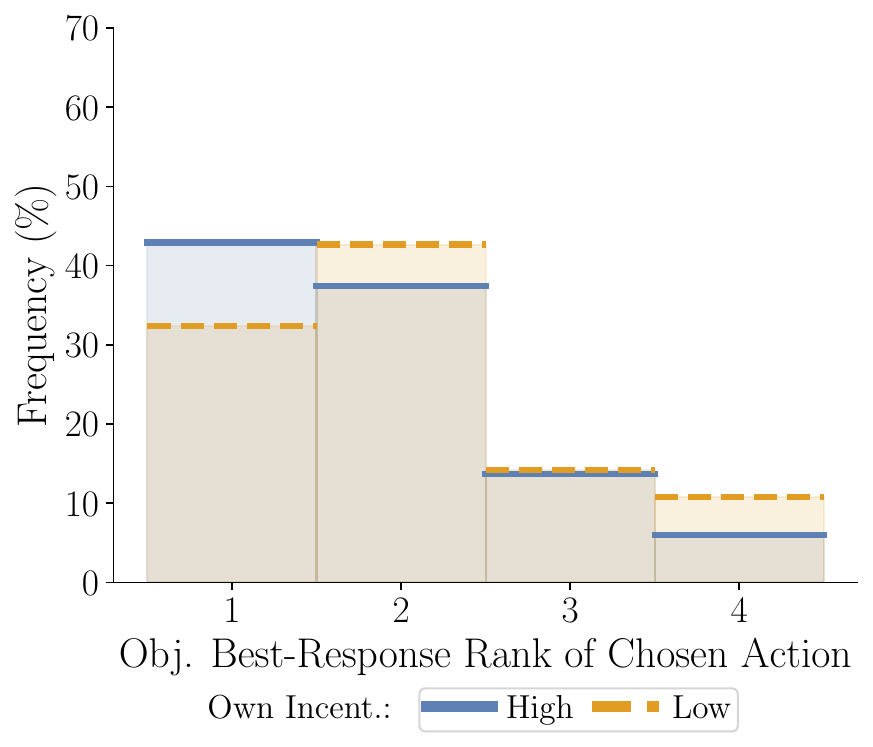}
            \caption{3-Step Game; Low Opp. Incentive Level}
            \label{figure:obr-pmf-iesds3-oppi0}
        \end{subfigure}
    \end{tabular}
    \begin{minipage}{1\linewidth}
        \caption{Incentive Level and Objective Best-Response (\hyref{hypothesis:best-responses}[Hypothesis][b])}
        \label{figure:obr-pmf}
        \emph{Notes}:
        The different panels exhibit the frequency of the objective best-response ranking of the action chosen by the participant for high and low own incentive levels, for different games (2-Step game, (a) and (b), and 3-Step game, (c) and (d)) and holding fixed opponent incentive level (High, (a) and (c), or Low, (b) and (d)).
        The objective best-response ranking of an action is $n$ if the action entails the $n$-th highest objective expected payoffs according to the reported beliefs; e.g. actions with objective rank 1 are those that maximise objective expected payoffs.
        2-Step and 3-Step games denote the different games in the experiment (see \hyref{figure:games}[Figure]).
    \end{minipage}
\end{figure}

\FloatBarrier \newpage

\subsection{Cumulative Frequency Figures}
\label{section:appendix:additional-figures-tables:cdf}

\begin{figure}[!ht]
    \centering\small\singlespacing
    \begin{tabular}{@{\extracolsep{0pt}}cc@{}}
        \begin{subfigure}{.45\linewidth}
            \includegraphics[width=\linewidth]{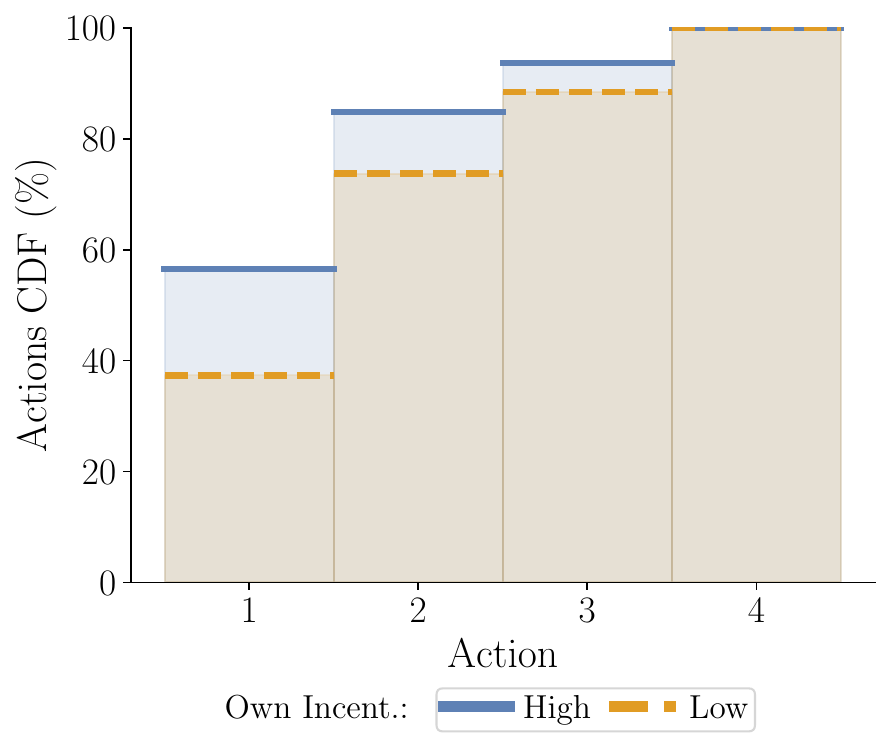}
            \caption{2-Step Game; High Opp. Incentive Level}
            \label{figure:action-sophistication-cdf-iesds2-oppi1}
        \end{subfigure}
        &
        \begin{subfigure}{.45\linewidth}
            \includegraphics[width=\linewidth]{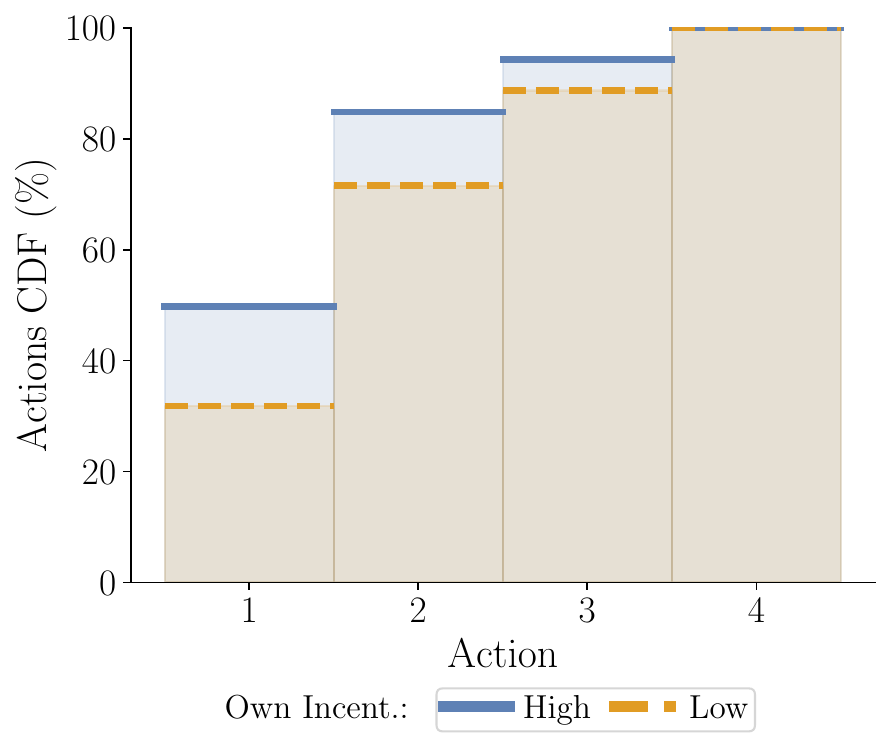}
            \caption{2-Step Game; Low Opp. Incentive Level}
            \label{figure:action-sophistication-cdf-iesds2-oppi0}
        \end{subfigure}
        \\
        \begin{subfigure}{.45\linewidth}
            \includegraphics[width=\linewidth]{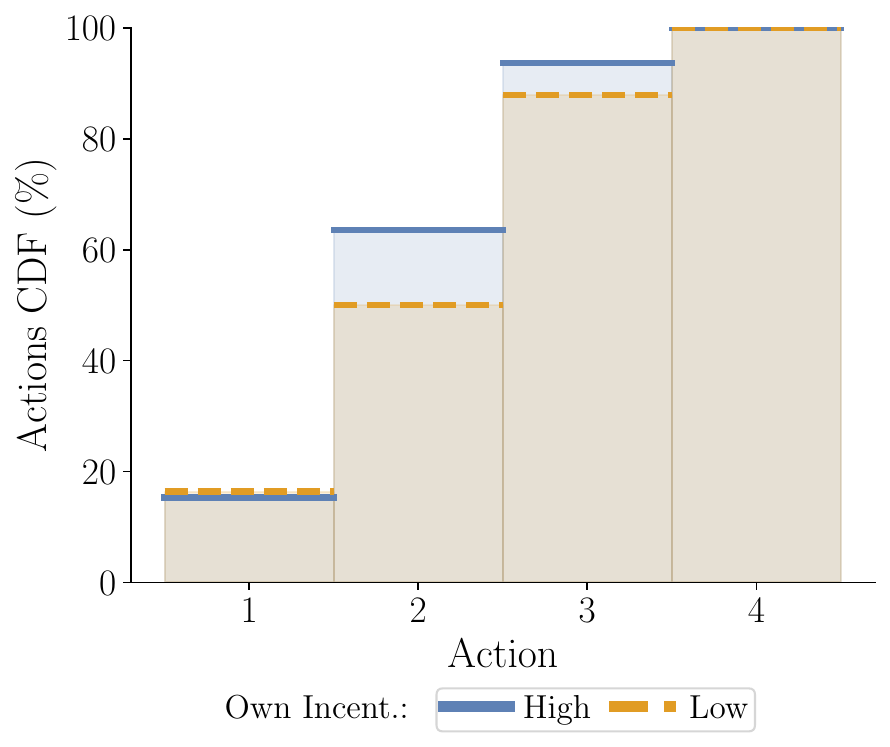}
            \caption{3-Step Game; High Opp. Incentive Level}
            \label{figure:action-sophistication-cdf-iesds3-oppi1}
        \end{subfigure}
        &
        \begin{subfigure}{.45\linewidth}
            \includegraphics[width=\linewidth]{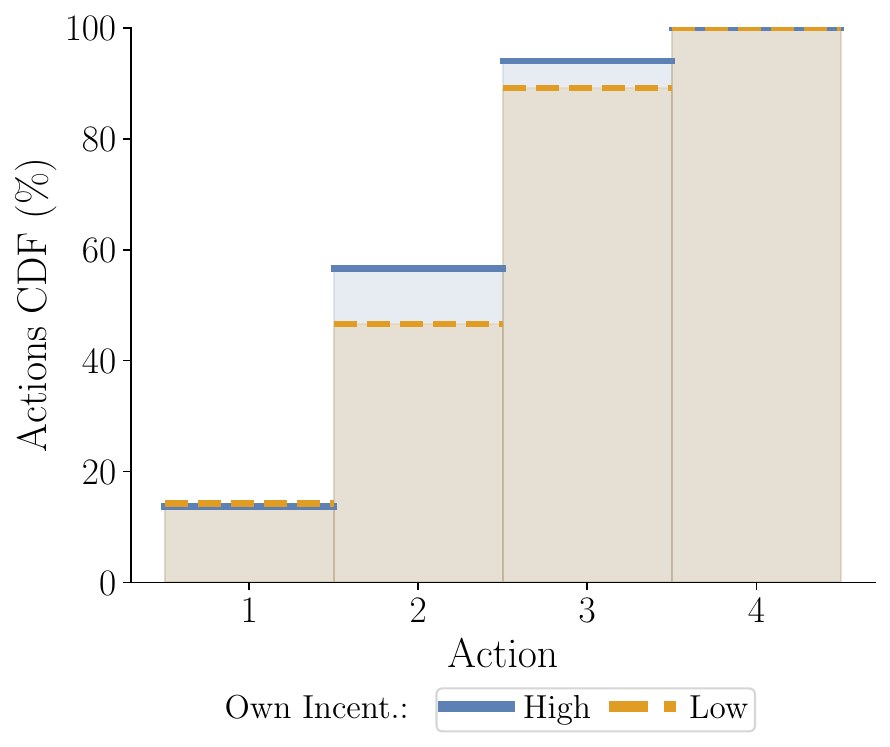}
            \caption{3-Step Game; Low Opp. Incentive Level}
            \label{figure:action-sophistication-cdf-iesds3-oppi0}
        \end{subfigure}
    \end{tabular}
    \begin{minipage}{1\linewidth}
        \caption{Incentive Level and Action Frequency: Own Incentives, Cumulative Frequency}
        \label{figure:action-sophistication-cdf-own}
        \emph{Notes}:
        The panels exhibit the action cumulative frequency according to the strategic ranking defined in \hyref{section:hypotheses-design}[Section] for high and low own incentive levels, for different games (2-Step game, (a) and (b), and 3-Step game, (c) and (d)) and holding fixed opponent incentive level (High, (a) and (c), or Low, (b) and (d)).
        2-Step and 3-Step games denote the different games in the experiment (see \hyref{figure:games}[Figure]).
    \end{minipage}
\end{figure}

\begin{figure}[!ht]
    \centering\small\singlespacing
    \begin{tabular}{@{\extracolsep{0pt}}cc@{}}
        \begin{subfigure}{.45\linewidth}
            \includegraphics[width=\linewidth]{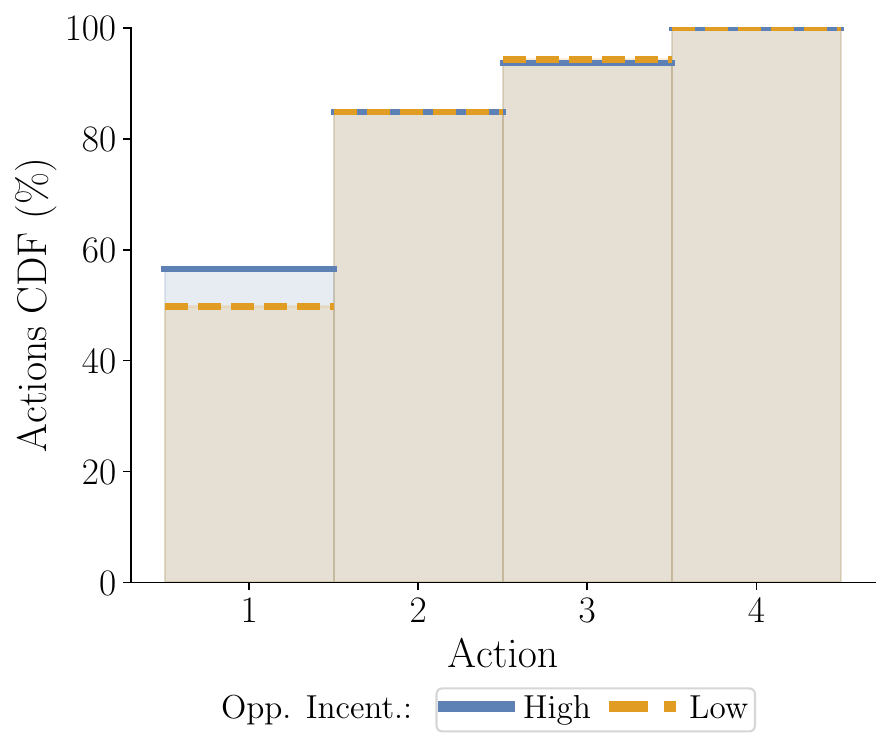}
            \caption{2-Step Game; High Own Incentive Level}
            \label{figure:action-sophistication-cdf-iesds2-owni1}
        \end{subfigure}
        &
        \begin{subfigure}{.45\linewidth}
            \includegraphics[width=\linewidth]{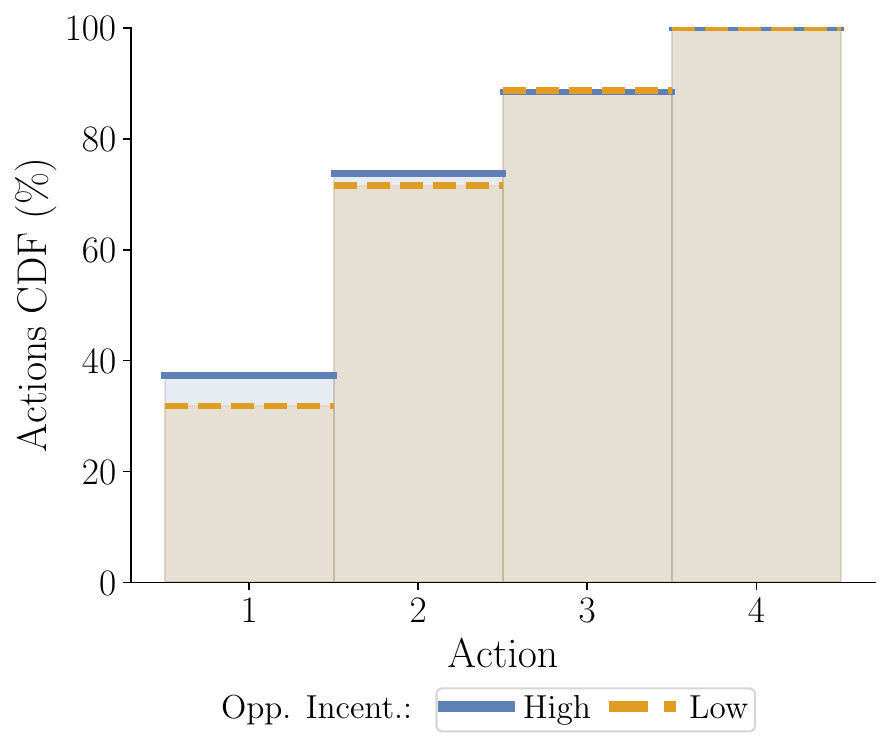}
            \caption{2-Step Game; Low Own Incentive Level}
            \label{figure:action-sophistication-cdf-iesds2-owni0}
        \end{subfigure}
        \\
        \begin{subfigure}{.45\linewidth}
            \includegraphics[width=\linewidth]{Figures/figure-action-sophistication-cdf-iesds3-oppi1.pdf}
            \caption{3-Step Game; High Own Incentive Level}
            \label{figure:action-sophistication-cdf-iesds3-oppi1}
        \end{subfigure}
        &
        \begin{subfigure}{.45\linewidth}
            \includegraphics[width=\linewidth]{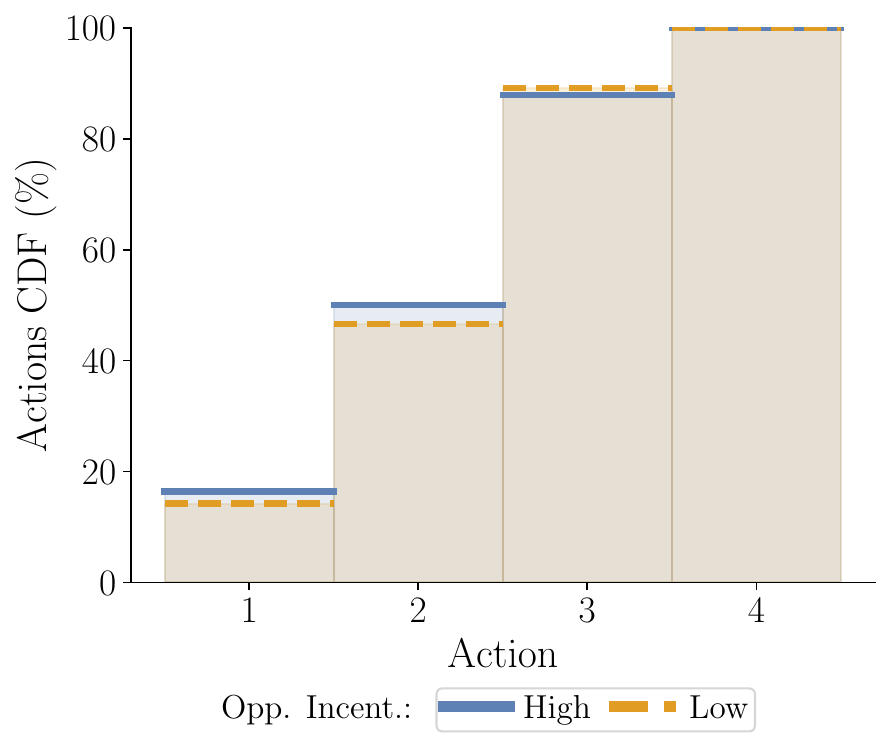}
            \caption{3-Step Game; Low Own Incentive Level}
            \label{figure:action-sophistication-cdf-iesds3-owni0}
        \end{subfigure}
    \end{tabular}
    \begin{minipage}{1\linewidth}
        \caption{Incentive Level and Action Frequency: Opponent Incentives, Cumulative Frequency}
        \label{figure:action-sophistication-cdf-opp}
        \emph{Notes}:
        The panels exhibit the action cumulative frequency according to the strategic ranking defined in \hyref{section:hypotheses-design}[Section] for high and low opponent incentive levels, for different games (2-Step game, (a) and (b), and 3-Step game, (c) and (d)) and holding fixed the participant's own incentive level (High, (a) and (c), or Low, (b) and (d)).
        2-Step and 3-Step games denote the different games in the experiment (see \hyref{figure:games}[Figure]).
    \end{minipage}
\end{figure}

\begin{figure}[!ht]
    \centering\small\singlespacing
    \begin{tabular}{@{\extracolsep{0pt}}cc@{}}
        \begin{subfigure}{.45\linewidth}
            \includegraphics[width=\linewidth]{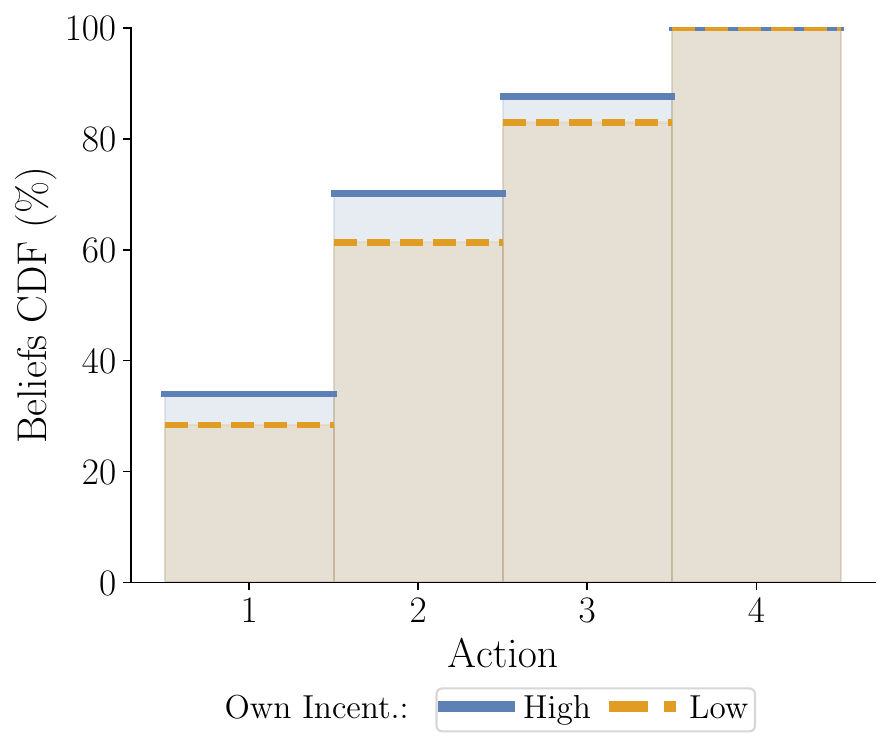}
            \caption{2-Step Game; High Opp. Incentive Level}
            \label{figure:belief-sophistication-cdf-iesds2-oppi1}
        \end{subfigure}
        &
        \begin{subfigure}{.45\linewidth}
            \includegraphics[width=\linewidth]{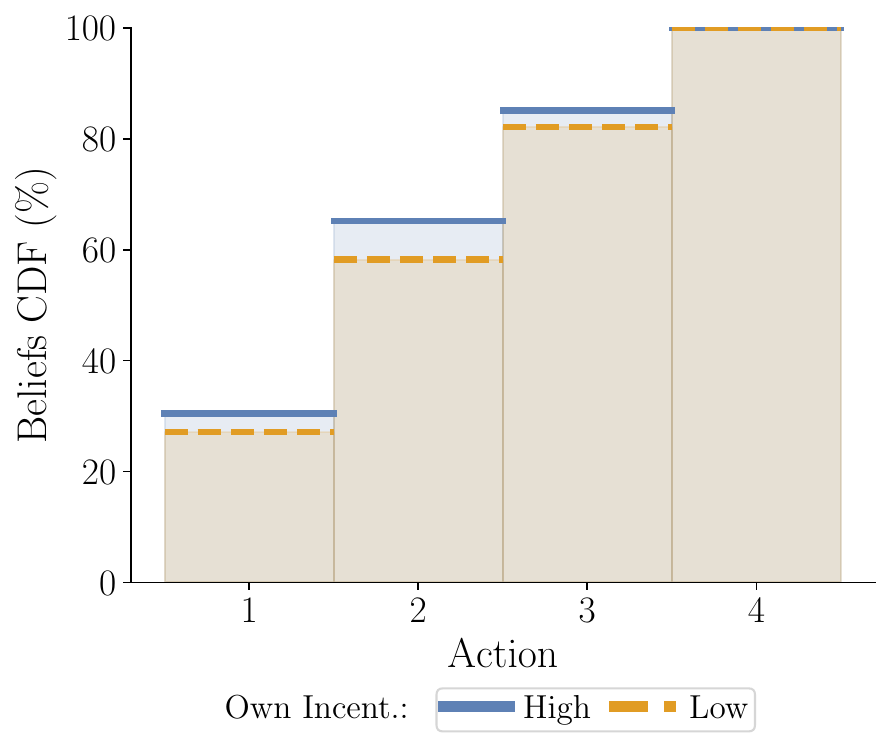}
            \caption{2-Step Game; Low Opp. Incentive Level}
            \label{figure:belief-sophistication-cdf-iesds2-oppi0}
        \end{subfigure}
        \\
        \begin{subfigure}{.45\linewidth}
            \includegraphics[width=\linewidth]{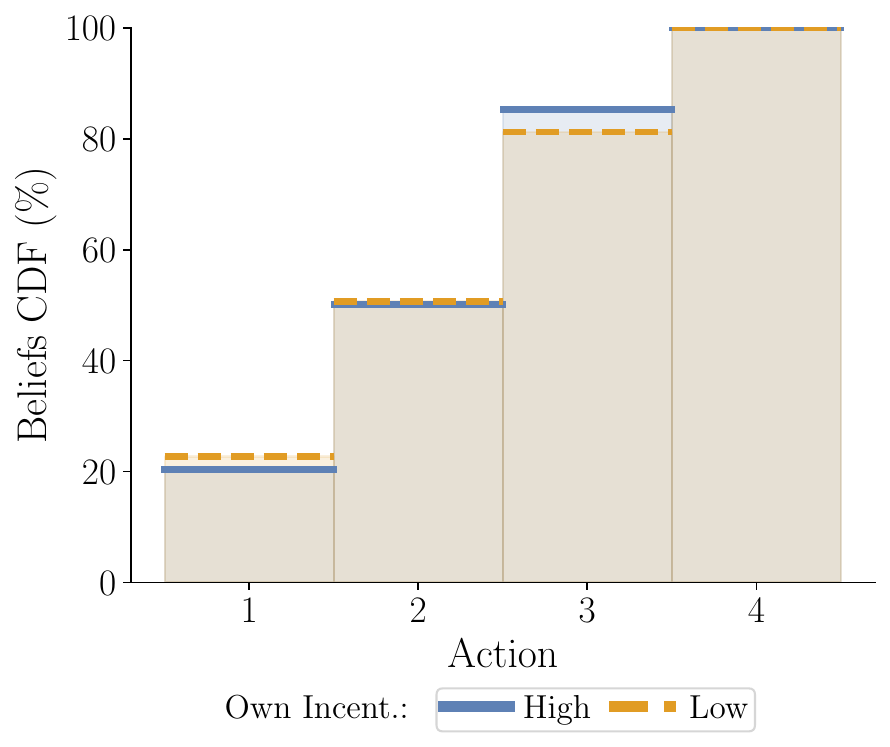}
            \caption{3-Step Game; High Opp. Incentive Level}
            \label{figure:belief-sophistication-cdf-iesds3-oppi1}
        \end{subfigure}
        &
        \begin{subfigure}{.45\linewidth}
            \includegraphics[width=\linewidth]{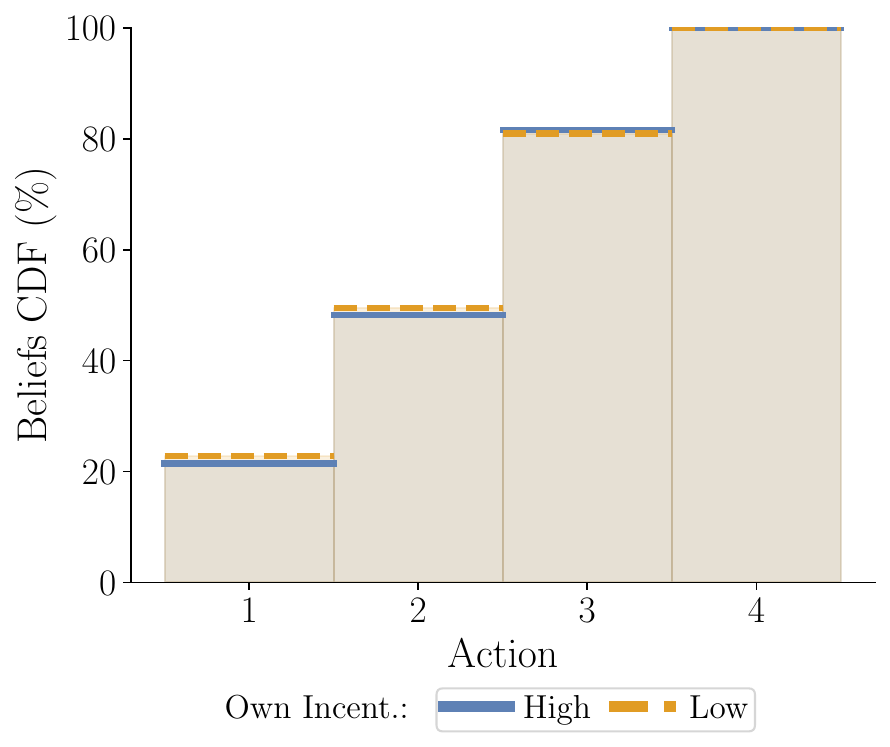}
            \caption{3-Step Game; Low Opp. Incentive Level}
            \label{figure:belief-sophistication-cdf-iesds3-oppi0}
        \end{subfigure}
    \end{tabular}
    \begin{minipage}{1\linewidth}
        \caption{Incentive Level and Belief CDFs: Own Incentives}
        \label{figure:belief-sophistication-cdf-own}
        \emph{Notes}:
        The different panels exhibit the cumulative mean reported beliefs according to the strategic ranking defined in \hyref{section:hypotheses-design}[Section] for high and low own incentive levels, for different games (2-Step game, (a) and (b), and 3-Step game, (c) and (d)) and holding fixed opponent incentive level (High, (a) and (c), or Low, (b) and (d)).
        2-Step and 3-Step games denote the different games in the experiment (see \hyref{figure:games}[Figure]).
    \end{minipage}
\end{figure}

\begin{figure}[!ht]
    \centering\small\singlespacing
    \begin{tabular}{@{\extracolsep{0pt}}cc@{}}
        \begin{subfigure}{.45\linewidth}
            \includegraphics[width=\linewidth]{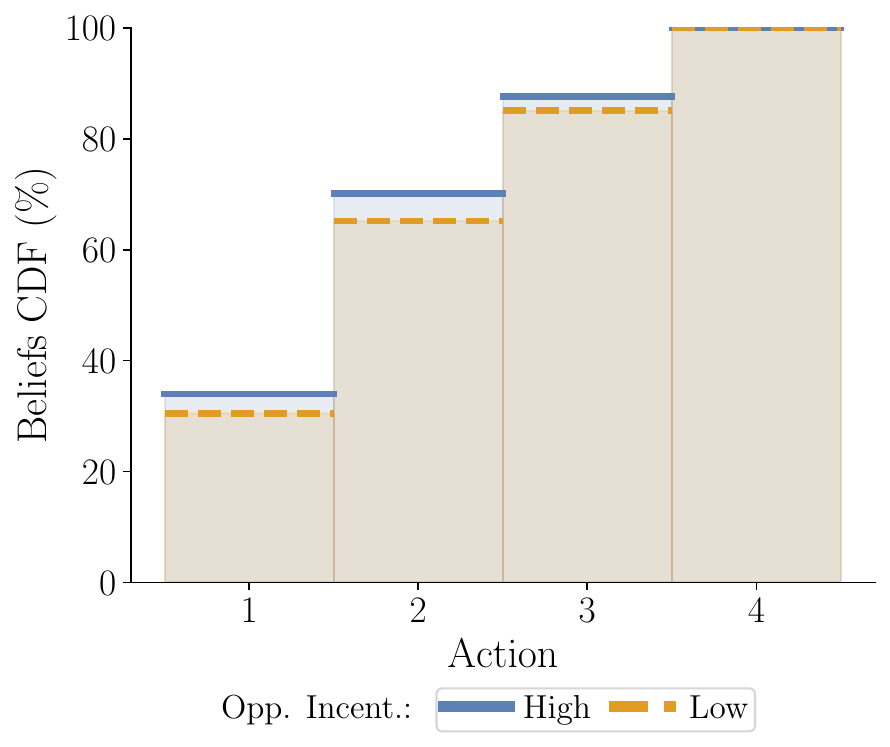}
            \caption{2-Step Game; High Own Incentive Level}
            \label{figure:belief-sophistication-cdf-iesds2-owni1}
        \end{subfigure}
        &
        \begin{subfigure}{.45\linewidth}
            \includegraphics[width=\linewidth]{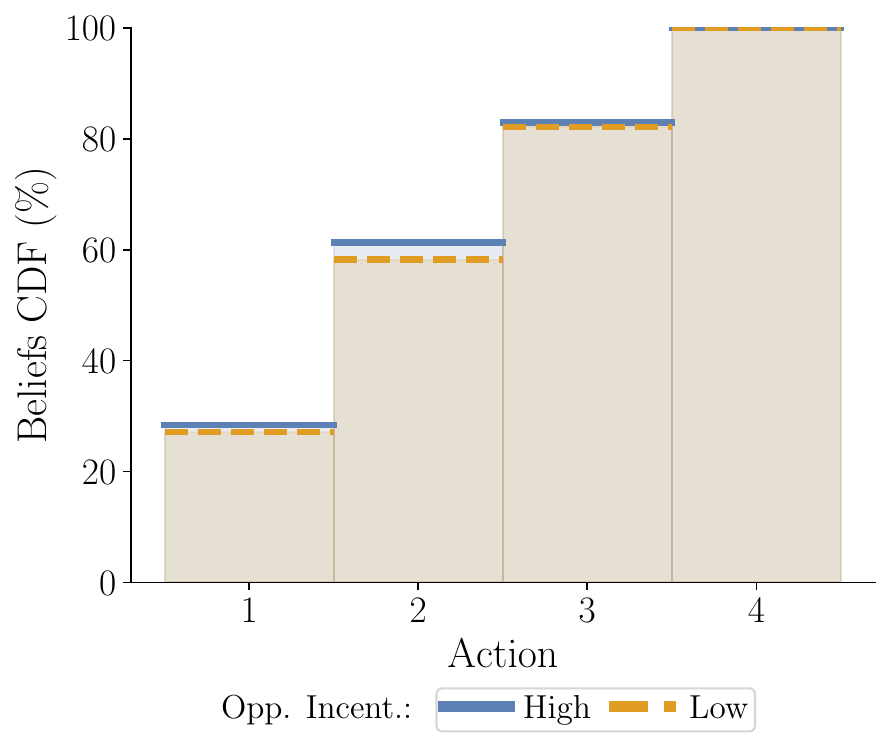}
            \caption{2-Step Game; Low Own Incentive Level}
            \label{figure:belief-sophistication-cdf-iesds2-owni0}
        \end{subfigure}
        \\
        \begin{subfigure}{.45\linewidth}
            \includegraphics[width=\linewidth]{Figures/figure-belief-sophistication-cdf-iesds3-oppi1.pdf}
            \caption{3-Step Game; High Own Incentive Level}
            \label{figure:belief-sophistication-cdf-iesds3-oppi1}
        \end{subfigure}
        &
        \begin{subfigure}{.45\linewidth}
            \includegraphics[width=\linewidth]{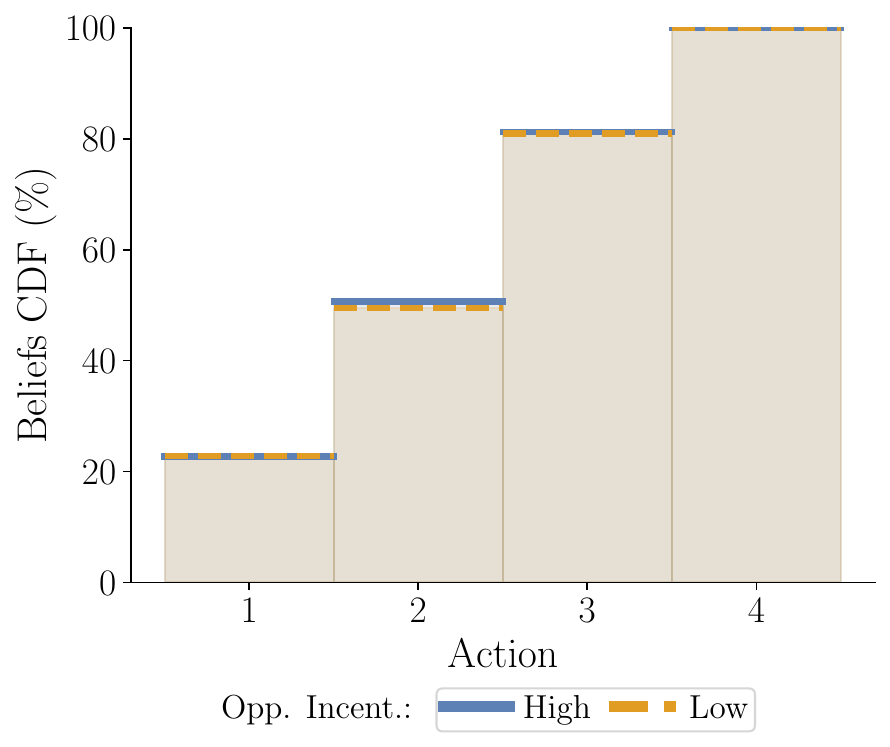}
            \caption{3-Step Game; Low Own Incentive Level}
            \label{figure:belief-sophistication-cdf-iesds3-owni0}
        \end{subfigure}
    \end{tabular}
    \begin{minipage}{1\linewidth}
        \caption{Incentive Level and Belief CDFs: Opponent Incentives}
        \label{figure:belief-sophistication-cdf-opp}
        \emph{Notes}:
        The different panels exhibit the cumulative mean reported beliefs according to the strategic ranking defined in \hyref{section:hypotheses-design}[Section] for high and low opponent incentive levels, for different games (2-Step game, (a) and (b), and 3-Step game, (c) and (d)) and holding fixed own incentive level (High, (a) and (c), or Low, (b) and (d)).
        2-Step and 3-Step games denote the different games in the experiment (see \hyref{figure:games}[Figure]).
    \end{minipage}
\end{figure}

\FloatBarrier \newpage

\subsection{Less Uniform Beliefs}
\label{section:appendix:additional-figures-tables:beliefs-less-uniform}

\begin{table}[!ht]\setstretch{1.1}
    \centering\small\singlespacing
	\begin{tabular}{l@{\extracolsep{4pt}}cc@{}}
\hline\hline
& \multicolumn{2}{c}{Distance to Uniform}  \\
\cline{2-3} 
& 2 Steps & 3 Steps \\
& (1) & (2) \\
\hline
Own Incent. & 7.41$^{***}$ & 3.76$^{*}$ \\
 & (2.42) & (2.13) \\ [.25em]
Opp. Incent. & 7.24$^{***}$ & 1.56 \\
 & (2.37) & (2.14) \\ [.25em]

Controls    & Yes & Yes \\
\hline
R$^2$ & 0.06 & 0.04 \\
N & 837 & 877 \\
\hline\hline
\end{tabular}
    \begin{minipage}{1\linewidth}
        \caption{Incentive Level and Belief Distance to Uniform Distribution}
        \label{table:belief-uniform}
        \emph{Notes}:
        This table shows the results for the regression specified in equation (\ref{equation:dominance}), considering as dependent variable the L1 distance between reported belief and the uniform distribution.
        High Own/Opponent Incentives correspond to indicators for whether the participant and their opponent face a high incentive level.
        2-Step and 3-Step games denote the different games in the experiment (see \hyref{figure:games}[Figure]).
        Controls refer to the participants' socio-demographic characteristics.
        Heterocedasticity-robust standard errors are given in parentheses; $^*$, $^{**}$, and $^{***}$ denote p-values $< 0.1$, $<0.05$, and $<0.01$, respectively.
    \end{minipage}
\end{table}

\FloatBarrier \newpage

\subsection{Actions, Beliefs, and Response Time}
\label{section:appendix:additional-figures-tables:rt}

\begin{table}[ht!]\setstretch{1.1}
    \centering\small\singlespacing
	\begin{tabular}{l@{\extracolsep{4pt}}cccccccc@{}}
\hline\hline
& \multicolumn{2}{c}{Action 1} & \multicolumn{2}{c}{Action 2} & \multicolumn{2}{c}{Action 3} & \multicolumn{2}{c}{Action 4}  \\
\cline{2-3} \cline{4-5} \cline{6-7} \cline{8-9} 
& 2 Steps & 3 Steps & 2 Steps & 3 Steps & 2 Steps & 3 Steps & 2 Steps & 3 Steps \\
& (1) & (2) & (3) & (4) & (5) & (6) & (7) & (8) \\
\hline
Log(Time) & 17.89$^{***}$ & 1.96 & -5.70$^{***}$ & 12.53$^{***}$ & -6.20$^{***}$ & -8.48$^{***}$ & -5.99$^{***}$ & -6.00$^{***}$ \\
 & (2.10) & (1.75) & (2.17) & (1.97) & (1.48) & (1.92) & (1.27) & (1.21) \\ [.25em]

Controls    & Yes & Yes & Yes & Yes & Yes & Yes & Yes & Yes \\
\hline
R$^2$ & 0.10 & 0.02 & 0.02 & 0.07 & 0.05 & 0.05 & 0.07 & 0.05 \\
N & 837 & 877 & 837 & 877 & 837 & 877 & 837 & 877 \\
\hline\hline
\end{tabular}
    \begin{minipage}{1\linewidth}
        \caption{Action Frequency and Response Time}
        \label{table:action-rt}
        \emph{Notes}:
        This table shows the results for the association between choice of a particular action $a_n$ and response time (in seconds) and incentive level treatments, controlling for individual characteristics.
        Different columns refer to linear probability models relating the choice of an action in a game (2-Step game in columns (1), (3), (5), and (7); 3-Step game in columns (2), (4), (6), and (8)).
        Errors across regressions are correlated by construction and the table is to be taken as describing measures of association supporting the patterns described in \hyref{figure:action-rt-mnl}[Figure].
        2 Steps and 3 Steps denote the different games in the experiment (see \hyref{figure:games}[Figure]).
        Controls refer to the participants' socio-demographic characteristics.
        Heterocedasticity-robust standard errors are given in parentheses; $^*$, $^{**}$, and $^{***}$ denote p-values $< 0.1$, $<0.05$, and $<0.01$, respectively.
    \end{minipage}
\end{table}

\begin{table}[ht!]\setstretch{1.1}
    \centering\small\singlespacing
	\begin{tabular}{l@{\extracolsep{4pt}}cccccccc@{}}
\hline\hline
& \multicolumn{2}{c}{Action 1} & \multicolumn{2}{c}{Action 2} & \multicolumn{2}{c}{Action 3} & \multicolumn{2}{c}{Action 4}  \\
\cline{2-3} \cline{4-5} \cline{6-7} \cline{8-9} 
& 2 Steps & 3 Steps & 2 Steps & 3 Steps & 2 Steps & 3 Steps & 2 Steps & 3 Steps \\
& (1) & (2) & (3) & (4) & (5) & (6) & (7) & (8) \\
\hline
Log(Time) & 5.16$^{***}$ & -1.50$^{**}$ & 5.37$^{***}$ & 2.10$^{***}$ & -4.57$^{***}$ & 4.91$^{***}$ & -5.96$^{***}$ & -5.51$^{***}$ \\
 & (0.65) & (0.59) & (0.67) & (0.52) & (0.50) & (0.66) & (0.39) & (0.47) \\ [.25em]

Controls    & Yes & Yes & Yes & Yes & Yes & Yes & Yes & Yes \\
\hline
R$^2$ & 0.09 & 0.03 & 0.10 & 0.04 & 0.13 & 0.08 & 0.23 & 0.16 \\
N & 837 & 877 & 837 & 877 & 837 & 877 & 837 & 877 \\
\hline\hline
\end{tabular}
    \begin{minipage}{1\linewidth}
        \caption{Beliefs and Response Time}
        \label{table:belief-rt}
        \emph{Notes}:
        This table shows the results for the association between belief $b_n$ in opponents choosing a particular action $a_n$ and response time (in seconds) and incentive level treatments, controlling for individual characteristics.
        Different columns refer to linear regressions relating the reported beliefs referring to a particular action in a given game (2-Step game in columns (1), (3), (5), and (7); 3-Step game in columns (2), (4), (6), and (8)).
        Errors across regressions are correlated by construction and the table is to be taken as describing measures of association supporting the patterns described in \hyref{figure:belief-rt-kernel}[Figure].
        2 Steps and 3 Steps denote the different games in the experiment (see \hyref{figure:games}[Figure]).
        Controls refer to the participants' socio-demographic characteristics.
        Heterocedasticity-robust standard errors are given in parentheses; $^*$, $^{**}$, and $^{***}$ denote p-values $< 0.1$, $<0.05$, and $<0.01$, respectively.
    \end{minipage}
\end{table}

\FloatBarrier \newpage

\subsection{Quantal Response Estimation}
\label{section:appendix:additional-figures-tables:qre}

\begin{figure}[ht!]
    \centering\small\singlespacing
    \begin{tabular}{@{\extracolsep{0pt}}cc@{}}
        \begin{subfigure}{.45\linewidth}
            \includegraphics[width=\linewidth]{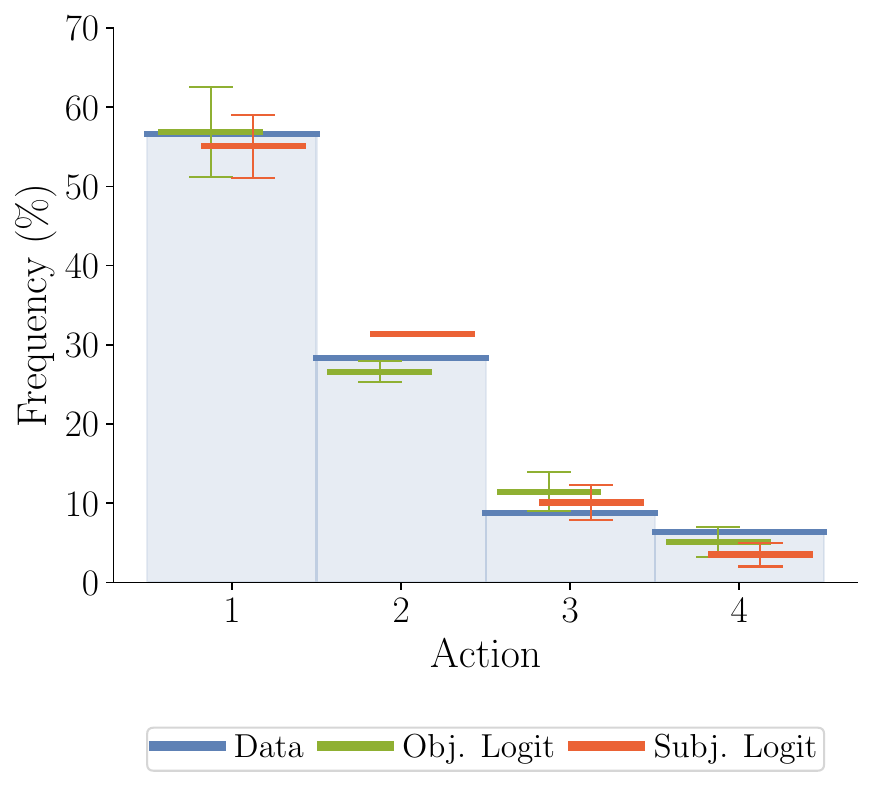}
            \caption{2-Step Game; High Incentive Level}
            \label{figure:lqre-fit-iesds2-owni1-oppi1}
        \end{subfigure}
        &
        \begin{subfigure}{.45\linewidth}
            \includegraphics[width=\linewidth]{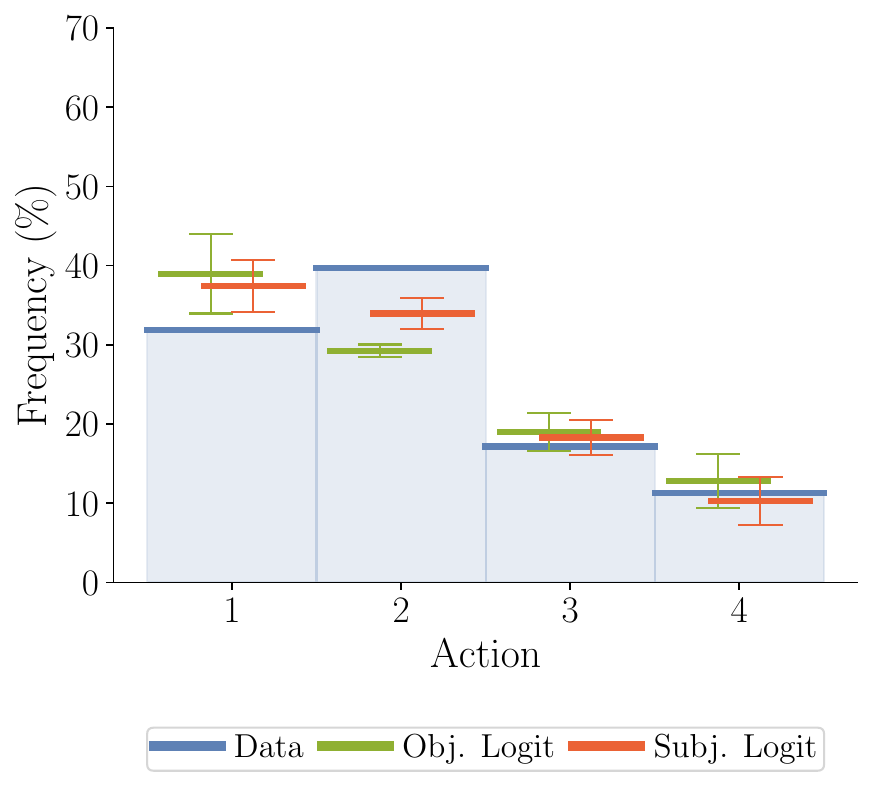}
            \caption{2-Step Game; Low Incentive Level}
            \label{figure:lqre-fit-iesds2-owni0-oppi0}
        \end{subfigure}
        \\
        \begin{subfigure}{.45\linewidth}
            \includegraphics[width=\linewidth]{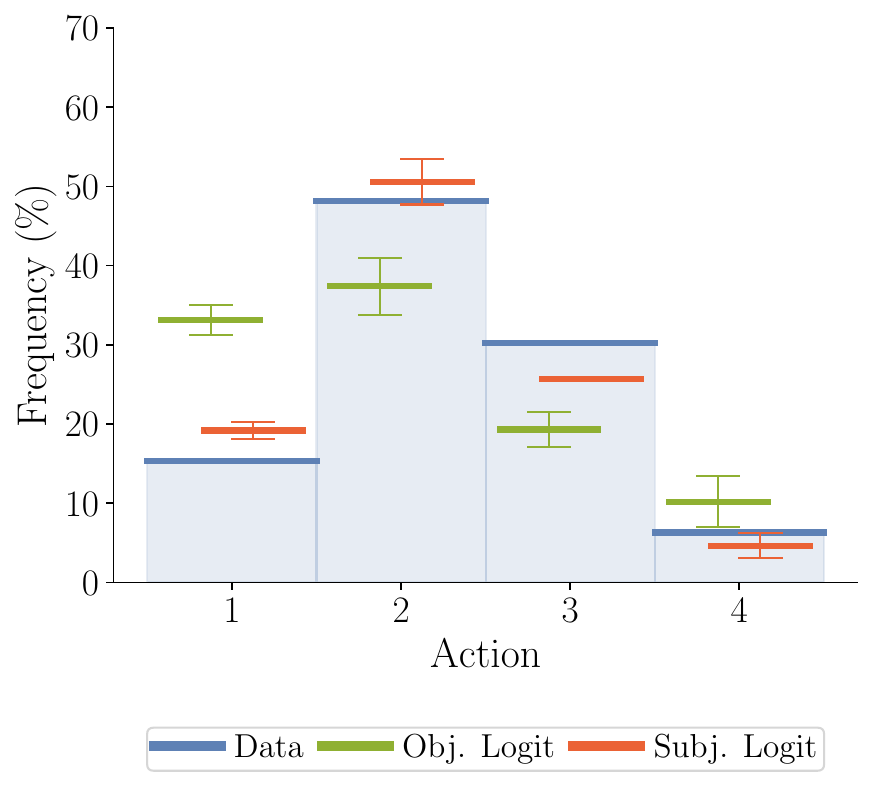}
            \caption{3-Step Game; High Incentive Level}
            \label{figure:lqre-fit-iesds3-owni1-oppi1}
        \end{subfigure}
        &
        \begin{subfigure}{.45\linewidth}
            \includegraphics[width=\linewidth]{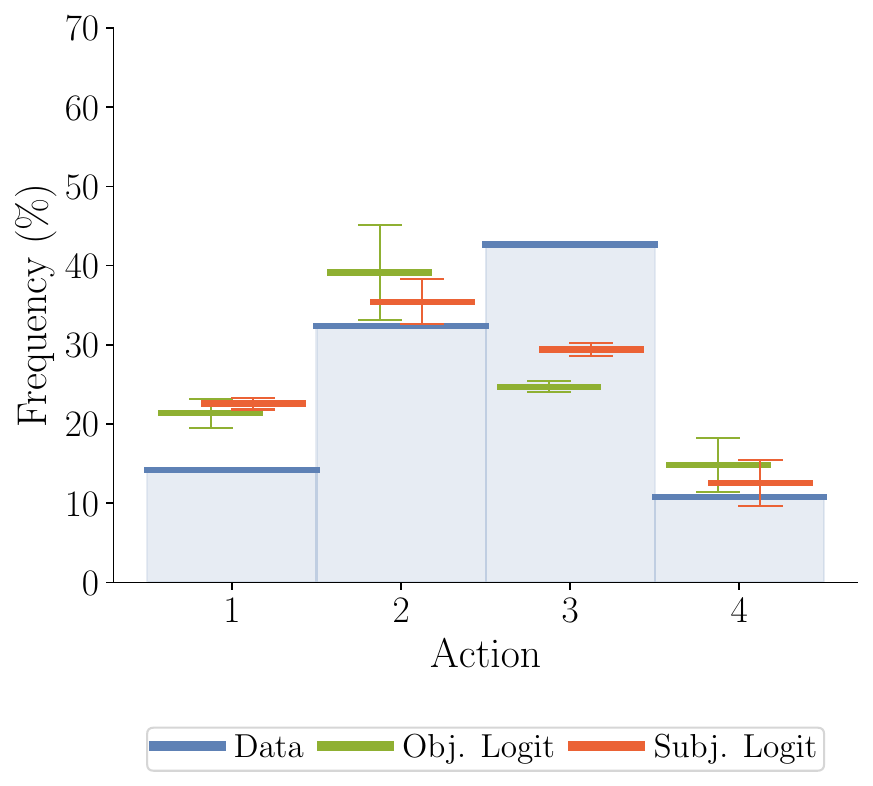}
            \caption{3-Step Game; Low Incentive Level}
            \label{figure:logit-fit-iesds3-owni0-oppi0}
        \end{subfigure}
    \end{tabular}
    \begin{minipage}{1\linewidth}
        \caption{Logit Quantal Response}
        \label{figure:logit-fit}
        \emph{Notes}:
        The figure displays the empirical frequency of choices in the data (blue), the predicted choice frequencies obtained via logit quantal response using either objective opponent choice frequencies (green) or reported beliefs about opponent's choices (red), both of which estimated using maximum likelihood estimation.
        Different panels correspond to different incentive levels (High, (a) and (c), and Low, (b) and (d)) and to different games (2-Step game, (a) and (b), and 3-Step game, (c) and (d)).
        We restrict observations to cases in which own and opponents' incentive levels are the same.
        2-Step and 3-Step games denote the different games in the experiment (see \hyref{figure:games}[Figure]).
        Whiskers represent 95\% confidence intervals.
    \end{minipage}
\end{figure}

\FloatBarrier \newpage
\subsection{Sample Characteristics}
\label{section:appendix:additional-figures-tables:sample-demographics}
~

\begin{table}[!ht]\setstretch{1.1}
    \centering\small\singlespacing
	\begin{tabular}{@{\extracolsep{4pt}}cccc@{}}
\hline\hline
\multicolumn{1}{c}{Pool}  &  \multicolumn{1}{c}{Avg. Age}  &  \multicolumn{1}{c}{University Degree (\%)}  &  \multicolumn{1}{c}{Post-Graduate Degree (\%)} \\
\cline{1-1} \cline{2-2} \cline{3-3} \cline{4-4}
 (1) & (2) & (3) & (4) \\
\hline
MTurk & 36.83 & 81.06 & 11.51 \\
Prolific & 42.32 & 84.77 & 21.93 \\
\hline\hline
\end{tabular}
    \begin{minipage}{1\linewidth}
        \caption{Sample Socio-Demographic Characteristics}
        \label{table:sample-sample-avg-dem}
        \emph{Notes}:
        This table reports socio-demographic characteristics by participant pool: the average age (column (2)), the percentage of participants who completed a bachelor's degree (column (3)) and a postgraduate degree (column (4)).
    \end{minipage}
\end{table}

\begin{table}[!ht]\setstretch{1.1}
    \centering\small\singlespacing
	\begin{tabular}{l@{\extracolsep{4pt}}cc@{}}
\hline\hline
\multicolumn{1}{c}{Education Field}  &  \multicolumn{1}{c}{MTurk}  &  \multicolumn{1}{c}{Prolific} \\
\cline{1-1} \cline{2-2} \cline{3-3}
 \multicolumn{1}{c}{(1)} & \multicolumn{1}{c}{(2)} & \multicolumn{1}{c}{(3)} \\
\hline
Agriculture, Forestry, Fisheries, and Veterinary & 6 & 3 \\
Arts and Humanities & 128 & 113 \\
Business, Administration, and Law & 146 & 163 \\
Computer Science, Information, and Communication Technologies & 130 & 138 \\
Economics & 29 & 23 \\
Education & 42 & 49 \\
Engineering, Manufacturing, and Construction & 34 & 55 \\
Generic & 107 & 77 \\
Health and Welfare & 54 & 96 \\
Mathematics and Statistics & 21 & 25 \\
Natural Sciences & 44 & 42 \\
Services (Transport, Hygiene and Health, Security, and Other) & 39 & 38 \\
Social Sciences and Journalism & 54 & 58 \\
\hline\hline
\end{tabular}
    \begin{minipage}{1\linewidth}
        \caption{Sample Characteristics: Field of Education}
        \label{table:sample-sample-avg-demeducfield}
        \emph{Notes}:
        This table reports the number of participants who self-reported each field of study in the experiment run on Amazon Mechanical Turk (column (2)) and the experiment run on Prolific (column (3)).
    \end{minipage}
\end{table}

\FloatBarrier \newpage
\subsection{Incentive-Interaction Effects}
\label{section:appendix:additional-figures-tables:interaction}
~

\begin{table}[!ht]\setstretch{1.1}
    \centering\small\singlespacing
	\begin{tabular}{l@{\extracolsep{4pt}}cccc@{}}
\hline\hline
& \multicolumn{2}{c}{Dominance Play} & \multicolumn{2}{c}{Dominated Play}  \\
\cline{2-3} \cline{4-5} 
& 2 Steps & 3 Steps & 2 Steps & 3 Steps \\
& (1) & (2) & (3) & (4) \\
\hline
Own Incent. & 19.68$^{***}$ & -1.27 & -14.88$^{***}$ & -4.60 \\
 & (4.85) & (3.39) & (4.09) & (2.86) \\ [.25em]
Opp. Incent. & 5.78 & 1.58 & -2.45 & 1.30 \\
 & (4.75) & (3.44) & (4.41) & (3.07) \\ [.25em]
Own Incent. x Opp Incent. & -0.01 & 0.32 & 3.79 & -1.47 \\
 & (6.91) & (4.86) & (5.76) & (3.89) \\ [.25em]

Controls    & Yes & Yes & Yes & Yes \\
\hline
R$^2$ & 0.06 & 0.02 & 0.05 & 0.03 \\
N & 837 & 877 & 837 & 877 \\
\hline\hline
\end{tabular}
    \begin{minipage}{1\linewidth}
        \caption{Incentive Level, Dominance, and Dominated Play (\hyref{hypothesis:action-sophistication}[Hypothesis]): Incentive Interaction}
        \label{table:action-dominance-interaction}
        \emph{Notes}:
        This table reports estimates from the regression in \hyref{table:action-dominance}[Table] augmented to include the interaction between own and opponents' incentive levels.
        The dependent variable is an indicator for whether the participant chose the dominance-solution action (columns (1) and (2)) or a strictly dominated action (columns (3) and (4)). 
        High Own/Opponent Incentives are indicators for whether the participant and their opponent face a high incentive level. 
        2 Steps and 3 Steps denote the two games in the experiment; see \hyref{figure:games}[Figure]. 
        Controls refer to participants' socio-demographic characteristics.
        Heterocedasticity-robust standard errors are reported in parentheses; $^{*}$, $^{**}$, and $^{***}$ denote $p<0.10$, $p<0.05$, and $p<0.01$, respectively.
    \end{minipage}
\end{table}

\begin{table}[!ht]\setstretch{1.1}
    \centering\small\singlespacing
	\begin{tabular}{l@{\extracolsep{4pt}}cccc@{}}
\hline\hline
& \multicolumn{2}{c}{Subjective BR} & \multicolumn{2}{c}{Objective BR}  \\
\cline{2-3} \cline{4-5} 
& 2 Steps & 3 Steps & 2 Steps & 3 Steps \\
& (1) & (2) & (3) & (4) \\
\hline
Own Incent. & 20.48$^{***}$ & 12.40$^{**}$ & 19.68$^{***}$ & 10.88$^{**}$ \\
 & (4.93) & (4.96) & (4.85) & (4.76) \\ [.25em]
Opp. Incent. & 5.24 & 0.27 & 5.78 & 0.85 \\
 & (4.96) & (4.79) & (4.75) & (4.58) \\ [.25em]
Own Incent. x Opp Incent. & 0.70 & 6.45 & -0.01 & 3.68 \\
 & (6.92) & (6.79) & (6.91) & (6.61) \\ [.25em]

Controls    & Yes & Yes & Yes & Yes \\
\hline
R$^2$ & 0.07 & 0.05 & 0.06 & 0.04 \\
N & 837 & 877 & 837 & 877 \\
\hline\hline
\end{tabular}
    \begin{minipage}{1\linewidth}
        \caption{Incentive Level and Best-Response Rate (\hyref{hypothesis:best-responses}[Hypothesis][a]): Incentive Interaction}
        \label{table:action-br-interaction}
        \emph{Notes}: 
        This table reports estimates from the regression in \hyref{table:action-br}[Table] augmented to include the interaction between own and opponents' incentive levels. 
        The dependent variable is an indicator for whether the participant best responds to their reported beliefs --- that is, chooses the action that maximises subjective expected payoff according to those beliefs (columns (1) and (2)) --- or best responds to the empirical frequency of opponents' actions --- that is, an objective best response (columns (3) and (4)). 
        High Own/Opponent Incentives correspond to indicators for whether the participant and their opponent face a high incentive level.
        2 Steps and 3 Steps denote the different games in the experiment (see \hyref{figure:games}[Figure]).
        Controls refer to the participants' socio-demographic characteristics.
        Heterocedasticity-robust standard errors are given in parentheses; $^*$, $^{**}$, and $^{***}$ denote p-values $< 0.1$, $<0.05$, and $<0.01$, respectively.
    \end{minipage}
\end{table}

\begin{table}[!ht]\setstretch{1.1}
    \centering\small\singlespacing
	\begin{tabular}{l@{\extracolsep{4pt}}cccc@{}}
\hline\hline
& \multicolumn{2}{c}{Belief in}  &  \multicolumn{2}{c}{Belief in} \\
& \multicolumn{2}{c}{Dominance Play} & \multicolumn{2}{c}{Dominated Play}  \\
\cline{2-3} \cline{4-5} 
& 2 Steps & 3 Steps & 2 Steps & 3 Steps \\
& (1) & (2) & (3) & (4) \\
\hline
Own Incent. & 3.57$^{**}$ & -1.35 & -7.17$^{***}$ & -0.72 \\
 & (1.61) & (1.38) & (1.78) & (1.17) \\ [.25em]
Opp. Incent. & 0.95 & -0.26 & -3.26$^{*}$ & -0.15 \\
 & (1.49) & (1.37) & (1.82) & (1.20) \\ [.25em]
Own Incent. x Opp Incent. & 2.20 & -0.90 & -0.99 & -3.33$^{**}$ \\
 & (2.40) & (1.96) & (2.54) & (1.64) \\ [.25em]

Controls    & Yes & Yes & Yes & Yes \\
\hline
R$^2$ & 0.06 & 0.02 & 0.09 & 0.05 \\
N & 837 & 877 & 837 & 877 \\
\hline\hline
\end{tabular}
    \begin{minipage}{1\linewidth}
        \caption{Incentive Level and Belief in Opponent Sophistication (\hyref{hypothesis:belief-sophistication}[Hypotheses][a-b]): Incentive Interaction}
        \label{table:belief-dominance-interaction}
        \emph{Notes}:
        This table reports estimates from the regression in \hyref{table:belief-dominance}[Table] augmented to include the interaction between own and opponents' incentive levels. 
        The dependent variable is the reported belief in dominance play, $b_{i,1}$, (columns (1) and (2)) or the reported belief in strictly dominated play, $b_{i,3}+b_{i,4}$ in column (3) and $b_{i,4}$ in column (4).
        High Own/Opponent Incentives correspond to indicators for whether the participant and their opponent face a high incentive level.
        2 Steps and 3 Steps denote the different games in the experiment (see \hyref{figure:games}[Figure]).
        Controls refer to the participants' socio-demographic characteristics.
        Heterocedasticity-robust standard errors are given in parentheses; $^*$, $^{**}$, and $^{***}$ denote p-values $< 0.1$, $<0.05$, and $<0.01$, respectively.
    \end{minipage}
\end{table}

\begin{table}[!ht]\setstretch{1.1}
    \centering\small\singlespacing
	\begin{tabular}{l@{\extracolsep{4pt}}cccccc@{}}
\hline\hline
& \multicolumn{4}{c}{|Belief - Opponent Action Frequency|} & \multicolumn{2}{c}{|Subj. - Obj. Payoff|} \\
& \multicolumn{2}{c}{Dominance Play}  &  \multicolumn{2}{c}{Dominated Play} & & \\
\cline{2-3} \cline{4-5} \cline{6-7} 
& 2 Steps & 3 Steps & 2 Steps & 3 Steps & 2 Steps & 3 Steps \\
& (1) & (2) & (3) & (4) & (5) & (6) \\
\hline
Own Incent. & 0.67 & -0.26 & -3.91$^{***}$ & -0.26 & -3.45$^{***}$ & -1.41 \\
 & (1.13) & (1.04) & (1.10) & (0.87) & (1.24) & (1.21) \\ [.25em]
Opp. Incent. & 18.82$^{***}$ & -0.69 & 7.16$^{***}$ & 2.94$^{***}$ & 10.18$^{***}$ & 6.48$^{***}$ \\
 & (1.08) & (1.07) & (1.33) & (0.97) & (1.28) & (1.23) \\ [.25em]
Own Incent. x Opp Incent. & -4.94$^{***}$ & -0.76 & -2.85 & -2.86$^{**}$ & -2.87 & -0.77 \\
 & (1.73) & (1.51) & (1.76) & (1.28) & (1.82) & (1.64) \\ [.25em]

Controls    & Yes & Yes & Yes & Yes & Yes & Yes \\
\hline
R$^2$ & 0.34 & 0.03 & 0.13 & 0.05 & 0.17 & 0.09 \\
N & 837 & 877 & 837 & 877 & 837 & 877 \\
\hline\hline
\end{tabular}
    \begin{minipage}{1\linewidth}
        \caption{Incentive Level and Belief Accuracy (\hyref{hypothesis:belief-sophistication}[Hypothesis][c]): Incentive Interaction}
        \label{table:belief-accuracy-interaction}
        \emph{Notes}:
        This table reports estimates from the regression in \hyref{table:belief-accuracy}[Table] augmented to include the interaction between own and opponents' incentive levels. 
        The dependent variable is the absolute difference between reported belief in dominance play by opponents and its realised frequency, $|b_{i,1}-\overline \sigma_{-i,1}|$, (columns (1) and (2)), or the analogous absolute difference for strictly dominated play, $|b_{i,3}+b_{i,4}-\overline \sigma_{-i,3}-\overline \sigma_{-i,4}|$ in column (3) and $|b_{i,4}-\overline \sigma_{-i,4}|$ in column (4). 
        In columns (5) and (6) the dependent variable is the L1-norm of the difference between subjective expected payoffs (according to reported beliefs) and objective expected payoffs (according to observed action frequencies).
        High Own/Opponent Incentives correspond to indicators for whether the participant and their opponent face a high incentive level.
        2 Steps and 3 Steps denote the different games in the experiment (see \hyref{figure:games}[Figure]).
        Controls refer to the participants' socio-demographic characteristics.
        Heterocedasticity-robust standard errors are given in parentheses; $^*$, $^{**}$, and $^{***}$ denote p-values $< 0.1$, $<0.05$, and $<0.01$, respectively.
    \end{minipage}
\end{table}

\begin{table}[!ht]\setstretch{1.1}
    \centering\small\singlespacing
	\begin{tabular}{l@{\extracolsep{4pt}}cc@{}}
\hline\hline
& \multicolumn{2}{c}{Distance to Uniform}  \\
\cline{2-3} 
& 2 Steps & 3 Steps \\
& (1) & (2) \\
\hline
Own Incent. & 5.29 & 0.86 \\
 & (3.36) & (3.12) \\ [.25em]
Opp. Incent. & 5.16 & -1.24 \\
 & (3.27) & (2.97) \\ [.25em]
Own Incent. x Opp Incent. & 4.21 & 5.56 \\
 & (4.90) & (4.28) \\ [.25em]

Controls    & Yes & Yes \\
\hline
R$^2$ & 0.06 & 0.05 \\
N & 837 & 877 \\
\hline\hline
\end{tabular}
    \begin{minipage}{1\linewidth}
        \caption{Incentive Level and Belief Distance to Uniform Distribution: Incentive Interaction}
        \label{table:belief-uniform-interaction}
        \emph{Notes}:
        This table reports estimates from the regression in \hyref{table:belief-uniform}[Table] augmented to include the interaction between own and opponents' incentive levels. 
        The dependent variable is the L1 distance between reported belief and the uniform distribution. 
        High Own/Opponent Incentives correspond to indicators for whether the participant and their opponent face a high incentive level.
        2 Steps and 3 Steps denote the different games in the experiment (see \hyref{figure:games}[Figure]).
        Controls refer to the participants' socio-demographic characteristics.
        Heterocedasticity-robust standard errors are given in parentheses; $^*$, $^{**}$, and $^{***}$ denote p-values $< 0.1$, $<0.05$, and $<0.01$, respectively.
    \end{minipage}
\end{table}

\begin{table}[!ht]\setstretch{1.1}
    \centering\small\singlespacing
	\begin{tabular}{l@{\extracolsep{4pt}}cc@{}}
\hline\hline
& \multicolumn{2}{c}{Log(Time)}  \\
\cline{2-3} 
& 2 Steps & 3 Steps \\
& (1) & (2) \\
\hline
Own Incent. & 0.34$^{***}$ & 0.38$^{***}$ \\
 & (0.08) & (0.08) \\ [.25em]
Opp. Incent. & -0.02 & 0.20$^{***}$ \\
 & (0.08) & (0.07) \\ [.25em]
Own Incent. x Opp Incent. & 0.08 & -0.03 \\
 & (0.11) & (0.11) \\ [.25em]

Controls    & Yes & Yes \\
\hline
R$^2$ & 0.10 & 0.13 \\
N & 837 & 877 \\
\hline\hline
\end{tabular}
    \begin{minipage}{1\linewidth}
        \caption{Incentive Level and Response Time (\hyref{hypothesis:response-time}[Hypothesis][a]): Incentive Interaction}
        \label{table:rt-interaction}
        \emph{Notes}:
        This table reports estimates from the regression in \hyref{table:rt}[Table] augmented to include the interaction between own and opponents' incentive levels. 
        The dependent variable is the participant's log response time (in seconds).
        High Own/Opponent Incentives correspond to indicators for whether the participant and their opponent face a high incentive level.
        2 Steps and 3 Steps denote the different games in the experiment (see \hyref{figure:games}[Figure]).
        Controls refer to the participants' socio-demographic characteristics.
        Heterocedasticity-robust standard errors are given in parentheses; $^*$, $^{**}$, and $^{***}$ denote p-values $< 0.1$, $<0.05$, and $<0.01$, respectively.
    \end{minipage}
\end{table}

\begin{table}[!ht]\setstretch{1.1}
    \centering\small\singlespacing
	\begin{tabular}{l@{\extracolsep{4pt}}cccc@{}}
\hline\hline
& \multicolumn{2}{c}{Objective BR} & \multicolumn{2}{c}{Log(Obj. Exp. Payoff}  \\
\cline{2-3} \cline{4-5} 
& 2 Steps & 3 Steps & 2 Steps & 3 Steps \\
& (1) & (2) & (3) & (4) \\
\hline
Log(Time) & 15.77$^{***}$ & 11.31$^{***}$ & 0.10$^{***}$ & 0.06$^{***}$ \\
 & (2.18) & (2.05) & (0.01) & (0.01) \\ [.25em]
Own Incent. & 14.39$^{***}$ & 6.56 & 0.08$^{***}$ & 0.02 \\
 & (4.81) & (4.71) & (0.03) & (0.02) \\ [.25em]
Opp. Incent. & 6.04 & -1.38 & -0.13$^{***}$ & -0.04$^{*}$ \\
 & (4.65) & (4.53) & (0.03) & (0.02) \\ [.25em]
Own Incent. x Opp Incent. & -1.27 & 4.02 & -0.01 & 0.02 \\
 & (6.69) & (6.50) & (0.04) & (0.03) \\ [.25em]

Controls    & Yes & Yes & Yes & Yes \\
\hline
R$^2$ & 0.12 & 0.08 & 0.16 & 0.10 \\
N & 837 & 877 & 837 & 877 \\
\hline\hline
\end{tabular}
    \begin{minipage}{1\linewidth}
        \caption{Response Time, Best Responses, and Payoffs  (\hyref{hypothesis:response-time}[Hypothesis][b]): Incentive Interaction}
        \label{table:obr-oeu-interaction}
        \emph{Notes}:
        This table reports estimates from the regression in \hyref{table:obr-oeu}[Table] augmented to include the interaction between own and opponents' incentive levels. 
        It examines the relation between, on the one hand, log response times (in seconds), and on the other objective best responses to the empirical frequency of opponents' actions (columns (1) and (2)) and log expected payoffs, where expectations are taken also with respect to the empirical frequency of opponents' actions (columns (3) and (4)).
        High Own/Opponent Incentives correspond to indicators for whether the participant and their opponent face a high incentive level.
        2 Steps and 3 Steps denote the different games in the experiment (see \hyref{figure:games}[Figure]).
        Controls refer to the participants' socio-demographic characteristics.
        Heterocedasticity-robust standard errors are given in parentheses; $^*$, $^{**}$, and $^{***}$ denote p-values $< 0.1$, $<0.05$, and $<0.01$, respectively.
    \end{minipage}
\end{table}

\FloatBarrier \newpage

\section{Robustness Checks}
\label{section:appendix:robustness}

\subsection{Controlling for Attention and Understanding}
\label{section:appendix:robustness:comp}
~

\begin{table}[!h]\setstretch{1.1}
    \centering\small\singlespacing
	\begin{tabular}{l@{\extracolsep{4pt}}cccc@{}}
\hline\hline
& \multicolumn{2}{c}{Dominance Play} & \multicolumn{2}{c}{Dominated Play}  \\
\cline{2-3} \cline{4-5} 
& 2 Steps & 3 Steps & 2 Steps & 3 Steps \\
& (1) & (2) & (3) & (4) \\
\hline
Own Incent. & 18.29$^{***}$ & -1.57 & -10.63$^{***}$ & -6.48$^{***}$ \\
 & (3.77) & (2.80) & (2.99) & (2.12) \\ [.25em]
Opp. Incent. & 8.00$^{**}$ & 1.16 & -0.02 & 1.14 \\
 & (3.93) & (2.73) & (3.02) & (2.06) \\ [.25em]
Own Incent. x Comprehension Mistakes & 0.38 & 0.17 & -0.65$^{**}$ & 0.37 \\
 & (0.40) & (0.35) & (0.27) & (0.26) \\ [.25em]
Opp. Incent. x Comprehension Mistakes & -0.60 & 0.19 & -0.12 & -0.20 \\
 & (0.59) & (0.34) & (0.40) & (0.27) \\ [.25em]
Comprehension Mistakes & 0.25 & -0.28 & 0.51$^{**}$ & -0.09 \\
 & (0.36) & (0.31) & (0.25) & (0.30) \\ [.25em]

Controls    & Yes & Yes & Yes & Yes \\
\hline
R$^2$ & 0.07 & 0.02 & 0.06 & 0.03 \\
N & 837 & 877 & 837 & 877 \\
\hline\hline
\end{tabular}
    \begin{minipage}{1\linewidth}
        \caption{Incentive Level, Dominance, and Dominated Play (\hyref{hypothesis:action-sophistication}[Hypothesis]): Attention and Understanding}
        \label{table:action-dominance-comp}
        \emph{Notes}:
        This table reports estimates from the regression in \hyref{table:action-dominance}[Table] augmented to include the number of comprehension-quiz mistakes and its interactions with the treatment indicators. 
        The dependent variable is an indicator for whether the participant chose the dominance-solution action (columns (1) and (2)) or a strictly dominated action (columns (3) and (4)). 
        High Own/Opponent Incentives are indicators for whether the participant and their opponent face a high incentive level. 
        2 Steps and 3 Steps denote the two games in the experiment; see \hyref{figure:games}[Figure]. 
        Controls refer to participants' socio-demographic characteristics.
        Heterocedasticity-robust standard errors are reported in parentheses; $^{*}$, $^{**}$, and $^{***}$ denote $p<0.10$, $p<0.05$, and $p<0.01$, respectively.
    \end{minipage}
\end{table}

\begin{table}[!ht]\setstretch{1.1}
    \centering\small\singlespacing
	\begin{tabular}{l@{\extracolsep{4pt}}cccc@{}}
\hline\hline
& \multicolumn{2}{c}{Subjective BR} & \multicolumn{2}{c}{Objective BR}  \\
\cline{2-3} \cline{4-5} 
& 2 Steps & 3 Steps & 2 Steps & 3 Steps \\
& (1) & (2) & (3) & (4) \\
\hline
Own Incent. & 19.56$^{***}$ & 18.25$^{***}$ & 18.29$^{***}$ & 15.23$^{***}$ \\
 & (3.79) & (3.89) & (3.77) & (3.97) \\ [.25em]
Opp. Incent. & 9.30$^{**}$ & 5.42 & 8.00$^{**}$ & 4.29 \\
 & (3.94) & (3.92) & (3.93) & (3.99) \\ [.25em]
Own Incent. x Comprehension Mistakes & 0.28 & -0.82 & 0.38 & -0.86 \\
 & (0.45) & (0.61) & (0.40) & (0.72) \\ [.25em]
Opp. Incent. x Comprehension Mistakes & -1.24$^{*}$ & -0.67 & -0.60 & -0.51 \\
 & (0.66) & (0.66) & (0.59) & (0.73) \\ [.25em]
Comprehension Mistakes & -0.27 & 0.38 & 0.25 & 1.09 \\
 & (0.21) & (0.67) & (0.36) & (0.70) \\ [.25em]

Controls    & Yes & Yes & Yes & Yes \\
\hline
R$^2$ & 0.08 & 0.05 & 0.07 & 0.05 \\
N & 837 & 877 & 837 & 877 \\
\hline\hline
\end{tabular}
    \begin{minipage}{1\linewidth}
        \caption{Incentive Level and Best-Response Rate (\hyref{hypothesis:best-responses}[Hypothesis][a]): Attention and Understanding}
        \label{table:action-br-comp}
        \emph{Notes}:
        This table reports estimates from the regression in \hyref{table:action-br}[Table] augmented to include the number of comprehension-quiz mistakes and its interactions with the treatment indicators. 
        The dependent variable is an indicator for whether the participant best responds to their reported beliefs --- that is, chooses the action that maximises subjective expected payoff according to those beliefs (columns (1) and (2)) --- or best responds to the empirical frequency of opponents' actions --- that is, an objective best response (columns (3) and (4)). 
        High Own/Opponent Incentives correspond to indicators for whether the participant and their opponent face a high incentive level.
        2 Steps and 3 Steps denote the different games in the experiment (see \hyref{figure:games}[Figure]).
        Controls refer to the participants' socio-demographic characteristics.
        Heterocedasticity-robust standard errors are given in parentheses; $^*$, $^{**}$, and $^{***}$ denote p-values $< 0.1$, $<0.05$, and $<0.01$, respectively.
    \end{minipage}
\end{table}

\begin{table}[!ht]\setstretch{1.1}
    \centering\small\singlespacing
	\begin{tabular}{l@{\extracolsep{4pt}}cccc@{}}
\hline\hline
& \multicolumn{2}{c}{Belief in}  &  \multicolumn{2}{c}{Belief in} \\
& \multicolumn{2}{c}{Dominance Play} & \multicolumn{2}{c}{Dominated Play}  \\
\cline{2-3} \cline{4-5} 
& 2 Steps & 3 Steps & 2 Steps & 3 Steps \\
& (1) & (2) & (3) & (4) \\
\hline
Own Incent. & 4.23$^{***}$ & -1.96$^{*}$ & -7.70$^{***}$ & -2.42$^{***}$ \\
 & (1.34) & (1.12) & (1.42) & (0.93) \\ [.25em]
Opp. Incent. & 3.20$^{**}$ & -0.60 & -4.17$^{***}$ & -2.06$^{**}$ \\
 & (1.34) & (1.13) & (1.45) & (0.96) \\ [.25em]
Own Incent. x Comprehension Mistakes & 0.11 & 0.05 & 0.02 & -0.01 \\
 & (0.22) & (0.16) & (0.18) & (0.14) \\ [.25em]
Opp. Incent. x Comprehension Mistakes & -0.35$^{*}$ & -0.04 & 0.14 & 0.08 \\
 & (0.21) & (0.17) & (0.23) & (0.16) \\ [.25em]
Comprehension Mistakes & 0.03 & -0.02 & 0.02 & -0.03 \\
 & (0.06) & (0.17) & (0.09) & (0.15) \\ [.25em]

Controls    & Yes & Yes & Yes & Yes \\
\hline
R$^2$ & 0.06 & 0.02 & 0.09 & 0.05 \\
N & 837 & 877 & 837 & 877 \\
\hline\hline
\end{tabular}
    \begin{minipage}{1\linewidth}
        \caption{Incentive Level and Belief in Opponent Sophistication (\hyref{hypothesis:belief-sophistication}[Hypotheses][a-b]): Attention and Understanding}
        \label{table:belief-dominance-comp}
        \emph{Notes}:
        This table reports estimates from the regression in \hyref{table:belief-dominance}[Table] augmented to include the number of comprehension-quiz mistakes and its interactions with the treatment indicators. 
        The dependent variable is the reported belief in dominance play, $b_{i,1}$, (columns (1) and (2)) or the reported belief in strictly dominated play, $b_{i,3}+b_{i,4}$ in column (3) and $b_{i,4}$ in column (4). 
        High Own/Opponent Incentives correspond to indicators for whether the participant and their opponent face a high incentive level.
        2 Steps and 3 Steps denote the different games in the experiment (see \hyref{figure:games}[Figure]).
        Controls refer to the participants' socio-demographic characteristics.
        Heterocedasticity-robust standard errors are given in parentheses; $^*$, $^{**}$, and $^{***}$ denote p-values $< 0.1$, $<0.05$, and $<0.01$, respectively.
    \end{minipage}
\end{table}

\begin{table}[!ht]\setstretch{1.1}
    \centering\small\singlespacing
	\begin{tabular}{l@{\extracolsep{4pt}}cccccc@{}}
\hline\hline
& \multicolumn{4}{c}{|Belief - Opponent Action Frequency|} & \multicolumn{2}{c}{|Subj. - Obj. Payoff|} \\
& \multicolumn{2}{c}{Dominance Play}  &  \multicolumn{2}{c}{Dominated Play} & & \\
\cline{2-3} \cline{4-5} \cline{6-7} 
& 2 Steps & 3 Steps & 2 Steps & 3 Steps & 2 Steps & 3 Steps \\
& (1) & (2) & (3) & (4) & (5) & (6) \\
\hline
Own Incent. & -2.83$^{***}$ & -0.66 & -5.88$^{***}$ & -1.89$^{***}$ & -5.59$^{***}$ & -2.21$^{**}$ \\
 & (0.98) & (0.86) & (0.96) & (0.73) & (0.99) & (0.93) \\ [.25em]
Opp. Incent. & 15.56$^{***}$ & -0.72 & 5.57$^{***}$ & 1.49$^{**}$ & 8.17$^{***}$ & 5.59$^{***}$ \\
 & (1.00) & (0.87) & (1.01) & (0.76) & (1.00) & (0.96) \\ [.25em]
Own Incent. x Comprehension Mistakes & 0.30$^{**}$ & -0.00 & 0.15$^{*}$ & 0.05 & 0.21$^{**}$ & 0.14 \\
 & (0.15) & (0.12) & (0.09) & (0.11) & (0.09) & (0.13) \\ [.25em]
Opp. Incent. x Comprehension Mistakes & 0.27 & -0.12 & 0.06 & 0.00 & 0.20$^{*}$ & 0.17 \\
 & (0.17) & (0.12) & (0.17) & (0.12) & (0.12) & (0.14) \\ [.25em]
Comprehension Mistakes & -0.12$^{**}$ & 0.07 & -0.04 & -0.02 & -0.04 & -0.15 \\
 & (0.06) & (0.13) & (0.06) & (0.12) & (0.06) & (0.13) \\ [.25em]

Controls    & Yes & Yes & Yes & Yes & Yes & Yes \\
\hline
R$^2$ & 0.34 & 0.03 & 0.13 & 0.04 & 0.17 & 0.09 \\
N & 837 & 877 & 837 & 877 & 837 & 877 \\
\hline\hline
\end{tabular}
    \begin{minipage}{1\linewidth}
        \caption{Incentive Level and Belief Accuracy (\hyref{hypothesis:belief-sophistication}[Hypothesis][c]): Attention and Understanding}
        \label{table:belief-accuracy-comp}
        \emph{Notes}:
        This table reports estimates from the regression in \hyref{table:belief-accuracy}[Table] augmented to include the number of comprehension-quiz mistakes and its interactions with the treatment indicators. 
        The dependent variable is the absolute difference between reported belief in dominance play by opponents and its realised frequency, $|b_{i,1}-\overline \sigma_{-i,1}|$, (columns (1) and (2)), or the analogous absolute difference for strictly dominated play, $|b_{i,3}+b_{i,4}-\overline \sigma_{-i,3}-\overline \sigma_{-i,4}|$ in column (3) and $|b_{i,4}-\overline \sigma_{-i,4}|$ in column (4). 
        In columns (5) and (6) the dependent variable is the L1-norm of the difference between subjective expected payoffs (according to reported beliefs) and objective expected payoffs (according to observed action frequencies).
        High Own/Opponent Incentives correspond to indicators for whether the participant and their opponent face a high incentive level.
        2 Steps and 3 Steps denote the different games in the experiment (see \hyref{figure:games}[Figure]).
        Controls refer to the participants' socio-demographic characteristics.
        Heterocedasticity-robust standard errors are given in parentheses; $^*$, $^{**}$, and $^{***}$ denote p-values $< 0.1$, $<0.05$, and $<0.01$, respectively.
    \end{minipage}
\end{table}

\begin{table}[!ht]\setstretch{1.1}
    \centering\small\singlespacing
	\begin{tabular}{l@{\extracolsep{4pt}}cc@{}}
\hline\hline
& \multicolumn{2}{c}{Distance to Uniform}  \\
\cline{2-3} 
& 2 Steps & 3 Steps \\
& (1) & (2) \\
\hline
Own Incent. & 6.46$^{**}$ & 3.27 \\
 & (2.77) & (2.44) \\ [.25em]
Opp. Incent. & 6.98$^{**}$ & 1.81 \\
 & (2.89) & (2.43) \\ [.25em]
Own Incent. x Comprehension Mistakes & 0.26 & 0.17 \\
 & (0.42) & (0.40) \\ [.25em]
Opp. Incent. x Comprehension Mistakes & 0.05 & -0.09 \\
 & (0.55) & (0.39) \\ [.25em]
Comprehension Mistakes & -0.27$^{*}$ & -0.18 \\
 & (0.16) & (0.38) \\ [.25em]

Controls    & Yes & Yes \\
\hline
R$^2$ & 0.06 & 0.04 \\
N & 837 & 877 \\
\hline\hline
\end{tabular}
    \begin{minipage}{1\linewidth}
        \caption{Incentive Level and Belief Distance to Uniform Distribution: Attention and Understanding}
        \label{table:belief-uniform-comp}
        \emph{Notes}:
        This table reports estimates from the regression in \hyref{table:belief-uniform}[Table] augmented to include the number of comprehension-quiz mistakes and its interactions with the treatment indicators.  
        The dependent variable is the L1 distance between reported belief and the uniform distribution. 
        High Own/Opponent Incentives correspond to indicators for whether the participant and their opponent face a high incentive level.
        2 Steps and 3 Steps denote the different games in the experiment (see \hyref{figure:games}[Figure]).
        Controls refer to the participants' socio-demographic characteristics.
        Heterocedasticity-robust standard errors are given in parentheses; $^*$, $^{**}$, and $^{***}$ denote p-values $< 0.1$, $<0.05$, and $<0.01$, respectively.
    \end{minipage}
\end{table}

\begin{table}[!ht]\setstretch{1.1}
    \centering\small\singlespacing
	\begin{tabular}{l@{\extracolsep{4pt}}cc@{}}
\hline\hline
& \multicolumn{2}{c}{Log(Time)}  \\
\cline{2-3} 
& 2 Steps & 3 Steps \\
& (1) & (2) \\
\hline
Own Incent. & 0.34$^{***}$ & 0.40$^{***}$ \\
 & (0.06) & (0.06) \\ [.25em]
Opp. Incent. & 0.04 & 0.21$^{***}$ \\
 & (0.06) & (0.06) \\ [.25em]
Own Incent. x Comprehension Mistakes & 0.01 & -0.01 \\
 & (0.01) & (0.01) \\ [.25em]
Opp. Incent. x Comprehension Mistakes & -0.01 & -0.01 \\
 & (0.01) & (0.01) \\ [.25em]
Comprehension Mistakes & -0.02$^{***}$ & 0.00 \\
 & (0.00) & (0.01) \\ [.25em]

Controls    & Yes & Yes \\
\hline
R$^2$ & 0.12 & 0.13 \\
N & 837 & 877 \\
\hline\hline
\end{tabular}
    \begin{minipage}{1\linewidth}
        \caption{Incentive Level and Response Time (\hyref{hypothesis:response-time}[Hypothesis][a]): Attention and Understanding}
        \label{table:rt-comp}
        \emph{Notes}:
        This table reports estimates from the regression in \hyref{table:rt}[Table] augmented to include the number of comprehension-quiz mistakes and its interactions with the treatment indicators. 
        The dependent variable is the participant's log response time (in seconds). 
        High Own/Opponent Incentives correspond to indicators for whether the participant and their opponent face a high incentive level.
        2 Steps and 3 Steps denote the different games in the experiment (see \hyref{figure:games}[Figure]).
        Controls refer to the participants' socio-demographic characteristics.
        Heterocedasticity-robust standard errors are given in parentheses; $^*$, $^{**}$, and $^{***}$ denote p-values $< 0.1$, $<0.05$, and $<0.01$, respectively.
    \end{minipage}
\end{table}

\begin{table}[!ht]\setstretch{1.1}
    \centering\small\singlespacing
	\begin{tabular}{l@{\extracolsep{4pt}}cccc@{}}
\hline\hline
& \multicolumn{2}{c}{Objective BR} & \multicolumn{2}{c}{Log(Obj. Exp. Payoff}  \\
\cline{2-3} \cline{4-5} 
& 2 Steps & 3 Steps & 2 Steps & 3 Steps \\
& (1) & (2) & (3) & (4) \\
\hline
Log(Time) & 16.51$^{***}$ & 12.28$^{***}$ & 0.10$^{***}$ & 0.06$^{***}$ \\
 & (2.32) & (2.59) & (0.01) & (0.01) \\ [.25em]
Own Incent. & 12.65$^{***}$ & 10.73$^{***}$ & 0.07$^{***}$ & 0.05$^{***}$ \\
 & (3.96) & (3.95) & (0.02) & (0.02) \\ [.25em]
Opp. Incent. & 7.31$^{*}$ & 2.16 & -0.13$^{***}$ & -0.03$^{*}$ \\
 & (3.79) & (3.88) & (0.02) & (0.02) \\ [.25em]
Log(Time) x Comprehension Mistakes & -0.01 & -0.32 & -0.00 & -0.00 \\
 & (0.25) & (0.56) & (0.00) & (0.00) \\ [.25em]
Own Incent. x Comprehension Mistakes & 0.26 & -0.77 & 0.00 & -0.00$^{*}$ \\
 & (0.52) & (0.65) & (0.00) & (0.00) \\ [.25em]
Opp. Incent. x Comprehension Mistakes & -0.45 & -0.52 & -0.00 & -0.00 \\
 & (0.54) & (0.66) & (0.00) & (0.00) \\ [.25em]
Comprehension Mistakes & 0.56 & 2.44 & 0.00 & 0.01 \\
 & (0.73) & (2.37) & (0.00) & (0.01) \\ [.25em]

Controls    & Yes & Yes & Yes & Yes \\
\hline
R$^2$ & 0.13 & 0.08 & 0.16 & 0.10 \\
N & 837 & 877 & 837 & 877 \\
\hline\hline
\end{tabular}
    \begin{minipage}{1\linewidth}
        \caption{Response Time, Best Responses, and Payoffs  (\hyref{hypothesis:response-time}[Hypothesis][b]): Attention and Understanding}
        \label{table:obr-oeu-comp}
        \emph{Notes}:
        This table reports estimates from the regression in \hyref{table:obr-oeu}[Table] augmented to include the number of comprehension-quiz mistakes and its interactions with the treatment indicators and log response time. 
        It examines the relation between, on the one hand, log response times (in seconds), and on the other objective best responses to the empirical frequency of opponents' actions (columns (1) and (2)) and log expected payoffs, where expectations are taken also with respect to the empirical frequency of opponents' actions (columns (3) and (4)).
        High Own/Opponent Incentives correspond to indicators for whether the participant and their opponent face a high incentive level.
        2 Steps and 3 Steps denote the different games in the experiment (see \hyref{figure:games}[Figure]).
        Controls refer to the participants' socio-demographic characteristics.
        Heterocedasticity-robust standard errors are given in parentheses; $^*$, $^{**}$, and $^{***}$ denote p-values $< 0.1$, $<0.05$, and $<0.01$, respectively.
    \end{minipage}
\end{table}

\FloatBarrier \newpage

\subsection{Participant-Pool-Specific Effects}
\label{section:appendix:robustness:pool}
~

\begin{table}[!ht]\setstretch{1.1}
    \centering\small\singlespacing
	\begin{tabular}{l@{\extracolsep{4pt}}cccc@{}}
\hline\hline
& \multicolumn{2}{c}{Dominance Play} & \multicolumn{2}{c}{Dominated Play}  \\
\cline{2-3} \cline{4-5} 
& 2 Steps & 3 Steps & 2 Steps & 3 Steps \\
& (1) & (2) & (3) & (4) \\
\hline
Own Incent. & 20.73$^{***}$ & -6.14$^{*}$ & -14.20$^{***}$ & -9.08$^{***}$ \\
 & (4.85) & (3.65) & (4.09) & (3.18) \\ [.25em]
Opp. Incent. & 3.12 & 0.59 & -0.27 & 0.01 \\
 & (4.91) & (3.69) & (3.99) & (3.15) \\ [.25em]
Own Incent. x Prolific & -2.02 & 9.70$^{**}$ & 2.32 & 7.28$^{*}$ \\
 & (6.86) & (4.84) & (5.72) & (3.95) \\ [.25em]
Opp. Incent. x Prolific & 5.18 & 2.44 & -0.39 & 1.33 \\
 & (6.87) & (4.90) & (5.59) & (3.93) \\ [.25em]
Prolific & 0.59 & -6.02 & -7.44 & -8.64$^{**}$ \\
 & (6.62) & (4.82) & (6.01) & (4.22) \\ [.25em]

Controls    & Yes & Yes & Yes & Yes \\
\hline
R$^2$ & 0.06 & 0.03 & 0.06 & 0.03 \\
N & 837 & 877 & 837 & 877 \\
\hline\hline
\end{tabular}
    \begin{minipage}{1\linewidth}
        \caption{Incentive Level, Dominance, and Dominated Play (\hyref{hypothesis:action-sophistication}[Hypothesis]): Participant Pool}
        \label{table:action-dominance-pool}
        \emph{Notes}:
        This table reports estimates from the regression in \hyref{table:action-dominance}[Table] augmented to include a Prolific indicator and its interactions with the treatment indicators. 
        The dependent variable is an indicator for whether the participant chose the dominance-solution action (columns (1) and (2)) or a strictly dominated action (columns (3) and (4)). 
        High Own/Opponent Incentives are indicators for whether the participant and their opponent face a high incentive level. 
        2 Steps and 3 Steps denote the two games in the experiment; see \hyref{figure:games}[Figure]. 
        Controls refer to participants' socio-demographic characteristics.
        Heterocedasticity-robust standard errors are reported in parentheses; $^{*}$, $^{**}$, and $^{***}$ denote $p<0.10$, $p<0.05$, and $p<0.01$, respectively.
    \end{minipage}
\end{table}

\begin{table}[!ht]\setstretch{1.1}
    \centering\small\singlespacing
	\begin{tabular}{l@{\extracolsep{4pt}}cccc@{}}
\hline\hline
& \multicolumn{2}{c}{Subjective BR} & \multicolumn{2}{c}{Objective BR}  \\
\cline{2-3} \cline{4-5} 
& 2 Steps & 3 Steps & 2 Steps & 3 Steps \\
& (1) & (2) & (3) & (4) \\
\hline
Own Incent. & 21.41$^{***}$ & 14.90$^{***}$ & 20.73$^{***}$ & 16.27$^{***}$ \\
 & (4.89) & (4.99) & (4.85) & (4.78) \\ [.25em]
Opp. Incent. & -1.14 & 2.78 & 3.12 & 4.62 \\
 & (4.93) & (4.95) & (4.91) & (4.78) \\ [.25em]
Own Incent. x Prolific & -0.96 & 1.80 & -2.02 & -6.79 \\
 & (6.82) & (6.84) & (6.86) & (6.69) \\ [.25em]
Opp. Incent. x Prolific & 13.12$^{*}$ & 1.54 & 5.18 & -3.89 \\
 & (6.87) & (6.85) & (6.87) & (6.69) \\ [.25em]
Prolific & -1.35 & -5.67 & 0.59 & 7.53 \\
 & (6.85) & (6.84) & (6.62) & (6.59) \\ [.25em]

Controls    & Yes & Yes & Yes & Yes \\
\hline
R$^2$ & 0.07 & 0.05 & 0.06 & 0.05 \\
N & 837 & 877 & 837 & 877 \\
\hline\hline
\end{tabular}
    \begin{minipage}{1\linewidth}
        \caption{Incentive Level and Best-Response Rate (\hyref{hypothesis:best-responses}[Hypothesis][a]): Participant Pool}
        \label{table:action-br-pool}
        \emph{Notes}:
        This table reports estimates from the regression in \hyref{table:action-br}[Table] augmented to include a Prolific indicator and its interactions with the treatment indicators. 
        The dependent variable is an indicator for whether the participant best responds to their reported beliefs --- that is, chooses the action that maximises subjective expected payoff according to those beliefs (columns (1) and (2)) --- or best responds to the empirical frequency of opponents' actions --- that is, an objective best response (columns (3) and (4)). 
        High Own/Opponent Incentives correspond to indicators for whether the participant and their opponent face a high incentive level.
        2 Steps and 3 Steps denote the different games in the experiment (see \hyref{figure:games}[Figure]).
        Controls refer to the participants' socio-demographic characteristics.
        Heterocedasticity-robust standard errors are given in parentheses; $^*$, $^{**}$, and $^{***}$ denote p-values $< 0.1$, $<0.05$, and $<0.01$, respectively.
    \end{minipage}
\end{table}

\begin{table}[!ht]\setstretch{1.1}
    \centering\small\singlespacing
	\begin{tabular}{l@{\extracolsep{4pt}}cccc@{}}
\hline\hline
& \multicolumn{2}{c}{Belief in}  &  \multicolumn{2}{c}{Belief in} \\
& \multicolumn{2}{c}{Dominance Play} & \multicolumn{2}{c}{Dominated Play}  \\
\cline{2-3} \cline{4-5} 
& 2 Steps & 3 Steps & 2 Steps & 3 Steps \\
& (1) & (2) & (3) & (4) \\
\hline
Own Incent. & 3.76$^{**}$ & -2.18 & -7.82$^{***}$ & -2.08$^{*}$ \\
 & (1.55) & (1.35) & (1.78) & (1.26) \\ [.25em]
Opp. Incent. & 1.99 & -1.96 & -1.93 & -1.87 \\
 & (1.60) & (1.37) & (1.83) & (1.26) \\ [.25em]
Own Incent. x Prolific & 1.91 & 0.71 & 0.21 & -0.74 \\
 & (2.32) & (1.89) & (2.53) & (1.66) \\ [.25em]
Opp. Incent. x Prolific & -0.03 & 2.39 & -3.51 & 0.05 \\
 & (2.31) & (1.94) & (2.51) & (1.65) \\ [.25em]
Prolific & 2.71 & -0.65 & -1.22 & 1.01 \\
 & (1.96) & (1.87) & (2.40) & (1.60) \\ [.25em]

Controls    & Yes & Yes & Yes & Yes \\
\hline
R$^2$ & 0.06 & 0.03 & 0.10 & 0.05 \\
N & 837 & 877 & 837 & 877 \\
\hline\hline
\end{tabular}
    \begin{minipage}{1\linewidth}
        \caption{Incentive Level and Belief in Opponent Sophistication (\hyref{hypothesis:belief-sophistication}[Hypotheses][a-b]): Participant Pool}
        \label{table:belief-dominance-pool}
        \emph{Notes}:
        This table reports estimates from the regression in \hyref{table:belief-dominance}[Table] augmented to include a Prolific indicator and its interactions with the treatment indicators. 
        The dependent variable is the reported belief in dominance play, $b_{i,1}$, (columns (1) and (2)) or the reported belief in strictly dominated play, $b_{i,3}+b_{i,4}$ in column (3) and $b_{i,4}$ in column (4).
        High Own/Opponent Incentives correspond to indicators for whether the participant and their opponent face a high incentive level.
        2 Steps and 3 Steps denote the different games in the experiment (see \hyref{figure:games}[Figure]).
        Controls refer to the participants' socio-demographic characteristics.
        Heterocedasticity-robust standard errors are given in parentheses; $^*$, $^{**}$, and $^{***}$ denote p-values $< 0.1$, $<0.05$, and $<0.01$, respectively.
    \end{minipage}
\end{table}

\begin{table}[!ht]\setstretch{1.1}
    \centering\small\singlespacing
	\begin{tabular}{l@{\extracolsep{4pt}}cccccc@{}}
\hline\hline
& \multicolumn{4}{c}{|Belief - Opponent Action Frequency|} & \multicolumn{2}{c}{|Subj. - Obj. Payoff|} \\
& \multicolumn{2}{c}{Dominance Play}  &  \multicolumn{2}{c}{Dominated Play} & & \\
\cline{2-3} \cline{4-5} \cline{6-7} 
& 2 Steps & 3 Steps & 2 Steps & 3 Steps & 2 Steps & 3 Steps \\
& (1) & (2) & (3) & (4) & (5) & (6) \\
\hline
Own Incent. & -2.53$^{**}$ & -0.48 & -7.00$^{***}$ & -1.69$^{*}$ & -5.94$^{***}$ & -1.11 \\
 & (1.18) & (0.97) & (1.29) & (1.01) & (1.37) & (1.25) \\ [.25em]
Opp. Incent. & 16.76$^{***}$ & -2.44$^{**}$ & 6.30$^{***}$ & 1.22 & 9.69$^{***}$ & 5.29$^{***}$ \\
 & (1.21) & (1.00) & (1.32) & (1.01) & (1.39) & (1.24) \\ [.25em]
Own Incent. x Prolific & 1.42 & -0.30 & 3.27$^{*}$ & -0.13 & 2.05 & -1.29 \\
 & (1.69) & (1.42) & (1.74) & (1.31) & (1.80) & (1.66) \\ [.25em]
Opp. Incent. x Prolific & -0.75 & 2.60$^{*}$ & -1.04 & 0.51 & -1.81 & 1.52 \\
 & (1.70) & (1.49) & (1.75) & (1.30) & (1.80) & (1.65) \\ [.25em]
Prolific & -0.74 & -0.91 & -2.51 & 0.13 & -1.55 & -0.76 \\
 & (1.40) & (1.45) & (1.67) & (1.24) & (1.73) & (1.64) \\ [.25em]

Controls    & Yes & Yes & Yes & Yes & Yes & Yes \\
\hline
R$^2$ & 0.33 & 0.03 & 0.14 & 0.04 & 0.17 & 0.09 \\
N & 837 & 877 & 837 & 877 & 837 & 877 \\
\hline\hline
\end{tabular}
    \begin{minipage}{1\linewidth}
        \caption{Incentive Level and Belief Accuracy (\hyref{hypothesis:belief-sophistication}[Hypothesis][c]): Participant Pool}
        \label{table:belief-accuracy-pool}
        \emph{Notes}:
        This table reports estimates from the regression in \hyref{table:belief-accuracy}[Table] augmented to include a Prolific indicator and its interactions with the treatment indicators. 
        The dependent variable is the absolute difference between reported belief in dominance play by opponents and its realised frequency, $|b_{i,1}-\overline \sigma_{-i,1}|$, (columns (1) and (2)), or the analogous absolute difference for strictly dominated play, $|b_{i,3}+b_{i,4}-\overline \sigma_{-i,3}-\overline \sigma_{-i,4}|$ in column (3) and $|b_{i,4}-\overline \sigma_{-i,4}|$ in column (4). 
        In columns (5) and (6) the dependent variable is the L1-norm of the difference between subjective expected payoffs (according to reported beliefs) and objective expected payoffs (according to observed action frequencies). 
        High Own/Opponent Incentives correspond to indicators for whether the participant and their opponent face a high incentive level.
        2 Steps and 3 Steps denote the different games in the experiment (see \hyref{figure:games}[Figure]).
        Controls refer to the participants' socio-demographic characteristics.
        Heterocedasticity-robust standard errors are given in parentheses; $^*$, $^{**}$, and $^{***}$ denote p-values $< 0.1$, $<0.05$, and $<0.01$, respectively.
    \end{minipage}
\end{table}

\begin{table}[!ht]\setstretch{1.1}
    \centering\small\singlespacing
	\begin{tabular}{l@{\extracolsep{4pt}}cc@{}}
\hline\hline
& \multicolumn{2}{c}{Distance to Uniform}  \\
\cline{2-3} 
& 2 Steps & 3 Steps \\
& (1) & (2) \\
\hline
Own Incent. & 8.72$^{***}$ & 5.83$^{**}$ \\
 & (2.91) & (2.91) \\ [.25em]
Opp. Incent. & 5.68$^{*}$ & 1.69 \\
 & (3.00) & (2.92) \\ [.25em]
Own Incent. x Prolific & -2.59 & -3.80 \\
 & (4.83) & (4.28) \\ [.25em]
Opp. Incent. x Prolific & 3.11 & -0.19 \\
 & (4.79) & (4.26) \\ [.25em]
Prolific & -0.96 & -3.46 \\
 & (4.22) & (4.25) \\ [.25em]

Controls    & Yes & Yes \\
\hline
R$^2$ & 0.06 & 0.05 \\
N & 837 & 877 \\
\hline\hline
\end{tabular}
    \begin{minipage}{1\linewidth}
        \caption{Incentive Level and Belief Distance to Uniform Distribution: Participant Pool}
        \label{table:belief-uniform-pool}
        \emph{Notes}:
        This table reports estimates from the regression in \hyref{table:belief-uniform}[Table] augmented to include a Prolific indicator and its interactions with the treatment indicators. 
        The dependent variable is the L1 distance between reported belief and the uniform distribution. 
        High Own/Opponent Incentives correspond to indicators for whether the participant and their opponent face a high incentive level.
        2 Steps and 3 Steps denote the different games in the experiment (see \hyref{figure:games}[Figure]).
        Controls refer to the participants' socio-demographic characteristics.
        Heterocedasticity-robust standard errors are given in parentheses; $^*$, $^{**}$, and $^{***}$ denote p-values $< 0.1$, $<0.05$, and $<0.01$, respectively.
    \end{minipage}
\end{table}

\begin{table}[!ht]\setstretch{1.1}
    \centering\small\singlespacing
	\begin{tabular}{l@{\extracolsep{4pt}}cc@{}}
\hline\hline
& \multicolumn{2}{c}{Log(Time)}  \\
\cline{2-3} 
& 2 Steps & 3 Steps \\
& (1) & (2) \\
\hline
Own Incent. & 0.47$^{***}$ & 0.39$^{***}$ \\
 & (0.07) & (0.08) \\ [.25em]
Opp. Incent. & 0.07 & 0.14$^{*}$ \\
 & (0.07) & (0.08) \\ [.25em]
Own Incent. x Prolific & -0.20$^{*}$ & -0.05 \\
 & (0.11) & (0.11) \\ [.25em]
Opp. Incent. x Prolific & -0.09 & 0.08 \\
 & (0.11) & (0.11) \\ [.25em]
Prolific & 0.12 & -0.12 \\
 & (0.11) & (0.10) \\ [.25em]

Controls    & Yes & Yes \\
\hline
R$^2$ & 0.11 & 0.13 \\
N & 837 & 877 \\
\hline\hline
\end{tabular}
    \begin{minipage}{1\linewidth}
        \caption{Incentive Level and Response Time (\hyref{hypothesis:response-time}[Hypothesis][a]): Participant Pool}
        \label{table:rt-pool}
        \emph{Notes}:
        This table reports estimates from the regression in \hyref{table:rt}[Table] augmented to include a Prolific indicator and its interactions with the treatment indicators. 
        The dependent variable is the participant's log response time (in seconds). 
        High Own/Opponent Incentives correspond to indicators for whether the participant and their opponent face a high incentive level.
        2 Steps and 3 Steps denote the different games in the experiment (see \hyref{figure:games}[Figure]).
        Controls refer to the participants' socio-demographic characteristics.
        Heterocedasticity-robust standard errors are given in parentheses; $^*$, $^{**}$, and $^{***}$ denote p-values $< 0.1$, $<0.05$, and $<0.01$, respectively.
    \end{minipage}
\end{table}

\begin{table}[!ht]\setstretch{1.1}
    \centering\small\singlespacing
	\begin{tabular}{l@{\extracolsep{4pt}}cccc@{}}
\hline\hline
& \multicolumn{2}{c}{Objective BR} & \multicolumn{2}{c}{Log(Obj. Exp. Payoff}  \\
\cline{2-3} \cline{4-5} 
& 2 Steps & 3 Steps & 2 Steps & 3 Steps \\
& (1) & (2) & (3) & (4) \\
\hline
Log(Time) & 14.10$^{***}$ & 13.32$^{***}$ & 0.10$^{***}$ & 0.07$^{***}$ \\
 & (3.29) & (2.79) & (0.02) & (0.01) \\ [.25em]
Own Incent. & 14.08$^{***}$ & 11.02$^{**}$ & 0.08$^{**}$ & 0.05$^{**}$ \\
 & (5.22) & (4.92) & (0.03) & (0.02) \\ [.25em]
Opp. Incent. & 2.14 & 2.75 & -0.15$^{***}$ & -0.02 \\
 & (4.80) & (4.70) & (0.03) & (0.02) \\ [.25em]
Log(Time) x Prolific & 3.34 & -3.86 & -0.00 & -0.01 \\
 & (4.40) & (4.07) & (0.03) & (0.02) \\ [.25em]
Own Incent. x Prolific & -0.15 & -4.69 & -0.01 & -0.04 \\
 & (7.06) & (6.85) & (0.04) & (0.03) \\ [.25em]
Opp. Incent. x Prolific & 6.50 & -4.10 & 0.02 & -0.03 \\
 & (6.70) & (6.64) & (0.04) & (0.03) \\ [.25em]
Prolific & -14.22 & 23.64 & 0.07 & 0.10 \\
 & (17.83) & (16.21) & (0.11) & (0.07) \\ [.25em]

Controls    & Yes & Yes & Yes & Yes \\
\hline
R$^2$ & 0.12 & 0.08 & 0.17 & 0.10 \\
N & 837 & 877 & 837 & 877 \\
\hline\hline
\end{tabular}
    \begin{minipage}{1\linewidth}
        \caption{Response Time, Best Responses, and Payoffs  (\hyref{hypothesis:response-time}[Hypothesis][b]): Participant Pool}
        \label{table:obr-oeu-pool}
        \emph{Notes}:
        This table reports estimates from the regression in \hyref{table:belief-uniform}[Table] augmented to include a Prolific indicator and its interactions with the treatment indicators and log response time. 
        It examines the relation between, on the one hand, log response times (in seconds), and on the other objective best responses to the empirical frequency of opponents' actions (columns (1) and (2)) and log expected payoffs, where expectations are taken also with respect to the empirical frequency of opponents' actions (columns (3) and (4)).
        High Own/Opponent Incentives correspond to indicators for whether the participant and their opponent face a high incentive level.
        2 Steps and 3 Steps denote the different games in the experiment (see \hyref{figure:games}[Figure]).
        Controls refer to the participants' socio-demographic characteristics.
        Heterocedasticity-robust standard errors are given in parentheses; $^*$, $^{**}$, and $^{***}$ denote p-values $< 0.1$, $<0.05$, and $<0.01$, respectively.
    \end{minipage}
\end{table}

\begin{table}[!ht]\setstretch{1.1}
	\centering\small\singlespacing
	\begin{tabular}{@{\extracolsep{4pt}}cccccccccccc@{}}
\hline\hline
\multicolumn{1}{c}{Game}  &  \multicolumn{2}{c}{Incentives}  &  \multicolumn{4}{c}{Action Frequency}  &  \multicolumn{4}{c}{Average Belief Distribution}  &  \multicolumn{1}{c}{Time} \\
\cline{1-1} \cline{2-3} \cline{4-7} \cline{8-11}  \cline{12-12}
Steps & Own & Opponent & 1 & 2 & 3 & 4 & 1 & 2 & 3 & 4 & (secs) \\
 (1) & (2) & (3) & (4) & (5) & (6) & (7) & (8) & (9) & (10) & (11) & (12) \\
\hline
2 & High & High & 57.28 & 27.18 & 8.74 & 6.80 & 33.01 & 35.65 & 18.92 & 12.42 & 106.7 \\
2 & High & Low & 51.89 & 33.96 & 8.49 & 5.66 & 27.94 & 36.91 & 20.15 & 15.01 & 89.9 \\
2 & Low & High & 35.92 & 35.92 & 14.56 & 13.59 & 26.77 & 31.69 & 23.01 & 18.53 & 59.7 \\
2 & Low & Low & 34.31 & 36.27 & 15.69 & 13.73 & 26.55 & 31.52 & 23.83 & 18.10 & 60.5 \\
3 & High & High & 14.42 & 49.04 & 29.81 & 6.73 & 18.40 & 31.05 & 36.35 & 14.19 & 117.7 \\
3 & High & Low & 13.00 & 44.00 & 36.00 & 7.00 & 21.28 & 25.89 & 33.96 & 18.86 & 104.9 \\
3 & Low & High & 19.64 & 33.04 & 31.25 & 16.07 & 21.94 & 27.49 & 32.02 & 18.55 & 71.0 \\
3 & Low & Low & 19.23 & 28.85 & 36.54 & 15.38 & 22.18 & 27.04 & 32.73 & 18.04 & 56.8 \\
\hline\hline
\end{tabular}
    \begin{minipage}{1\linewidth}
        \caption{Action Frequency, Average Beliefs Distributions, and Response Time: MTurk}
        \label{table:sample-avg-mturk}
        \emph{Notes}: 
        This table reports treatment-level summary statistics for the participants recruited on Amazon Mechanical Turk.
        Each participant is assigned to one treatment, defined by the game (2-Step or 3-Step game), own incentive level (high or low), and opponents' incentive level (high or low).
        Columns (4)--(7) report action frequencies, columns (8)--(11) report mean belief distributions, and column (12) reports mean response time in seconds.
        Participants play one game and make one incentivised decision.
    \end{minipage}
\end{table}

\begin{table}[!ht]\setstretch{1.1}
	\centering\small\singlespacing
	\begin{tabular}{@{\extracolsep{4pt}}cccccccccccc@{}}
\hline\hline
\multicolumn{1}{c}{Game}  &  \multicolumn{2}{c}{Incentives}  &  \multicolumn{4}{c}{Action Frequency}  &  \multicolumn{4}{c}{Average Belief Distribution}  &  \multicolumn{1}{c}{Time} \\
\cline{1-1} \cline{2-3} \cline{4-7} \cline{8-11}  \cline{12-12}
Steps & Own & Opponent & 1 & 2 & 3 & 4 & 1 & 2 & 3 & 4 & (secs) \\
 (1) & (2) & (3) & (4) & (5) & (6) & (7) & (8) & (9) & (10) & (11) & (12) \\
\hline
2 & High & High & 55.88 & 29.41 & 8.82 & 5.88 & 35.04 & 36.58 & 16.09 & 12.29 & 87.4 \\
2 & High & Low & 47.62 & 36.19 & 10.48 & 5.71 & 33.00 & 32.55 & 19.66 & 14.78 & 74.2 \\
2 & Low & High & 38.60 & 36.84 & 14.91 & 9.65 & 29.93 & 33.97 & 20.37 & 15.73 & 64.4 \\
2 & Low & Low & 29.41 & 43.14 & 18.63 & 8.82 & 27.69 & 30.70 & 24.12 & 17.50 & 70.5 \\
3 & High & High & 16.10 & 47.46 & 30.51 & 5.93 & 22.09 & 28.65 & 34.12 & 15.14 & 109.9 \\
3 & High & Low & 14.29 & 42.02 & 38.66 & 5.04 & 21.60 & 27.57 & 32.88 & 17.95 & 84.5 \\
3 & Low & High & 13.33 & 34.17 & 44.17 & 8.33 & 23.41 & 28.45 & 29.18 & 18.97 & 69.0 \\
3 & Low & Low & 9.00 & 36.00 & 49.00 & 6.00 & 23.42 & 26.48 & 30.00 & 20.11 & 52.3 \\
\hline\hline
\end{tabular}
    \begin{minipage}{1\linewidth}
        \caption{Action Frequency, Average Beliefs Distributions, and Response Time: Prolific}
        \label{table:sample-avg-prolific}
        \emph{Notes}:
        This table reports treatment-level summary statistics for the participants recruited on Prolific.
        Each participant is assigned to one treatment, defined by the game (2-Step or 3-Step game), own incentive level (high or low), and opponents' incentive level (high or low).
        Columns (4)--(7) report action frequencies, columns (8)--(11) report mean belief distributions, and column (12) reports mean response time in seconds.
        Participants play one game and make one incentivised decision.
    \end{minipage}
\end{table}

\FloatBarrier \newpage

\section{Behavioural Game Theory Model Predictions}
\label{section:appendix:theory-predictions}

\subsection{Quantal Response and Control Costs}
\label{section:appendix:theory-predictions:quantal-response}

Structural quantal response equilibrium \citep{McKelveyPalfrey1995GEB} corresponds to a fixed point $\sigma = q(\sigma)$ such that, for each $i\in I$, $q_i(\sigma_{-i})(a_i)=Q_i(u_i(\cdot,\sigma_{-i}))(a_i)=\mathbb P(a_i \in \argmax_{a_i'} u_i(a_i',\sigma_{-i})+\epsilon_{i,a_i})$, where $\epsilon_{i,a_i}$ is independently and identically drawn from some distribution with full support on the real line. 
In \hyref{section:hypotheses-design}[Section], we noted that, holding fixed $\sigma_{-i}$ and under mild conditions, $Q_i(\alpha u_i(\cdot,\sigma_{-i}))$ shifts in a first-order stochastic-dominance sense toward actions with higher expected payoff. 
A sufficient condition for this is that the distribution of $\epsilon_{i,a_i}$ admits a log-concave density, which is well-known to deliver such a result.

A related model, that of control costs \citep{MattssonWeibull2002GEB} based on additive perturbed utility \citep{FudenbergIijimaStrzalecki2015Ecta}, posits that 
$Q_i(u_i(\cdot,\sigma_{-i}))=\argmax_{\sigma_i}u_i(\sigma_i,\sigma_{-i}) - \sum_{a_i}c_i(\sigma_i(a_i))$, where $c_i:[0,1]\to \mathbb R$ is strictly increasing, convex, twice continuously differentiable, and satisfies $c_i'(0^+)=-\infty$. 
In this model, it is straightforward to verify that $Q_i(\alpha u_i(\cdot,\sigma_{-i}))$ shifts in a first-order stochastic dominance sense toward assigning higher probability to actions associated with a higher expected payoff whenever the elasticity of $c_i''$ is bounded above by $-1$, holding fixed $\sigma_{-i}$.

Both of these models nest logit quantal response equilibrium (LQRE), where $\epsilon_{i,a_i}$ follows an extreme value distribution or $c_i(x)=x \ln x - x \ln (1/|A_i|)$.
For this specific case and the games we consider, there is always a unique symmetric LQRE.
Moreover, one can derive clear comparative statics with respect to the incentive level parameter of both players.
From the first-order conditions obtained in its control cost formulation,
\begin{align}
    q_i(\sigma_{-i})(a_n)=q_i(\sigma_{-i})(a_{n-1})\exp(\alpha (u_i(a_n,\sigma_{-i})-u_i(a_{n-1},\sigma_{-i}))). \label{equation:LQRE}
\end{align}
Since $q_i(\sigma_i)=\sigma_i=\sigma_{-i}$ at any symmetric LQRE, one can express $\sigma_i(a_n)=f_n(\sigma_i(a_3))$ for $n=1,2,3,4$, and characterise equilibrium by the fixed-point condition: $F(\sigma_i(a_3))=\sum_{n=1}^4 f_n(\sigma_i(a_3))-1=0$.
Uniqueness follows from showing that $F$ crosses zero exactly once (indeed in the 2-Step game $F$ is monotone).
For equilibrium comparative statics in $\alpha$, it is sufficient to show that higher $\alpha$ leads to a lower equilibrium value of $\sigma_i(a_3)$, and then use the monotone-likelihood-ratio ordering implied by equation \eqref{equation:LQRE} to conclude that $\sigma_i(a_4)$ must also fall, so that $\sigma_i(a_1)+\sigma_i(a_2)$ rises. 
In the 2-Step game, $\sigma_i(a_1)$ is always increasing in $\alpha$, implying a first-order stochastic-dominance shift toward greater sophistication, whereas in the 3-Step game, the same is true whenever $\alpha$ is not too small.

\subsection{Level $k$ and Related Models}
\label{section:appendix:theory-predictions:level-k}

In a level-$k$ model \citep{Nagel1995AER,StahlWilson1994JEBO,StahlWilson1995GEB}, players choose according to their level of sophistication.
Specifically, a level-$k$ player chooses a strategy that best responds to opponents under the assumption that opponents are of level $k-1$, i.e., $\sigma_i^k\in \argmax_{\sigma_i \in \Sigma_i}u_i(\sigma_i,\sigma_{-i}^{k-1})$, with level-0 players uniformly randomising. 
In our 2-Step game, level-1 players choose $a_2$, while level-2 and higher players choose $a_1$. 
In contrast, in the 3-Step game, the level-1, level-2, and level-3-and-above actions correspond to, respectively, $a_3$, $a_2$, and $a_1$. 

The cognitive hierarchy model \citep{CamererHoChong2004QJE} assumes that players best-respond to the (truncated) distribution of types with lower sophistication than themselves, with level 0 typically uniformly randomising.
In the 2-Step game, a level-1 type would nevertheless choose $a_2$.
It is straightforward to show that there is a $\hat k$ such that for any $k\in \{1,...,\hat k\}$, type $k$ also chooses $a_2$, whereas higher types choose $a_1$. 
A similar observation applies for the 3-Step game.

A closely related strategy is that of dominance $k$ \citep{Costa-GomesCrawford2006AER}, according to which a dominance $k$ player performs $k$ rounds of (maximal) iterated elimination of strictly dominated strategies and then best responds to a uniform distribution over the surviving strategies of the opponents. 
In our games, actions taken by dominance $k$ players correspond exactly to those taken by level $k+1$ players.

Note that, in all of these models, players' chosen strategies are invariant to the incentive levels.

\subsection{Endogenous Depth of Reasoning}
\label{section:appendix:theory-predictions:edr}

A recent strand of work models cognitive effort as endogenous to the strategic environment. 
One such example is the endogenous depth of reasoning model by \citet{AlaouiPenta2016REStud}, which provides a way to endogenise level-$k$ reasoning as resulting from a cost-benefit analysis.

We briefly introduce the main elements of the model, adapted to our setting. 

First, the model posits that each player $i$ has a cost of (step) reasoning, $c_i: \mathbb N \to \mathbb R_+$. 
While the full model allows players to hold beliefs about their opponents' reasoning costs and higher-order beliefs about those costs, we focus on the simpler common-knowledge version, with fewer degrees of freedom. 

Second, players follow a path of reasoning, resulting from a cost-benefit analysis, denoted by $R_i=(a_{i,k}^i,a_{j,k-1}^i)_{k\geq 1}$, where $a_{j,k-1}^i$ denotes the (potentially mixed) action that, after $k$ steps of reasoning, player $i$ believes their opponent will take. 
In particular, the model determines that $a_{i,k}^i$ ($a_{j,k}^i$) is a best-response to $a_{j,k-1}^i$ ($a_{i,k-1}^i$) when there is a unique best-response; when otherwise, $a_{i,k}^i$ ($a_{j,k}^i$) corresponds to a uniform distribution over the set of best-responses to $a_{j,k-1}^i$ ($a_{i,k-1}^i$). 
The actions $(a_{i,0}^i,a_{j,0}^i)$, called player $i$'s anchor, are exogenously given, with $a_{j,0}^i$ representing player $i$'s starting belief about their opponent's choices and $a_{i,0}^i$ denoting the action that player $i$ chooses when not taking any step of reasoning. 
Because our games are symmetric, we assume that $a_{i,0}^i=a_{j,0}^i\equiv a_0^i$, and that these correspond either to a pure action or to the uniform distribution over actions.

Players' decision to stop reasoning follows a cost-benefit calculation and, for this purpose, a notion of the value of reasoning is needed.
In this paper, we follow the maximum-gain representation of the value of reasoning,\footnote{
    See \citet{AlaouiPentaJPE2022} for an axiomatic characterisation.
} which is given by
\begin{align}
    V_i(k):=\max_{a_i,a_j}u_i(a_i,a_j)-u_i(a_{i,k-1}^i,a_j). \label{eq:valuereasoning}
\end{align}
Finally, the model assumes that players follow a myopic stopping rule, deciding to stop reasoning after $\hat k_i$ steps:
\begin{align*}
    \hat k_i:=\min\{k \in \mathbb N_0 \; \vert\; c_i(k)\leq V_i(k)\text{ and }c_i(k+1)>V_i(k)\},
\end{align*}
where $\hat k_i$ is player $i$'s cognitive boundary. 
Both players are aware of the other's cognitive boundary, taking that into account in their reasoning process. 
We denote by $a_i^*$ the action that player $i$ ultimately chooses, that is, $$a_i^*\equiv a_{i,\hat k_i}^i := \arg \max_{a_i \in A_i}u_i(a_i,a_{j,\hat k_j^i}),\quad \hat k_j^i:= \min\{\hat k_i-1, \hat k_j\}.$$

A first observation is that actions along any reasoning path are weakly increasing in the strategic-sophistication ranking introduced in \hyref{section:hypotheses-design:design:games}[Section]. 
In other words, by letting $a_{i,k}=a_n$ and $a_{i,k'}=a_{n'}$, then it is immediate that $k'> k \implies n'\leq n$, with strict inequality whenever $n>1$. 
Moreover, higher incentive levels are equivalent, in the model, to proportionally lower reasoning costs, so the number of reasoning steps is increasing in incentives.

Finally, note that, under identical anchors and reasoning-cost distributions across participants (potentially heterogeneous across individuals), the model predicts that the frequencies of $a_3$ and $a_4$ should be the same in the 2-Step and 3-Step games, holding fixed incentive levels. 
This is because, under the maximum-gain representation, the value of reasoning when the tentative action is $a_3$ or $a_4$ is the same in the two games:
\[\max_{a_i,a_j}u_i(a_i,a_j)-u_i(a_4,a_j)=50,\qquad\max_{a_i,a_j}u_i(a_i,a_j)-u_i(a_3,a_j)=40.\]
Hence, any given anchor and cost configuration rationalising $a_3$ or $a_4$ in the 2-Step game would also rationalise it in the 3-Step game, and vice versa. 
This prediction is clearly violated in the data (cf. \hyref{table:sample-avg}[Table]), suggesting that alternative specifications of the value of reasoning, such as the expected value of sampling information \citep{AlaouiPentaJPE2022}, may be more appropriate here.

\subsection{Sequential Sampling Equilibrium}
\label{section:appendix:theory-predictions:sse}

Sequential sampling equilibrium \citep{Goncalves2022WP} is an equilibrium solution concept that generates a joint distribution of actions, beliefs, and response times based on sequential sampling models. 
It posits that players start from a uniform prior over the opponent's equilibrium action distribution, sequentially sample from it at a fixed cost, and optimally decide when to stop by trading-off the value of additional information against the cost of further sampling. 
Upon stopping, players best respond to the posterior beliefs they hold upon stopping. 
Because the final sample is random, stopping beliefs are also random, and this generates randomness in players' chosen strategies. 
A sequential sampling equilibrium corresponds to a fixed point on the distribution of chosen strategies.

\begin{figure}[!ht]
    \centering\small\singlespacing
    \begin{subfigure}{.45\linewidth}
        \includegraphics[width=\linewidth]{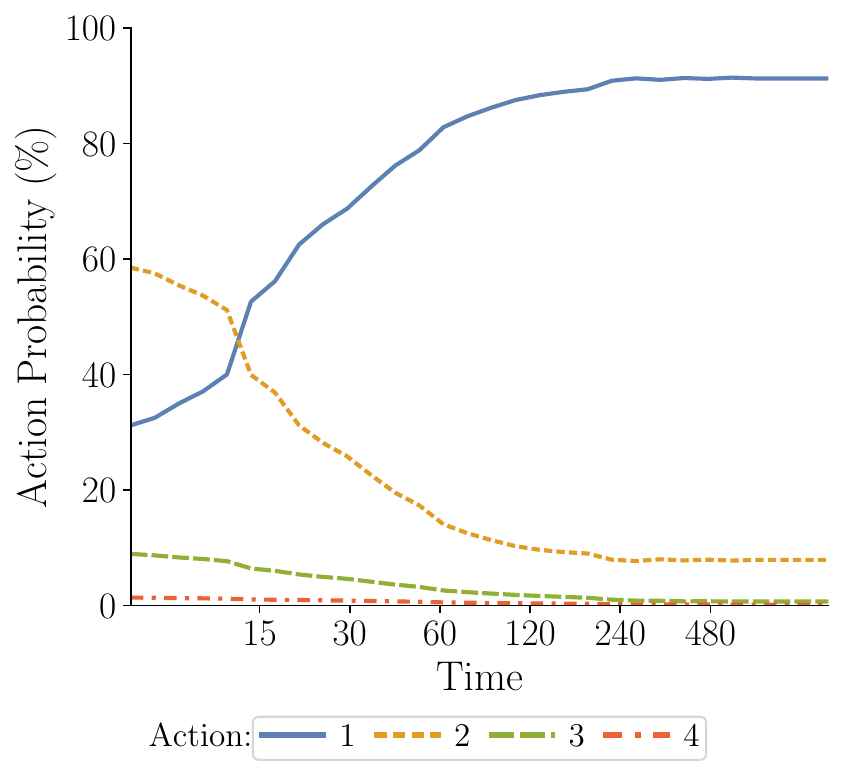}
        \caption{2-Step Game}
        \label{figure:action-rt-sse-iesds2-own01-opp01}
    \end{subfigure}
    \begin{subfigure}{.45\linewidth}
        \includegraphics[width=\linewidth]{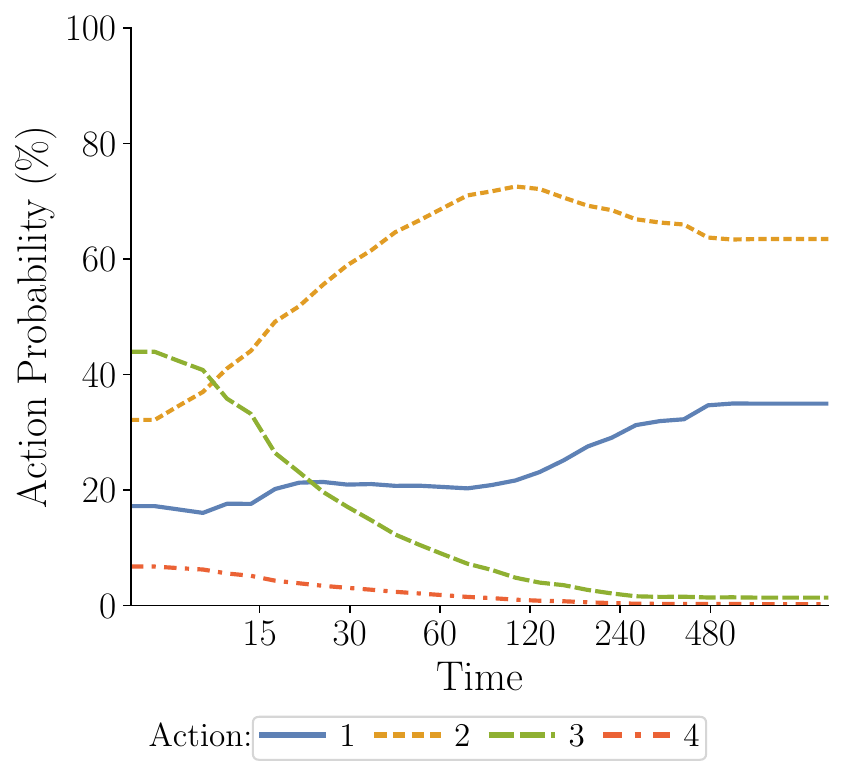}
        \caption{3-Step Game}
        \label{figure:action-rt-sse-iesds3-own01-opp01}
    \end{subfigure}
    \begin{minipage}{1\linewidth}
        \caption{Action Frequency and Response Time}
        \label{figure:action-rt-sse}
        \emph{Notes}:
        This figure shows the predicted relation between action frequency and response time as given by sequential sampling equilibrium under cost parameter $c=1/2$ and logit parameter $\lambda = 1/8$, assuming risk neutrality.
        2-Step and 3-Step games denote the different games in the experiment (see \hyref{figure:games}[Figure]).
    \end{minipage}
\end{figure}

\begin{figure}[!ht]
    \centering\small\singlespacing
    \begin{subfigure}{.45\linewidth}
        \includegraphics[width=\linewidth]{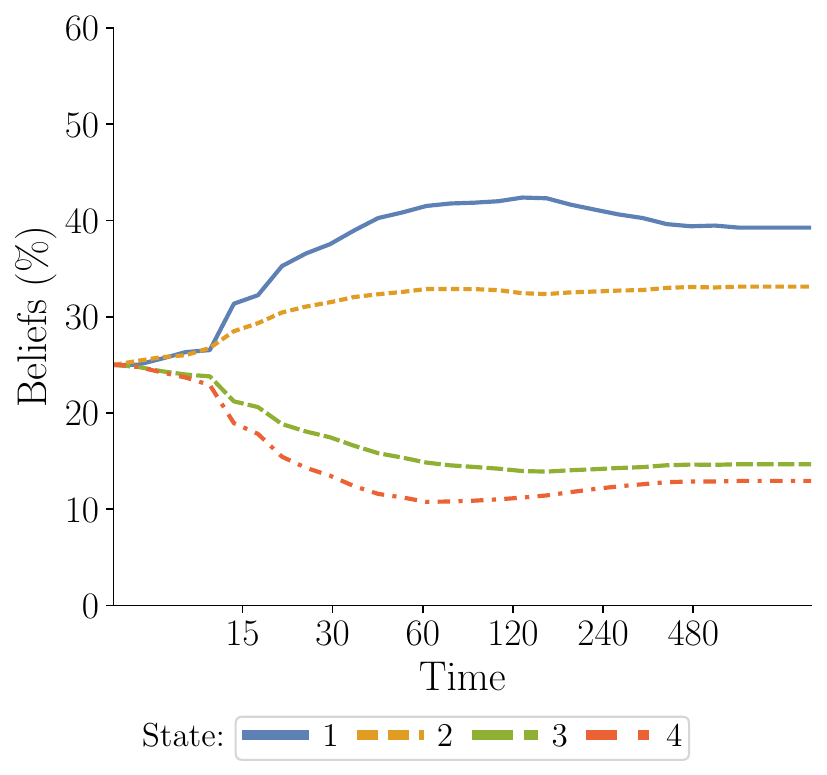}
        \caption{2-Step Game}
        \label{figure:belief-rt-sse-iesds2-own01-opp01}
    \end{subfigure}
    \begin{subfigure}{.45\linewidth}
        \includegraphics[width=\linewidth]{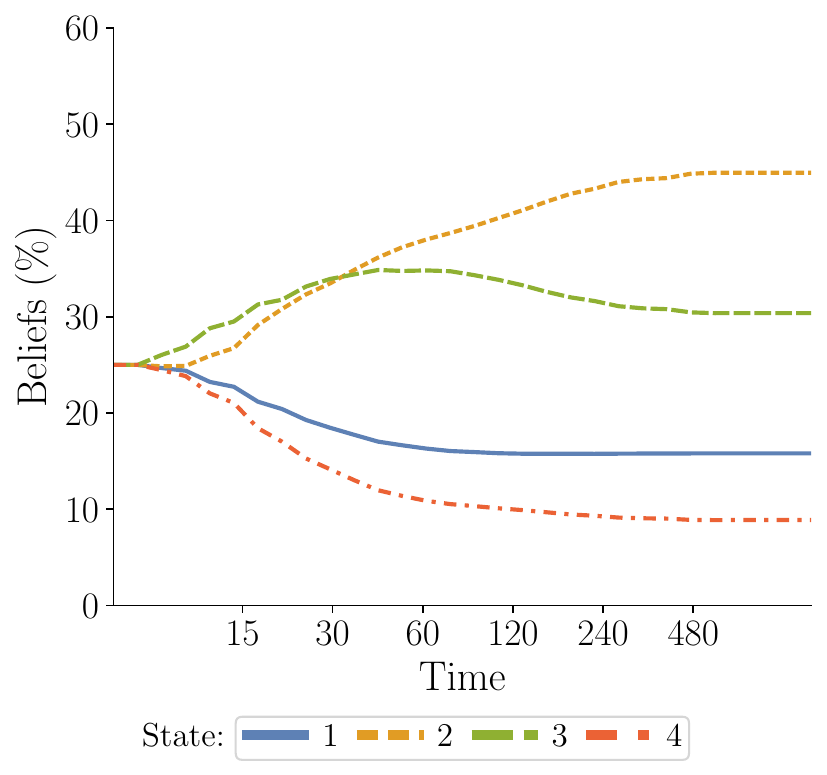}
        \caption{3-Step Game}
        \label{figure:belief-rt-sse-iesds3-own01-opp01}
    \end{subfigure}
    \begin{minipage}{1\linewidth}
        \caption{Beliefs and Response Time}
        \label{figure:belief-rt-sse}
        \emph{Notes}:
        This figure shows the predicted relation between belief distribution and response time as given by sequential sampling equilibrium under cost parameter $c=1/2$ and logit parameter $\lambda = 1/8$, assuming risk neutrality.
        2-Step and 3-Step games denote the different games in the experiment (see \hyref{figure:games}[Figure]).
    \end{minipage}
\end{figure}

\citet{Goncalves2022WP} shows that higher incentive levels --- or, equivalently, lower sampling costs --- lead to players taking longer and choosing actions with a higher level of $k$-rationalisability. 
To illustrate how it can capture other features of our data, we turn to numerical simulation. 
In \hyref{figure:action-rt-sse}[Figures] and \hyref{figure:belief-rt-sse}, we numerically solve the model allowing for logit best responses. 
The simulated patterns recover the main qualitative features of the relation between response time, choices, and beliefs in our data: the level-1 action is initially modal, but its frequency declines in favour of the level-2 action, while the frequency of strictly dominated actions falls quickly. 
Longer response times are also associated with beliefs that assign more probability to level-1 and level-2 actions and less probability to dominated play.

\FloatBarrier \newpage
~\newpage

\setcounter{page}{1}
\setcounter{section}{0}
\renewcommand{\thesection}{Online Appendix \Alph{section}}
\renewcommand{\thesubsection}{Online Appendix \Alph{section}.\arabic{subsection}}

\section{Instructions and Interface}
\label{section:o-appendix:screenshots}
Below we reproduce screenshots with the instructions, practice rounds, the only main (incentivised) round, and the final questionnaire.

\subsection{Prolific}

Below we reproduce screenshots with the instructions, practice rounds, the only main (incentivised) round, and the final questionnaire for the experiment run on Prolific.

\textbf{Instructions}\\
\includegraphics[page=1, width=.8\linewidth]{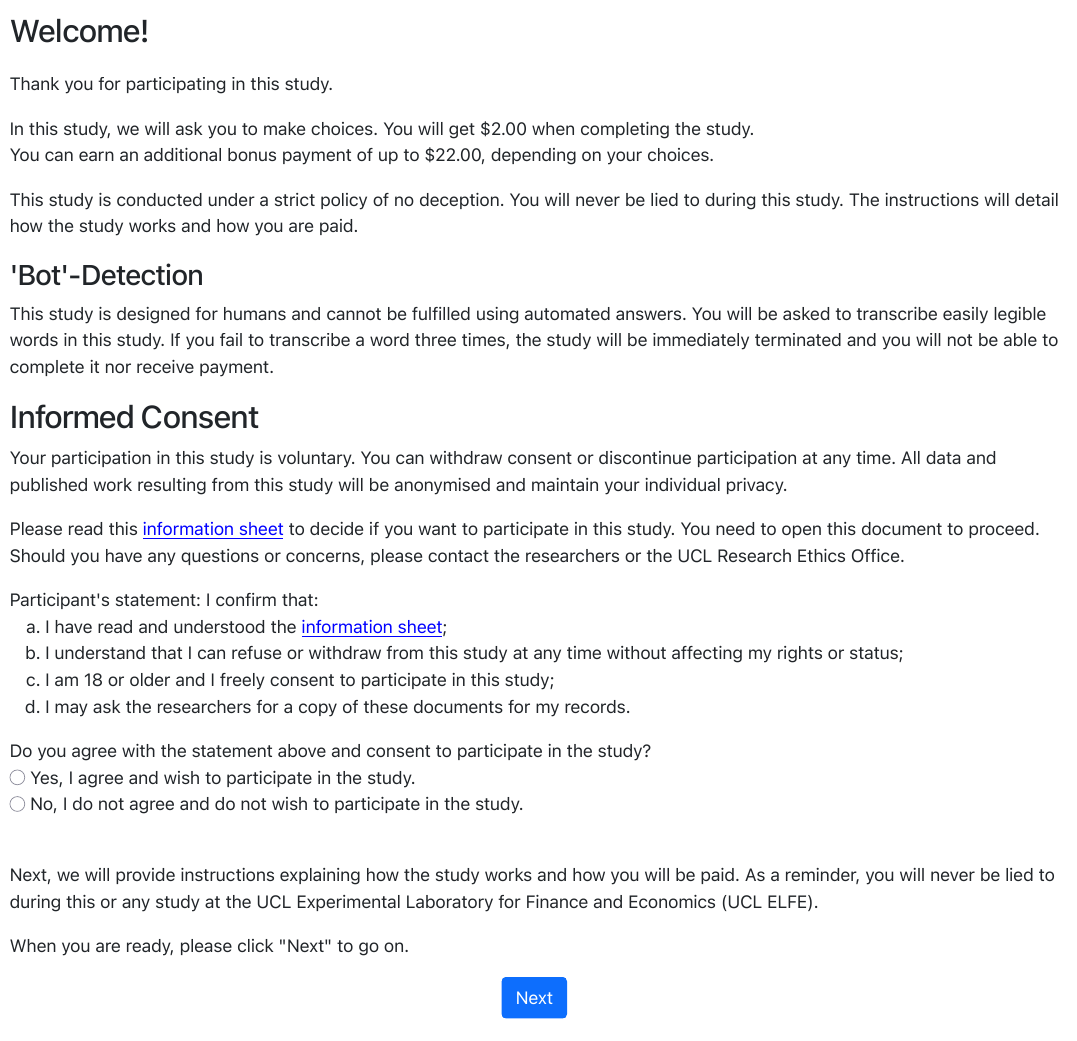}\\[3em]
\includegraphics[page=2, width=.8\linewidth]{Figures/instructions-prolific.pdf}\\[3em]
\includegraphics[page=3, width=.8\linewidth]{Figures/instructions-prolific.pdf}\\[3em]
\includegraphics[page=4, width=.8\linewidth]{Figures/instructions-prolific.pdf}

\newpage
\textbf{Captchas}\\
\includegraphics[page=5, width=.8\linewidth]{Figures/instructions-prolific.pdf}

\textbf{Practice Rounds}\\
\includegraphics[page=6, width=.8\linewidth]{Figures/instructions-prolific.pdf}\\[3em]
\includegraphics[page=7, width=.8\linewidth]{Figures/instructions-prolific.pdf}

\newpage
\textbf{Main Task}\\
\includegraphics[page=8, width=.8\linewidth]{Figures/instructions-prolific.pdf}\\[3em]
\includegraphics[page=9, width=.8\linewidth]{Figures/instructions-prolific.pdf}\\[3em]
\includegraphics[page=10, width=.8\linewidth]{Figures/instructions-prolific.pdf}

\newpage
\textbf{Questionnaire}\\
\includegraphics[page=11, width=.8\linewidth]{Figures/instructions-prolific.pdf}

\newpage

\subsection{Amazon Mechanical Turk}

Below we reproduce screenshots with the instructions, practice rounds, the only main (incentivised) round, and the final questionnaire for the experiment run on MTurk.

\textbf{Instructions}\\
\includegraphics[page=1, width=.8\linewidth]{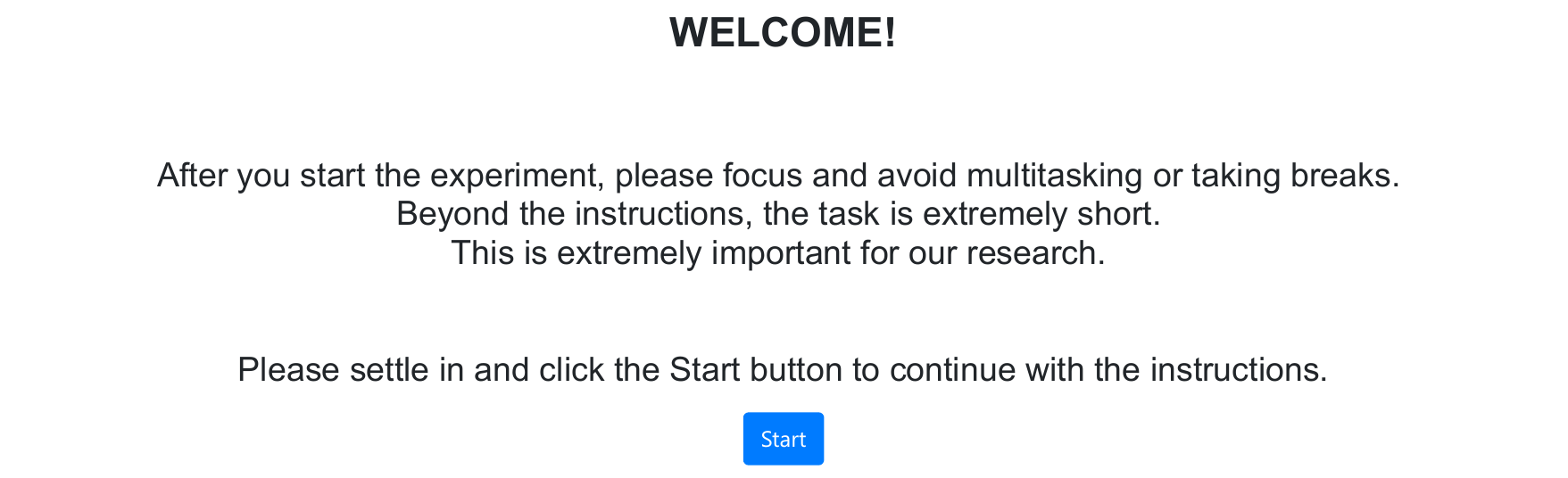}\\[3em]
\includegraphics[page=2, width=\linewidth]{Figures/instructions-mturk.pdf}\\[3em]
\includegraphics[page=3, width=\linewidth]{Figures/instructions-mturk.pdf}\\[3em]
\includegraphics[page=4, width=.85\linewidth]{Figures/instructions-mturk.pdf}\\[3em]
\includegraphics[page=5, width=.8\linewidth]{Figures/instructions-mturk.pdf}\\[3em]
\includegraphics[page=6, width=\linewidth]{Figures/instructions-mturk.pdf}\\[3em]
\includegraphics[page=7, width=\linewidth]{Figures/instructions-mturk.pdf}\\[3em]
\includegraphics[page=8, width=\linewidth]{Figures/instructions-mturk.pdf}\\[3em]
\includegraphics[page=9, width=\linewidth]{Figures/instructions-mturk.pdf}\\[3em]
\includegraphics[page=10, width=\linewidth]{Figures/instructions-mturk.pdf}\\[3em]
\includegraphics[page=11, width=\linewidth]{Figures/instructions-mturk.pdf}\\[3em]
\includegraphics[page=12, width=\linewidth]{Figures/instructions-mturk.pdf}

\newpage
\textbf{Captchas}\\
\includegraphics[page=13, width=\linewidth]{Figures/instructions-mturk.pdf}\\

\textbf{Practice Rounds}\\
\includegraphics[page=14, width=.8\linewidth]{Figures/instructions-mturk.pdf}\\
~\\[10pt]
\includegraphics[page=15, width=.8\linewidth]{Figures/instructions-mturk.pdf}\\

\newpage
\textbf{Main Task}\\
\includegraphics[page=16, width=.85\linewidth]{Figures/instructions-mturk.pdf}\\[3em]
\includegraphics[page=17, width=.85\linewidth]{Figures/instructions-mturk.pdf}\\[3em]
\includegraphics[page=18, width=.85\linewidth]{Figures/instructions-mturk.pdf}\\[3em]
\includegraphics[page=19, width=.85\linewidth]{Figures/instructions-mturk.pdf}\\[3em]

\newpage
\textbf{Questionnaire}\\
\includegraphics[page=20, width=.85\linewidth]{Figures/instructions-mturk.pdf}\\[3em]

\newpage
\textbf{Payment}\\
\includegraphics[page=21, width=.85\linewidth]{Figures/instructions-mturk.pdf}\\[3em]

\end{document}